\documentclass[manuscript,times]{aastex7}
\usepackage{amsmath}
\usepackage{appendix}

\definecolor{lightblue}{rgb}{0.2,0.4,0.8}
\definecolor{lila}{rgb}{0.5,0.2,0.9}

\begin{document}

\title{Energy Partitioning at the Termination Shock}

\author[0000-0002-9465-7470]{Judit Szente}
\affiliation{Boston University, MA, USA}
\email[show]{judithsz@bu.edu}

\author[0000-0002-9465-7470]{Bart van der Holst}
\affiliation{Boston University, MA, USA}
\email{bartvand@bu.edu}

\author[0000-0001-8459-2100]{Gabor Toth}
\affiliation{University of Michigan, Ann Arbor, MI 48109, USA}
\email{gtoth@umich.edu}

\author[0000-0002-8767-8273]{Merav Opher}
\affiliation{Boston University, MA, USA}
\email{mopher@bu.edu}


\begin{abstract}
We show new results of a global 3D magnetohydrodynamic (MHD) simulation of the
Boston University outer heliosphere model where we used a newly developed approach
that distributes the non-adiabatic shock heating among the cold protons, electrons,
and pickup-ions (PUIs), while maintaining total energy conservation of all ions. In previous simulations \citep{Bair:2025,vanderHolst:2026} non-adiabatic shock heating was distributed such that the resulting proton temperature jump at the termination shock (TS) was too high. Using a new methodology the simulation results improved with respect to the solar wind proton temperature jump observed at the TS by Voyager 2. Because we directly control the distribution of non-adiabatic heating at the TS, we have the opportunity to study the scenario of a non-adiabatic physical process heating cold plasma and PUIs in the TS and its effect on the global MHD solution. The results show that having a fraction of up to 8.5\% of the non-adiabatic shock heating going towards thermal solar wind protons at the TS reproduces the jump conditions observed along the V2 trajectory. This fraction depends on the thermal proton and PUI number densities. This information is key to understanding the physical processes that shape the heliosphere. As shown by \citet{Opher:2020}, PUIs significantly change the shape of the heliosphere; for example, the presence of hot
PUIs results in a thinner inner heliosheath. Our work provides the architecture of how
energy partitioning at shocks in kinetic simulations \citep{Giacalone:2021} can be
utilized in global MHD models.
\end{abstract}


\section{Introduction}
The interaction of the solar wind with the interstellar medium (ISM) results in a boundary, called the heliopause, that separates the solar wind from the ISM plasma. The termination shock (TS) is where the solar wind plasma slows down to the point that it becomes slower than the ambient speed of sound, resulting in a compressed region between the TS and the heliopause. The heated, compressed magnetized plasma, produced by the shock-jump, has been observed in-situ with the Voyager spacecrafts at (in the case of Voyager 1, V1 in December 2004) 94 astronomical units (AU) and (in the case of Voyager 2, V2 in August, 2007) at 84~AU \citep{Burlaga:2005Sci...309.2027B,Burlaga:2008Natur.454...75B,Burlaga:2015JPhCS.642a2003B,Richardson:2008Natur.454...63R}.

The V2 spacecraft measured multiple TS crossings; there is a consensus that at least 3 crossings have occurred, but there are suggestions of 5 or more crossings as well in the course of 2 days \citep{Decker:2008,Richardson:2008Natur.454...63R}. Possibly, multiple crossings were due to the geometric structure of the shock front being not smooth but rippled or due to short-term variability or non-stationarity of the solar wind. The shock structure has been well-observed in three crossing events, best in detail during the TS-3 event, with a thickness of 500,000~km: 320,000~km floor, 22,000~km ramp and 145,000~km overshoot. All three events showed quasi-perpendicular shocks. There was evidence of reformation during one crossing event. Non-collisional quasi-perpendicular shocks (which description is close to the termination shock at low latitudes) show temporal periodical behavior as its shock front steepens and then the ramp component rebuilds itself. Periodic behavior in the observed magnetic field by V2 and theoretical investigations - simulations confirmed that the apparent multiple crossing is due to the non-stationary nature of the TS \citep[see][and references therein]{Bale:2005SSRv..118..161B,Yang:2015}. 
Despite that V2 only measured the solar wind protons, \citet{Zieger:2015} reconstructed the possible upstream PUIs abundance and temperature based on V2 observations using a multi-fluid approach. \citet{Giacalone:2010GeoRL..3719104G, Giacalone:2021,Giacalone:2025ApJ...980...29G} studied the TS structure using 2D hybrid methods, while \citet{Swisdak:2023} also used a kinetic approach, reconstructing the solar wind plasma and PUIs energetics at the crossing. New Horizons (NH) will cross the shock and provide in-situ measurements of both the cold protons and the population of pickup ions (PUIs). Predictions for the NH crossing the termination shock and whether or not NH will encounter multiple TS crossings are shown in \citet{Gasser:2026}.

Here we use a multi-ion treatment where the solar wind plasma is treated in two components capturing the cold component as well as the PUIs, while the neutral H atoms are described as 4 populations of neutral H (as in \citet{Opher:2020}). Although this outer heliosphere model can model for kinetic neutrals \citep{2024ApJ...973...46C}, we use a fluid description for the neutral atoms for convenience despite that both the neutral-ion and neutral–neutral collisional mean free paths are either comparable to or larger than the size of the heliosphere, resulting in differences between kinetic and fluid modeling of the neutral atoms, see for instance \citet{2005A&A...439.1171A,2011ApJ...734...45A,2006JGRA..111.6110H,2026MNRAS.548ag664K}. We also include electrons with a separate pressure equation (as in \citet{Bair:2025}) and turbulence transport similar to \citet{vanderHolst:2026}. The turbulence energy is enhanced via the isotropization of newly generated PUIs. This increased turbulence can further heat the cold thermal protons and electrons.

In magnetohydrodynamic (MHD) shocks, the plasma is heated at the shock while complying with the conservation laws for energy, momentum, mass density, and the magnetic flux (the Rankine-Hugoniot relations). This results in more heating than would be described by the adiabatic compression of the plasma when crossing the shock front - this term we call non-adiabatic heating throughout this paper. One approach to address shock heating is with kinetic or hybrid models, see \citep{Giacalone:2021, Bera:2023}, but we utilize the solution presented by \citet{Toth:2024}: using a linear combination of volumetric entropy densities to distribute the non-adiabatic shock heating among the various pressure components in a deterministic manner, while maintaining total energy conservation. This provides a tool for studying shock heating of the multi-fluid plasma in an MHD setting; see Section~\ref{sec:method}. Physically, this means that we can control in the simulation how to distribute the non-adiabatic heating among plasma species in the multi-fluid MHD simulation. 

The multi-fluid simulation of \citet{Opher:2020} showed a thermal proton temperature jump at the termination shock that was a factor 4 to 5 too large, while the multi-fluid simulation of \citet{vanderHolst:2026} showed a temperature jump that was a factor 10 too large. Both simulations used hydrodynamic energy equations for the individual ion fluids, so there was non-adiabatic heating for both fluids, but without control of distribution of this heating component. \citet{Bera:2023} uses a multi-fluid model, in which the electrons and cold protons are described by a single temperature and the PUIs with a separate temperature but the same bulk speed as the thermal wind protons. The boundary conditions for the shock jump were obtained from kinetic simulations of collisionless shocks. In their work, the temperature jump obtained for the cold protons was between a factor of 20 and 40 (depending on the shock angle, Mach number, and boundary conditions for the PUIs), which is a factor of 2 higher than observed by V2. In our study, we used the multi-fluid MHD model developed in \citet{Opher:2020,Bair:2025,vanderHolst:2026} together with the newly developed approach that distributes the non-adiabatic shock heating among the cold protons, electrons and PUIs, while rigorously maintaining total energy conservation of all ions (thermal and PUIs) and electrons. At the TS, we partition the non-adiabatic shock heating among the PUIs and protons, whereas electrons receive only adiabatic heating at the shock transition. Using this new methodology, we aim to meet the observed temperature jump of a factor 10--20 in the cold proton temperature V2 measurements.

We use as in \citet{Opher:2020, Bair:2025, vanderHolst:2026} a steady state uniform solar wind solution. We provide a prediction for shock jump conditions for the complete angular distribution, including NH, while validating our model with the V2 crossing data. The energy partitioning at multi-fluid shocks is discussed in Section~\ref{sec:method}. Section~\ref{sec:model} describes the setup of the outer heliosphere model. The results of the simulations with and without the new shock conditions at the TS are presented in Section~\ref{sec:results}. We summarize our work in Section~\ref{sec:summary}

\section{Methodology}\label{sec:method}
In this section, we develop a generalization of the distribution of non-adiabatic shock heating among various pressure components \citep{Toth:2024} to multi-fluid plasma. We first present the total energy conservation needed to determine the total shock heating in Section \ref{sec:energy}. In Section \ref{sec:entropy}, we show using linear combinations of ion and electron entropies to channel non-adiabatic heating to the various species (in our case 2: solar wind cold protons and PUIs).

\subsection{Total energy conservation}\label{sec:energy}
The outer heliosphere model we use in this paper is based on multi-ion magnetohydrodynamics (MHD) described in \citet{Opher:2020}, where thermal solar wind protons and PUIs are separate fluids, interacting through charge exchange with neutral hydrogen atoms. \citet{Bair:2025} generalized this model to include the electron pressure, then \citet{vanderHolst:2026} incorporated incompressible turbulence. Plasma heating due to incompressible turbulence is used in a way similar to the \citet{Usmanov:2016} model; see the implementation and details in \citet{vanderHolst:2026}. These previous models did not yet address the treatment of non-adiabatic shock heating, meaning how non-adiabatic processes increase the volumetric entropy densities of the various species at shocks: In the simulations of \citet{Opher:2020,Bair:2025,vanderHolst:2026} too much non-adiabatic shock heating was channeled to the thermal protons, as dictated by the numerical scheme. In the present work, we use a general method that allows us to distribute the non-adiabatic shock heating among thermal protons, PUIs, and electrons while maintaining total energy conservation. While the PUIs velocity distribution function is strongly non-Maxwellian \citep{Malama:2006A&A...445..693M,Zhao:2019ApJ...879...32Z, Zirnstein:2022SSRv..218...28Z,Livadiotis:2024ApJ...968...66L}, unlike \citet{vanderHolst:2026} we follow a fluid-treatment and Maxwellian approximation.

The multi-fluid MHD equations for thermal solar wind protons, PUIs, and electrons are
\begin{equation}
  \frac{\partial \rho_i}{\partial t} + \nabla\cdot(\rho_i{\bf u}_i) = S_{\rho_i},\label{eq:rho}
\end{equation}
\begin{align}
  \frac{\partial \rho_1{\bf u}_1}{\partial t}
  &+ \nabla\cdot(\rho_1{\bf u}_1{\bf u}_1)
  + \nabla p_1 + \frac{q_1n_1}{en_e}\left[ \nabla p_e
  - \frac{1}{\mu_0}(\nabla\times{\bf B})\times{\bf B} \right] \nonumber \\
  &= q_1 n_1({\bf u}_1 -{\bf u}_+)\times{\bf B} 
  - \frac{\rho_1}{\rho}\nabla\cdot{\bf P}_w
  + {\bf F}_{\rm friction} + {\bf S}_{\rho_1{\bf u}_1}, \label{eq:momentum1}
\end{align}
\begin{align}
  \frac{\partial \rho_2{\bf u}_2}{\partial t}
  &+ \nabla\cdot(\rho_2{\bf u}_2{\bf u}_2)
  + \nabla p_2 + \frac{q_2n_2}{en_e}\left[ \nabla p_e
  - \frac{1}{\mu_0}(\nabla\times{\bf B})\times{\bf B} \right] \nonumber \\
  &= q_2 n_2({\bf u}_2 -{\bf u}_+)\times{\bf B} 
  - \frac{\rho_2}{\rho}\nabla\cdot{\bf P}_w
  - {\bf F}_{\rm friction} + {\bf S}_{\rho_2{\bf u}_2}, \label{eq:momentum2}
\end{align}
\begin{equation}
  \frac{\partial p_1}{\partial t}
  + \nabla\cdot\left( p_1 {\bf u}_1\right)
  + (\gamma_1-1)p_1\nabla\cdot{\bf u}_1=S_{p_1} + (\gamma_1-1)Q_1,\label{eq:swpressure}
\end{equation}
\begin{equation}
  \frac{\partial p_2}{\partial t}
  + \nabla\cdot\left( p_2 {\bf u}_2\right)
  + (\gamma_2-1)p_2\nabla\cdot{\bf u}_2=S_{p_2} - (\gamma_2-1)2S_w,\label{eq:puipressure}
\end{equation}
\begin{equation}
  \frac{\partial p_e}{\partial t}
  + \nabla\cdot\left( p_e{\bf u}_e\right)
  + (\gamma_e-1)p_e\nabla\cdot{\bf u}_e=S_{p_e} +(\gamma_e-1)Q_e,\label{eq:electronpressure}
\end{equation}
and for the magnetic field are
\begin{equation}
  \frac{\partial {\bf B}}{\partial t} - \nabla \times ({\bf u}_e \times {\bf B}) =  0, \qquad \nabla\cdot{\bf B}=0,\label{eq:magneticfield}
\end{equation}
where $i$ indexes the ions (1 for cold protons and 2 for PUIs), $\rho_i$, ${\bf u}_i$, and $p_i$ are the mass density, bulk velocity, and pressure for ion $i$, $p_e$ is the electron pressure, and ${\bf B}$ is the magnetic field, $\rho=\rho_1+\rho_2$ is the total mass density. The adiabatic index for both ions and electrons are $\gamma_i=\gamma_e=5/3$. The electron number density can be obtained from charge neutrality
\begin{equation}
    n_e = \sum_i Z_in_i,
\end{equation}
where $n_i=\rho_i/m_i$ is the ion number density, $m_i$ is the ion (in our case proton) mass, and $Z_i = q_i/e$ is the ion charge in units of elementary charge $e$. The charge weighted average ion velocity is then defined as
\begin{equation}
  {\bf u}_+ = \frac{1}{n_e}\sum_i Z_i n_i{\bf u}_i.
\end{equation}
The electron velocity is
\begin{equation}
    {\bf u}_e = {\bf u}_+ - \frac{{\bf j}}{en_e},
\end{equation}
where ${\bf j} = (\nabla\times{\bf B})/\mu_0$ is the current density and $\mu_0$ is the vacuum permeability. The temperatures $T_1, T_2, T_e$ for the solar wind protons, PUIs, and electrons, respectively, are determined by the equations of state $p_i=n_ik_{\rm B}T_i$ and $p_e=n_ek_{\rm B}T_e$, where $k_{\rm B}$ is the Boltzmann constant. In this work we do not focus on the heliosheath and due to the unipolar field, we do not have a heliospheric current sheet. This enables us to neglect the Hall term, resulting in ${\bf u}_e = {\bf u}_+$. The source terms in order are $S_{\rho_i}$, ${\bf S}_{\rho_i{\bf u}_i}$, $S_{p_i}$, and $S_{p_e}$ due to charge exchange, and we use an artificial friction force
\begin{equation}
   {\bf F}_{\rm friction} = \frac{\rho_2}{\tau_c}({\bf u}_2 - {\bf u}_1)\left( \frac{|{\bf u}_2 - {\bf u}_1|}{V_A} \right)^{\alpha_c}, 
\end{equation}
to limit the differential streaming (velocity difference) among the ions to the local Alfv\'en speed ${\bf V}_A={\bf B}/\sqrt{\mu_0\rho}$. We use $\tau_c=10^6\,s$ and $\alpha_c=4$ similar to \citet{Opher:2020}. In addition, we use turbulent pressure force and coronal heating due to turbulent dissipation, which source terms are described in more detail in \citet{vanderHolst:2026}. The turbulence pressure tensor ${\bf P}_w$ and the sources of plasma heating, $Q_1$ and $Q_e$, due to turbulent dissipation, are obtained from the incompressible turbulence equations that describe the advection, compression, reflection, and dissipation of the turbulence energy densities $w_\pm$, where $+$ is for the energy density propagating in the direction of the magnetic field and $-$ for the energy density propagation in the opposite direction. The equations presented here are the same as in \citet{Usmanov:2014,vanderHolst:2026}:
\begin{equation}
    \frac{\partial w_\pm}{\partial t} + \nabla\cdot\left[ w_\pm (
    {\bf u}\pm{\bf V}_{\rm A})\right] + \frac{1}{2}w_\pm\nabla\cdot{\bf u}
    = - (S \pm R)\frac{w_D}{2} -\Gamma_\pm w_\pm + S_w. \label{eq:icturbulence}
\end{equation}
The first two terms on the left represent the propagation of turbulence energy along the field line. The third term is the energy reduction in expanding flows. For transverse turbulence the shear flow and turbulence reflection contributions are, respectively,
\begin{equation}
    S=\frac{1}{2}\nabla\cdot{\bf u} - {\bf b}\cdot({\bf b}\cdot\nabla){\bf u},
    \qquad R = ({\bf b}\cdot\nabla)V_A.
\end{equation}
The residual energy density $w_D$ is, similar to \citet{Usmanov:2011}, assumed to be proportional to the total turbulent energy density
\begin{equation}
    w_D = \sigma_D w, \qquad w = w_+ + w_-,
\end{equation}
in which $\sigma_D=-1/3$. The second term on the right-hand side of Equation \ref{eq:icturbulence} is the dissipation of turbulence with cascade rate
\begin{equation}
    \Gamma_\pm = \frac{2\alpha}{L_\perp\sqrt{\rho}}\sqrt{w_\mp}.
\end{equation}
We use the same value $\alpha=0.128$ as \cite{Usmanov:2018}. The sum of the wave dissipation terms is channeled to the heating of solar wind protons and electrons
\begin{equation}
     \Gamma_+w_+ + \Gamma_-w_- = Q_1 + Q_e.
\end{equation}
We use the same energy partitioning as \citet{Usmanov:2014,vanderHolst:2026} with 60\% of wave dissipation channeled to thermal solar wind protons and the remainder to electrons. The time evolution of the correlation length $L_\perp$
\begin{equation}
  \frac{\partial \rho L_\perp}{\partial t} + \nabla\cdot(\rho L_\perp {\bf u}_) =
  \frac{\beta}{\alpha}\frac{\rho L_\perp}{w}(\Gamma_-w_-+\Gamma_+w_+ -2S_w) + L_\perp (S_{\rho_{\rm sw}} + S_{\rho_{\rm pui}})
    \label{eq:lperp}
\end{equation}
has a source term
\begin{equation}\label{eq:turbsource}
    S_w = \frac{1}{4}f_DV_A\frac{\delta\rho_{pui}}{\delta t}\Delta u,
\end{equation}
that also appears in the last term on the right-hand side of Equation \ref{eq:icturbulence} and as a sink on the right-hand side of the pui pressure equation \ref{eq:puipressure}. This source term is due to pitch-angle scattering by wave–particle interaction of newly born PUIs that may generate turbulence \citep{Williams1995,Zank:1996,Isenberg2003}.  The turbulence pressure is
\begin{equation}
    {\bf P}_{\rm w} = p_{{\rm w}\perp} I + (p_{{\rm w}\parallel}-p_{{\rm w}\perp}) {\bf b}{\bf b},
\end{equation}
where
$p_{{\rm w}\perp} = \frac{1}{2}(w + w_D)$ and $p_{{\rm w}\parallel} = \frac{1}{2}(w - w_D)$ for transverse turbulence. 

We define the total plasma energy density as
\begin{equation}
  E = \sum_i \left( \frac{\rho_iu_i^2}{2} + \frac{p_i}{\gamma_i -1} \right) + \frac{p_e}{\gamma_e-1} + \frac{B^2}{2\mu_0}.
\end{equation}
Then we obtain from the equations for ions, electrons, and magnetic field the near conservation form for the total plasma energy density (which does not include the energy density of turbulence and neutral atoms to obtain a fully conservative energy equation)
\begin{align}
    \frac{\partial E}{\partial t}
  &+ \nabla\cdot\left[\sum_i\left( \frac{\rho_iu_i^2}{2} + \frac{\gamma_i p_i}{\gamma_i -1}\right){\bf u}_i
  +\left(\frac{\gamma_e p_e}{\gamma_e-1}  + \frac{B^2}{\mu_0}\right) {\bf u}_e  - \frac{{\bf u}_e\cdot{\bf B}}{\mu_0}{\bf B} + {\bf P}_w\cdot{\bf u}\right] \nonumber \\
  &= {\bf P}_{\rm w}:\nabla{\bf u} + Q_1 + Q_e -2S_w + S_E -{\bf u}_e\cdot{\bf B}\nabla\cdot{\bf B}.
  \label{eq:totalenergy}
\end{align}
Here,
\begin{equation}
    S_E = \sum_i \left[ \frac{S_{p_i}}{\gamma_i-1} + {\bf u}_i\cdot{\bf S}_{\rho_i{\bf u}_i}-\frac{u_i^2}{2}S_{\rho_i} \right] + \frac{S_{p_e}}{\gamma_e-1},
\end{equation}
is the charge exchange source term for the total energy density equation; and ${\bf u} = \sum_i\rho_i{\bf u}_i/\rho$ is the center of mass ion velocity which is the same as the ${\bf u}_+$ for proton only ions. The first term of the right hand side of the total plasma energy equation \ref{eq:totalenergy} is at the discretized level the same as the last term on the left hand side and the first term on the right hand side of Equation \ref{eq:icturbulence} to ensure total energy conservation.

In this paper, we solve Equations \ref{eq:rho}, \ref{eq:momentum1}, and \ref{eq:momentum2} for the individual ion mass density and momentum (we do not solve for the sum mass density and momentum), Equation \ref{eq:magneticfield} for the magnetic field, Equations \ref{eq:icturbulence} and \ref{eq:lperp} for the incompressible turbulence equations, the hydrodynamic fluid equations for neutral atoms as discussed in \citet{Opher:2020}, Equation \ref{eq:totalenergy} for the total plasma energy density, and the equations for volumetric entropy density to be discussed in Section \ref{sec:entropy}. We solve these equations with the HLLE numerical scheme as described in \cite[][section 2.3.8]{Toth:2012}. We apply the generalized Koren slope limiter \citep{Koren:1993,Toth:2008} to the primitive variables $(\rho_i,{\bf u}_i,p_i,p_e,{\bf B},w_+,w_-,\rho L_\perp)$ and for the primitive variables of the neutral fluids $(\rho_n,{\bf u}_n, p_n)$. Note that the momentum and turbulence equations \ref{eq:momentum1}, \ref{eq:momentum2} and \ref{eq:icturbulence} are not in conservative form. In the momentum equations, $\nabla \cdot P_w$ is evaluated at the cell center from face-reconstructed $P_w$ and multiplied by the cell-centered $\rho_i/\rho$. Similarly, $\nabla p_e$ and $\nabla\times{\bf B}$ are cell-centered terms built from face-reconstructed values. The quantities $q_i n_i/(e n_e)$ and $\times{\bf B}$ are cell-centered. In the turbulence equation \ref{eq:icturbulence}, the divergence of the bulk velocity is discretized with the face reconstructed values of the bulk velocity consistent and multiplied by the cell-centered $w_\pm$. The face reconstructed values are consistent with the HLL scheme. We acknowledge that this particular discretization is not based on the Dal Maso-LeFloch-Murat theory \citep{DalMaso:1995}, so the jump conditions are not necessarily well-behaved. The goal of this paper is to exploit the total plasma energy conservation in combination with volumetric entropy density equations to arrive at a partitioning of the non-adiabatic shock heating among the various species of ions and electrons. This is the topic of the next subsection.

\subsection{Linear combination of entropy densities}\label{sec:entropy}

Following \citet{Toth:2024}, individual ion- and electron energies will be determined through linear combinations of volumetric entropy densities. We will use linear combinations with the first ion volumetric entropy density (cold protons) in all equations. To simplify the notation, we will use the index $N_i+1$ for the electron fluid. The volumetric entropy densities and their linear combinations can be written as
\begin{align}
    s_i &= \frac{p_i}{n_i^{\gamma_i-1}}, \qquad \qquad \text{for}\quad i=1,\ldots, N_i+1,\\
    s_{1i} &= W_i s_1 - (1-W_i)s_i,\qquad \text{for}\quad i=2,\ldots, N_i+1, \label{eq:entropydiff}
\end{align}
and their corresponding time evolution equations are
\begin{equation}
    \frac{\partial s_i}{\partial t} + \nabla\cdot({s_i \mathbf u}_i) = S_{s_i}, \label{eq:entropy}
\end{equation}
where
\begin{align}
    S_{s_1} &= \frac{1}{\rho_1^{\gamma_1-1}}
        \left[S_{p_1}+(\gamma_1-1)Q_1\right]-(\gamma_1-1)\frac{p_1}{\rho^{\gamma_1}}S_{\rho_1}, \\
    S_{s_2} &= \frac{1}{\rho_2^{\gamma_2-1}}
        \left[S_{p_2} -(\gamma_2-1)2S_w\right]-(\gamma_2-1)\frac{p_2}{\rho^{\gamma_2}}S_{\rho_2}, \\
    S_{s_e} &= \frac{1}{\rho_e^{\gamma_e-1}}
        S_{p_e}-(\gamma_e-1)\frac{p_e}{\rho^{\gamma_e}}S_{\rho_e}, 
\end{align}
are the entropy source terms for the solar wind protons, PUIs, and electrons, respectively, while the difference volumetric entropy densities are determined from these volumetric entropy density equations. Since all the individual entropy equations are in conservation form (except for source terms that do not contain derivatives) their linear combination is also well behaved, i.e. converge to a weak solution. A linear combination of conservation laws is a conservation law. This is true both at the analytic and discrete level. Conservation laws provide a converged weak solution across discontinuities. We note that the ion momentum equations \ref{eq:momentum1}$-$\ref{eq:momentum2} and the wave energy equations \ref{eq:icturbulence} contain some non-conservative terms with spatial derivatives, which means that for this complex system of equations the entropy-based method is still not guaranteed to converge to a weak solution. The weights $W_i$ for $i=2,\ldots,N_i+1$ are in the interval $[0,1)$. These equations can be rewritten in terms of the thermal energy densities $e_i=p_i/(\gamma_i-1)$ ($i=1,\ldots,N_i+1)$:
\begin{equation}
    s_{1i} = W_i c_1 e_1 - (1-W_i) c_i e_i,\qquad \text{for}\quad i=2,\ldots, N_i+1, \label{eq:s1i}
\end{equation}
where the factors $c_i$ are
\begin{equation}
   c_i = \frac{\gamma_i-1}{n_i^{\gamma_i-1}} \qquad
   \text{for}\quad i=1,\ldots, N_i+1.
\end{equation}
An additional equation requires that the sum of 
the thermal energies satisfies the total energy
equation:
\begin{equation}
    e_t := E - \sum_{i=1}^{N_i} \frac12\rho_i u_i^2 - \frac{B^2}{2\mu_0} = \sum_{i=1}^{N_i+1} e_i.  \label{eq:et}
\end{equation}
To express $e_i$ ($i=2,\ldots, N_i+1$) from $e_1$ using equation (\ref{eq:s1i}):
\begin{equation}
   e_i = \alpha_i e_1 - \beta_i \qquad\text{for}\quad i=2,\ldots, N_i+1, 
   \label{eq:ei}
\end{equation}
where we defined
\begin{equation}
   \alpha_i:=\frac{W_i c_1}{(1-W_i) c_i} 
   \qquad\text{and}\qquad
   \beta_i := \frac{s_{1i}}{(1 - W_i) c_i}.
\end{equation}
Next, substitute (\ref{eq:ei}) into the energy equation (\ref{eq:et})
\begin{equation}
  e_t = e_1 + \sum_{i=2}^{N_i+1} (\alpha_i e_1 - \beta_i),
\end{equation}
then solve for 
\begin{equation}
  e_1 = \frac{e_t + \sum_i \beta_i}
              {1 + \sum_i \alpha_i}, 
\end{equation}
and obtain the other energy densities from (\ref{eq:ei}).

Fixed weights can be used if the relative number density $n_i/n_1$ of the different ion fluids does not vary much. In general, however, one needs to adjust the weights to take into account the mass density differences. Since the volumetric entropy densities $s_i$ scale with $p_i$ and $n_i^{1-\gamma_i}$, the appropriate scaling in terms of the number density is
\begin{equation}
   W'_i = \frac{n_i^{2-\gamma_i}W_i}{n_i^{2-\gamma_i}W_i 
         + n_1^{2-\gamma_1}(1-W_i)}.
\end{equation}
For $\gamma_i<2$, which is the typical case (in ours 5/3), if $n_i/n_1$ is large, then $W'_i\approx 1$ and all heating goes into the $i$-th fluid, while for $n_i/n_1=0$ we get $W'_i=0$ and no heating goes into the $i$-th fluid, as expected. For $n_i=n_1$, the weights do not change: $W'_i=W_i$.

Finally, we consider an idealized case where the two fluids have the same ion mass, velocity, and temperature and $n_i/n_1$ is constant across the shock wave, so the separation into two fluids is completely artificial, testing whether they should behave the same as a single fluid. Setting $W_i=(1-W_i)=1/2$ means that we indeed want the non-adiabatic heating to be split proportionally between the two fluids for any $n_i/n_1$ ratio. In this case $W'_i/(1-W'_i)$ is proportional to $n_i^{2-\gamma_i}/n_1^{2-\gamma_1}$ similar to the ratio of volumetric entropy densities $s_i/s_1 = n_i^{2-\gamma_i}/n_1^{2-\gamma_1}$. Then the linear combination $s_{1i} = W'_i s_1 - (1-W'_i) s_i=0$ across the shock wave, which means that the non-adiabatic heating is split between the two fluids such that the proportionality of densities and pressures is maintained.

This energy partitioning among the various species is performed after the update with the fluxes and source terms of the equations described in the previous subsection. This is performed for each computational grid cell and does not modify the total energy density in each grid cell. We note that we do not update with a time-evolution equation based on the linear combination of volumetric entropy densities \ref{eq:entropydiff}, but instead solve for the volumetric entropy density equations \ref{eq:entropy}. Hence, we do not need to introduce source terms related to the spatial and temporal variation of $W_i'$. The weights $W_i'$ are used only in the subsequent algebraic energy-partitioning step, after the conservative entropy update.

The time evolution of the total plasma energy density equation \ref{eq:totalenergy} provides, when discretized, a change in the energy density $\Delta E$. The discretized version of the equations for the volumetric entropy densities \ref{eq:entropy} provide an adiabatic change in the thermal energy densities $\Delta_s e_i$. The change in the total thermal energy density, based on the update of the total plasma energy density, can then be written as
\begin{equation}
    \Delta e_t = \sum_{i=1}^{N_i} \Delta_s e_i + \Delta_s e_e + \Delta e_n,
\end{equation}
where $\Delta e_n$ is the non-adiabatic heating. For grid-resolved smooth solutions $\Delta e_n=0$, while for shocks $\Delta e_n\neq0$. Specializing the energy partitioning for our outer heliosphere equations with two ions, thermal protons and PUIs, setting $\gamma_1=\gamma_2=\gamma_e=5/3$, and channeling the non-adiabatic heating $\Delta e_n$ only to the ions, we obtain after some algebra the update for the ion thermal energy densities
\begin{align}
    \Delta e_1 &= \Delta_s e_1 + f_1 \Delta e_n, \label{eq:nonadiabsw} \\
    \Delta e_2 &= \Delta_s e_2 + f_2 \Delta e_n,
\end{align}
where the fractions
\begin{align}\label{eq:f}
    f_1 &= \frac{(1-W_2')c_2}{(1-W_2')c_2 + W_2'c_1} = \frac{1-W_2}{1-W_2 + W_2\frac{n_2}{n_1}}, \\
    f_2 &= \frac{W_2'c_1}{(1-W_2')c_2 + W_2'c_1} = \frac{W_2\frac{n_2}{n_1}}{1-W_2 + W_2\frac{n_2}{n_1}},
\end{align}
set the partitioning of non-adiabatic heating among the thermal protons and PUIs, respectively, shown in Figure~\ref{fig:f1}.

In our simulation, we used $W_2=0.984$, which corresponds to most of the heating going to the PUIs. This weight was obtained via a one-dimensional parameter study such that the model would recover the solar wind proton temperature jump as observed by Voyager 2. The exact percentage of transfer of non-adiabatic shock heating to PUIs will depend on the number densities, since the weights in the energy partitioning contain the ratio of PUI to thermal proton number density. For shocks near 1 AU, most of the non-adiabatic shock heating will, for instance, be transferred to the thermal protons instead of PUIs, due to the small PUI number density, while at the termination shock most of the non-adiabatic shock heating is transferred to the PUIs. In this paper, we do not channel non-adiabatic shock heating to the electrons, since (1) we have no measurement of the electron temperature and (2) we expect that most of the energy is channeled to the more energetic PUIs \citep{Zank:2010ApJ...708.1092Z}.

\section{Model Setup}\label{sec:model}
For the simulations, we used the Space Weather Modeling Framework's \citep[SWMF][]{Toth:2012} Outer Heliosphere multi-ion MHD model. The version number of SWMF/BATS-R-US \citep{Toth:2012} is "2025-03-27 87e38f6"; share is "2025-03-26 3e3b77e", srcBATL is "2025-03-24 61352b3", util is "2025-03-27 c798007" which we share for the sake of ensuring reproducibility. The Outer-Heliosphere model was first developed by \citet{Opher:2003ApJ...591L..61O}, further modified in \citet{Opher:2009Natur.462.1036O} to include multi-fluid neutrals, and extended to include multi-ion treatment in \citet{Opher:2020}. The cartesian grid extends from -1500 to 1500~AU in the X direction, and from -2000 to 2000~AU in the Y and Z directions. Z is the axis of the solar rotation. We use adaptive mesh refinement with self-similar grid blocks consisting of $8\times8\times8$ internal mesh cells. The smallest cell size is $0.0916$~AU by 0.122~AU by 0.122~AU near the inner boundary at 1~AU, while 46.9~AU by 62.5~AU by 62.5~AU in the very local interstellar medium (VLISM). At the termination shock, the grid resolution is 1~AU by 1.33~AU by 1.33~AU. In the inner 70~AU radius we used a non-conservative scheme, which means that we update using the individual pressure equations \ref{eq:swpressure}, \ref{eq:puipressure}, and \ref{eq:electronpressure} instead of the total plasma energy density and the volumetric entropy density equations \ref{eq:entropy} to reduce the numerical diffusion to better capture the steep gradients near the inner boundary, while in the rest of the domain the numerical scheme is conservative to capture the TS and to account for the redistribution of non-adiabatic energy at the termination shock. The model solves for thermal solar wind protons, pickup ions, electrons, and four neutral hydrogen populations following the numbering of \citet{Opher:2020}: (1) neutrals originating from the region between the bow wave and the heliopause, (2) neutrals created at the inner heliosheath, (3) neutrals created in the supersonic solar wind and (4) neutrals introduced from the interstellar medium. The inner boundary is at 1~AU, representing the average solar wind conditions: the solar wind is assumed to have a number density of 11~cm$^{-3}$, a proton temperature of 60,000~K and an electron temperature of 30,000~K. These inner boundary temperatures were taken from \citet{Adhikari2023}. The solar wind has a purely radial direction with a flow speed of 417.07~km/s. The magnetic field is a Parker spiral \citep{Parker1958} field with a magnitude such that at 1\,au the radial component is $B_r = 6.453$\,nT and the azimuthal component $B_\varphi=6.474$\,nT at the equator. To minimize numerical reconnection effects, this magnetic field is unipolar, so no heliospheric current is present in the supersonic solar wind \citep{Opher:2015ApJ...800L..28O}. The outer boundary is the VLISM, with a prescription based on case B of \citet{Opher:2020}. The plasma number density is 0.06~cm$^{-3}$, the proton and electron temperatures are both 6519~K. The interstellar wind's (X,Y,Z) velocity components at the outer boundaries are (26.3, 0.3, -2.3)~km/s, and the interstellar magnetic field is $(0.2336,-0.1434,-0.1645)$\,nT. The number density of interstellar neutral hydrogen is $0.18$\,cm$^{-3}$, while the velocity and temperature are taken the same as the interstellar plasma. 

\section{Results}\label{sec:results}
\begin{figure}[ht!]
\centering
\includegraphics[width=0.45\textwidth]{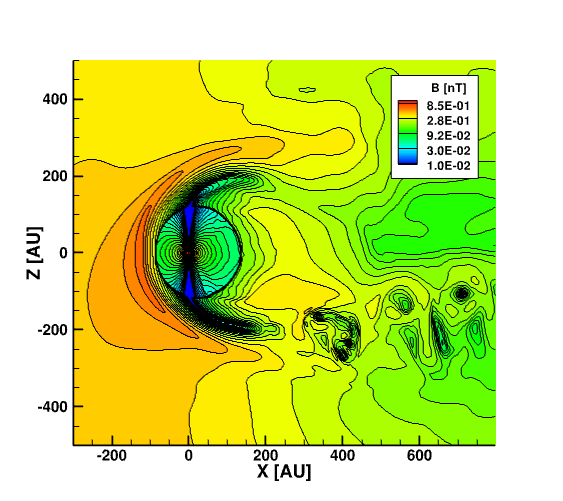}
\includegraphics[width=0.45\textwidth]{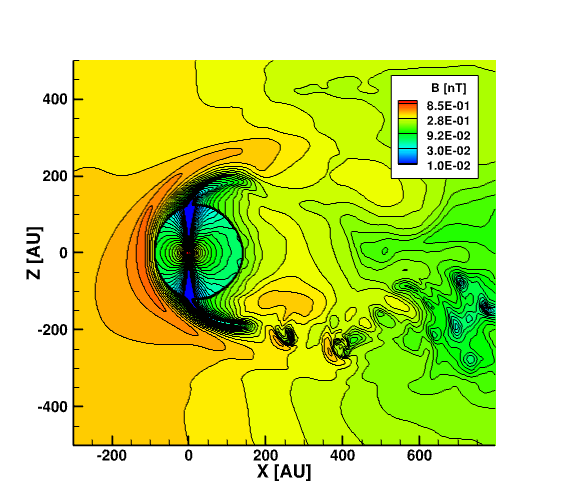}
\vspace{-2.5ex}
\caption{Magnetic field in the meridional plane: (left) shows without non-adiabatic PUI shock heating and (right) shows the latest results with non-adiabatic PUI shock heating. We observe minor differences outside of the TS region, in the turbulent tails.}
\label{fig:mhd_2d_b}
\end{figure}

We simulated the outer heliosphere for 1000 yr of physical time to achieve a near steady-state solution. The solution is not in perfect steady state due to the appearance of instabilities in the heliotail. In the noseward direction, the solution is in steady state. We present the simulation results through a series of cuts in the meridional plane and the synthetic data extracted along the trajectories of the V1, V2, and NH spacecrafts. In the upstream direction, the heliosphere's nose is in the -X direction, Z is heliographic north, and the system closes with Y in a right-handed orientation. Figure~\ref{fig:mhd_2d_b} shows the magnetic field solution for simulations performed without and with channeling the non-adiabatic shock heating to the PUIs, in the XZ or meridional plane. Their overall structure of the heliosphere shows the well-known croissant-shape of the BU model \citep{Opher:2020} with turbulent plasma blobs in the tail regions. The location of the termination shock varies from $x=-85$~AU in the direction of the nose to $x=135$~AU in the direction of the tail. The magnetic field strength in the supersonic solar wind is very similar in both solutions: it is higher in the direction of the equator than in the direction of the poles because of the Parker spiral, which is associated with the presence of a azimuthal magnetic field component away from the polar axis. The turbulent tail region shows minor differences in its eddies which is attributed to the changed thermal structure past the termination shock, as nothing else changed between the two simulations.

\begin{figure}[ht!]
\centering
\includegraphics[width=0.45\textwidth]{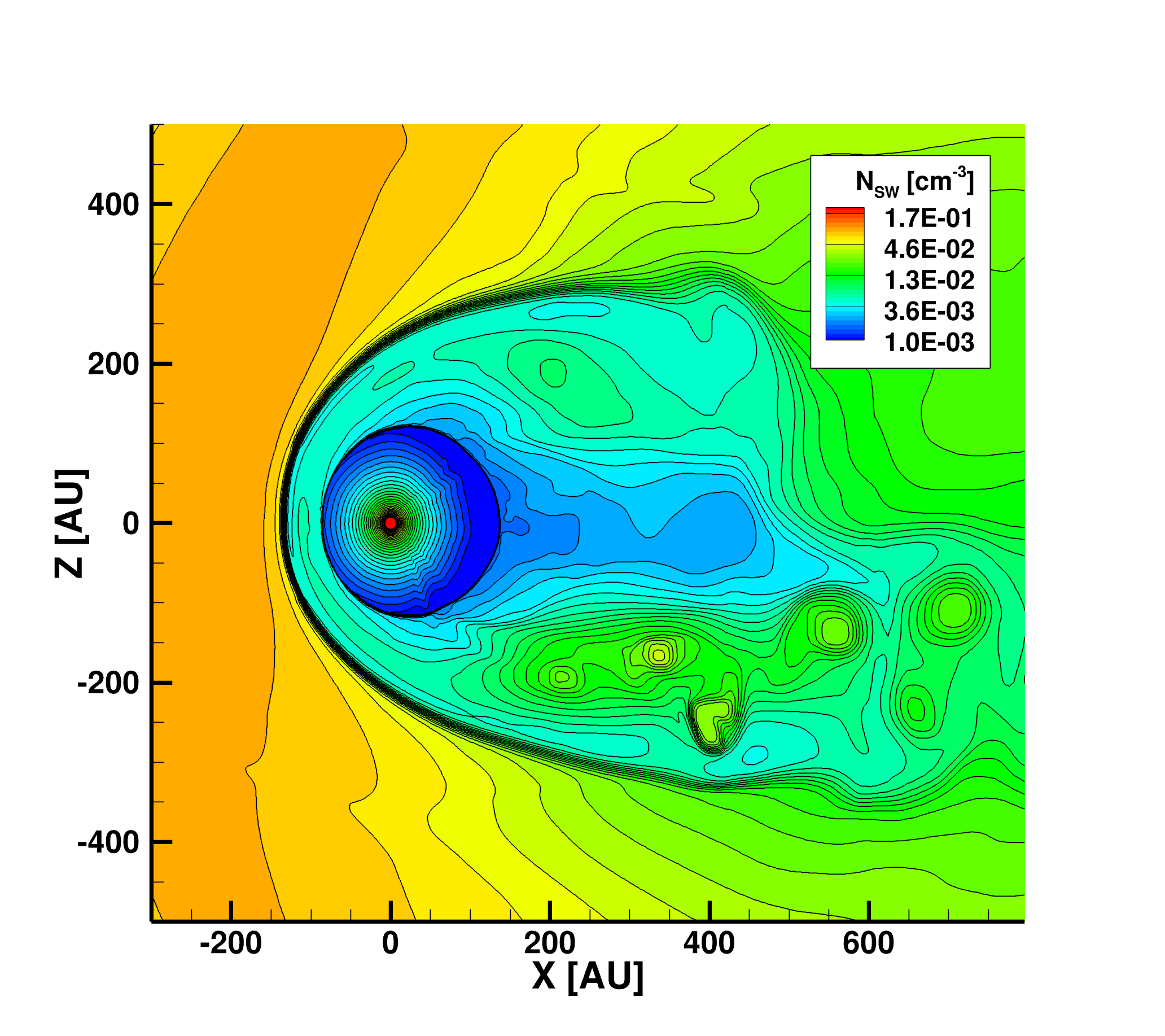}
\includegraphics[width=0.45\textwidth]{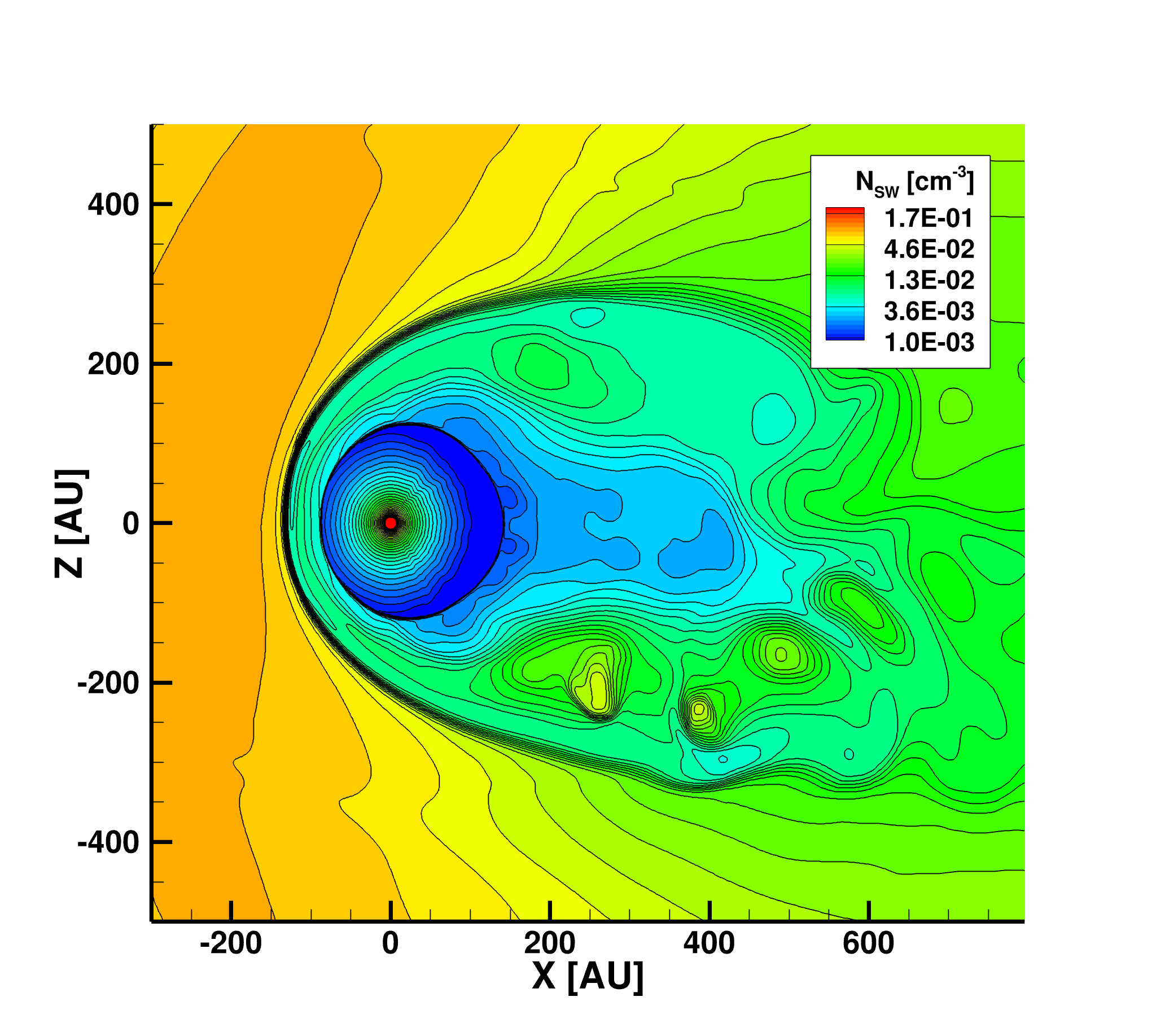}\\
\includegraphics[width=0.45\textwidth]{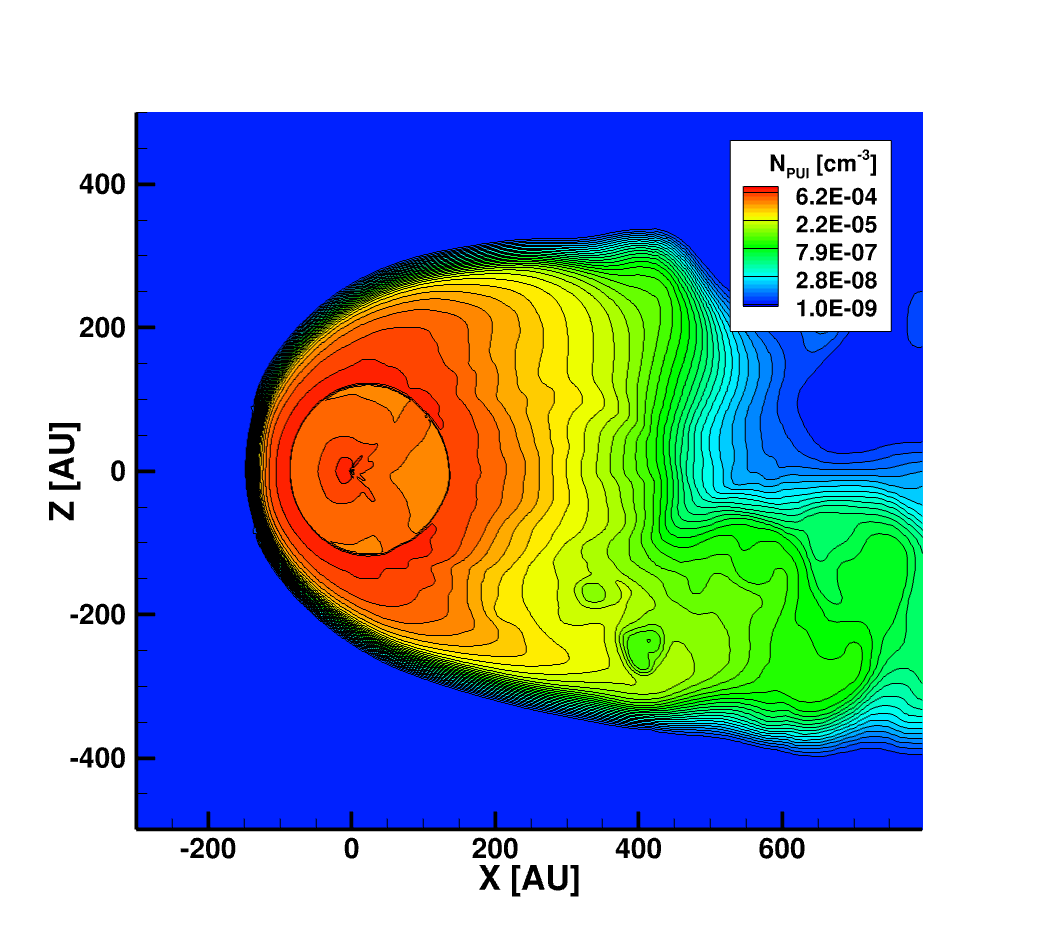}
\includegraphics[width=0.45\textwidth]{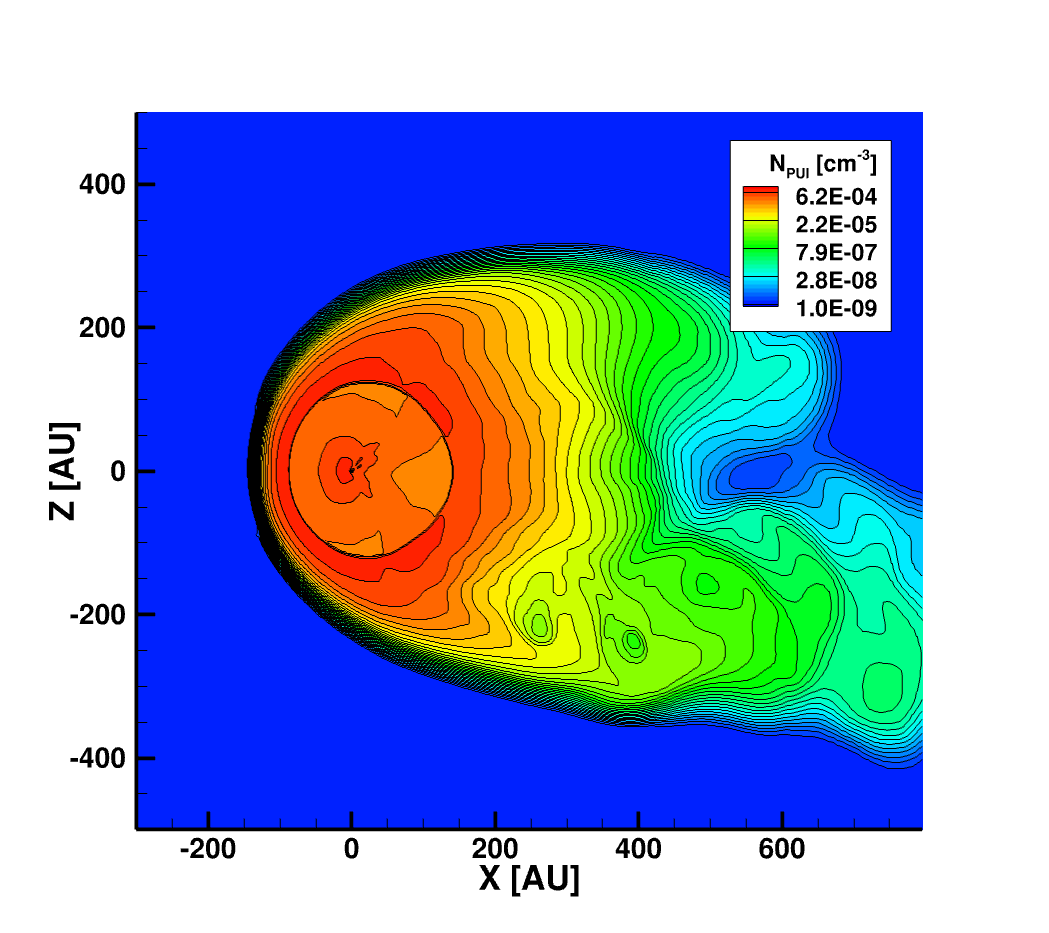}
\vspace{-2.5ex}
\caption{Number density in the meridional plane: (left) shows without non-adiabatic PUI shock-heating and (right) shows with non-adiabatic PUI shock heating for cold solar wind (top) and PUIs (bottom). The differences are minor in the large structures.}
\label{fig:mhd_2dn}
\end{figure}

In Figure \ref{fig:mhd_2dn}, we show the number densities of cold protons and PUIs. Whether we channel the non-adiabatic shock heating to the cold protons or PUIs has not much influence on the number density profiles. In both simulations, the cold proton number density does not have much latitudinal dependence in the supersonic solar wind. The number density of PUIs is higher in the direction of the nose where the neutral atoms of the VLISM enter than for the tail region.

\begin{figure}[ht!]
\centering
\includegraphics[width=0.45\textwidth]{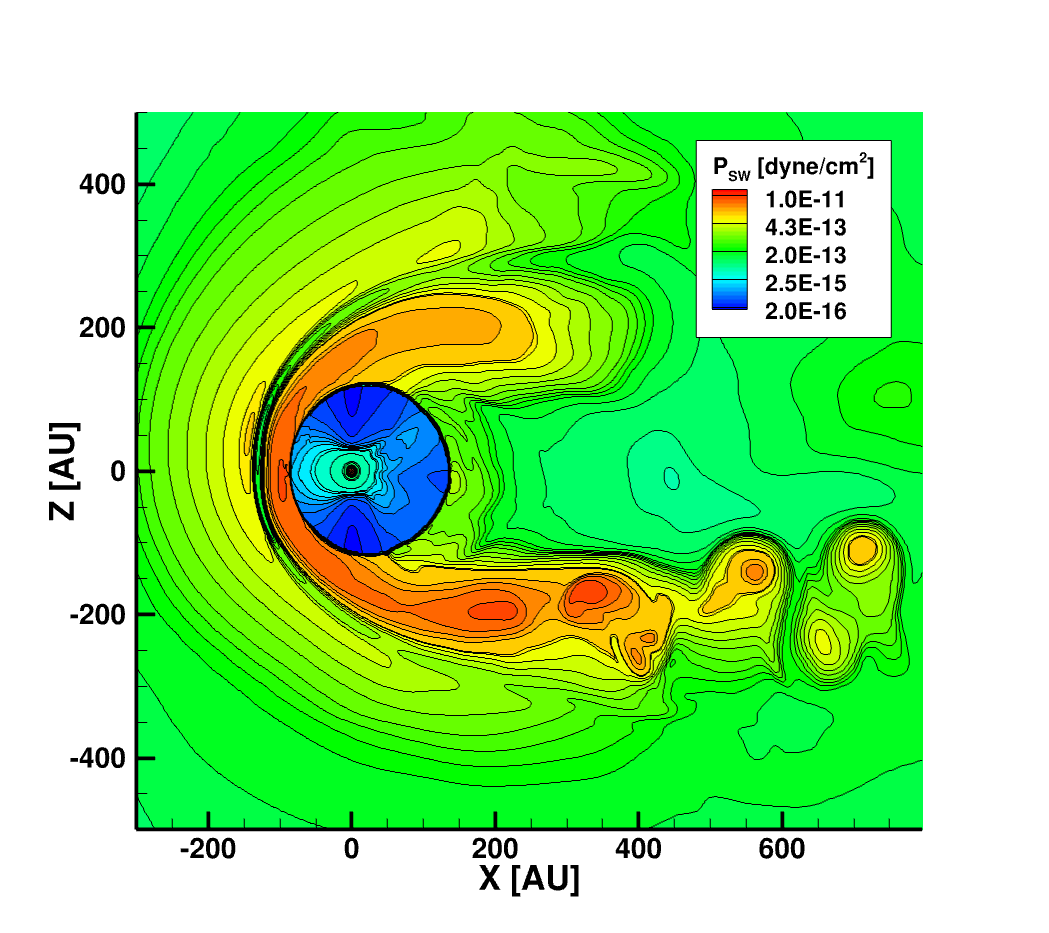}
\includegraphics[width=0.45\textwidth]{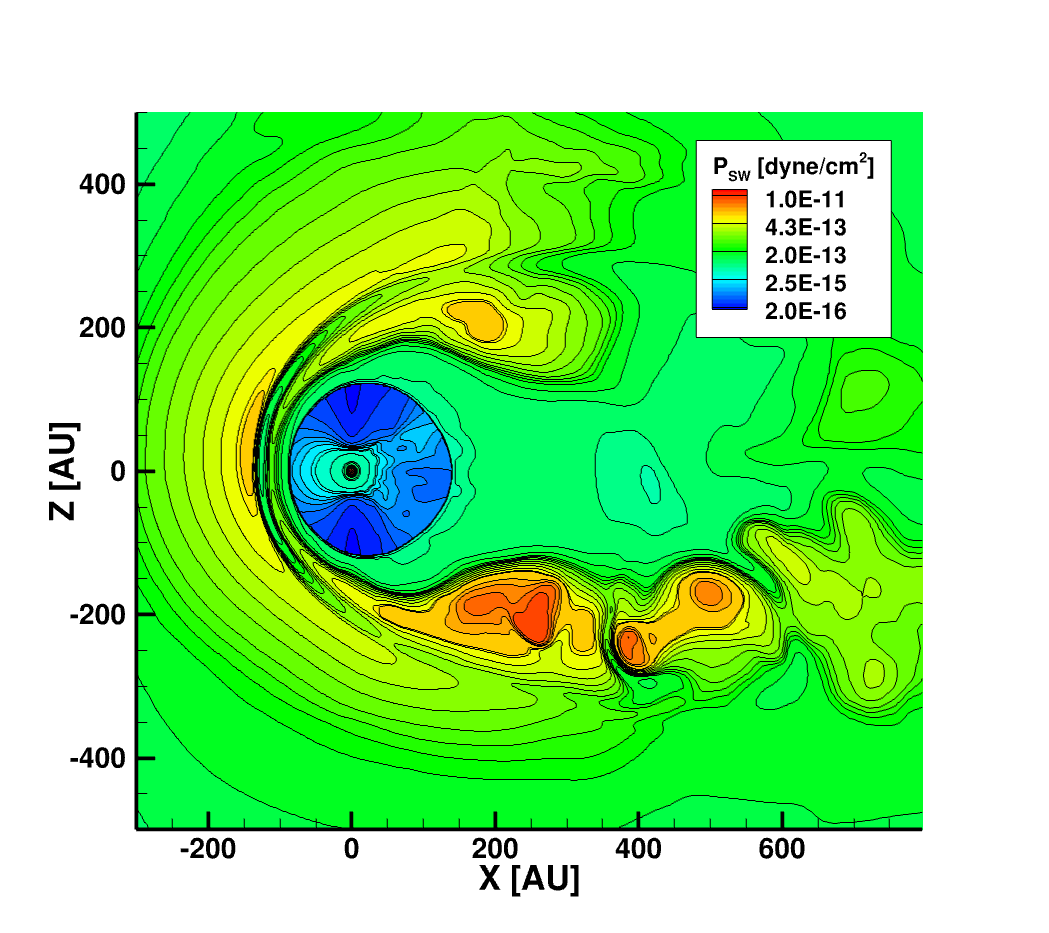}\\
\includegraphics[width=0.45\textwidth]{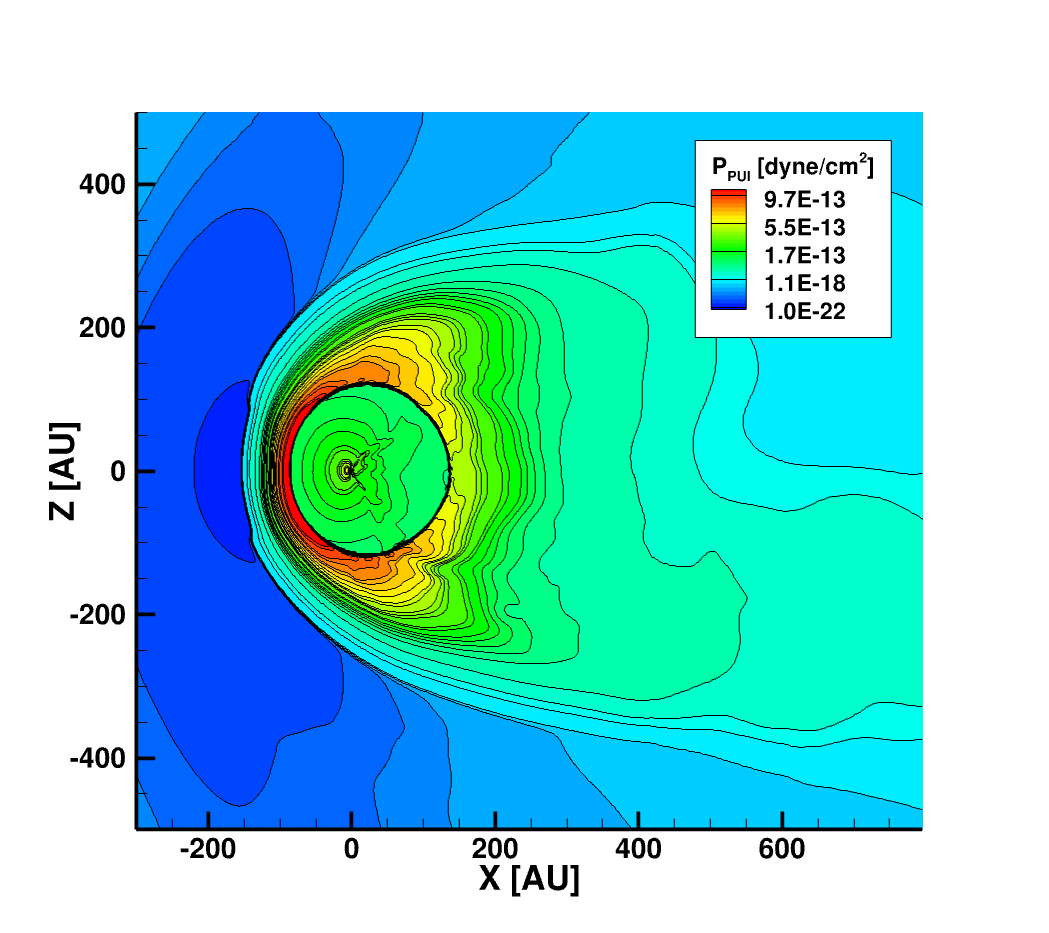}
\includegraphics[width=0.45\textwidth]{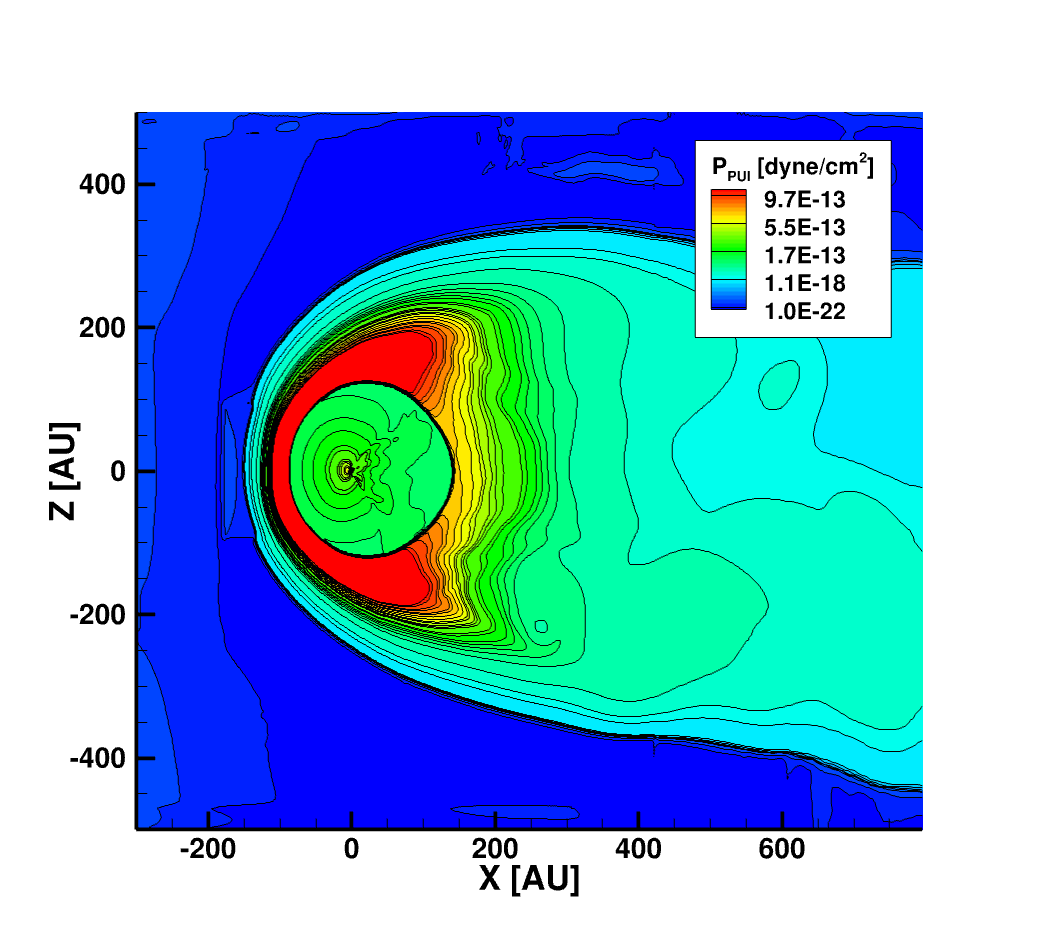}\\
\vspace{-2.5ex}
\caption{Pressure in the meridional plane without non-adiabatic PUI shock heating (left) and with non-adiabatic PUI shock heating (right) for cold solar wind (top) and PUIs (bottom).}
\label{fig:mhd_2dp}
\end{figure}

The pressures for the cold protons and the PUIs in Figure~\ref{fig:mhd_2dp} show very different behavior for both simulations. Although they are very similar in the supersonic solar wind, in the heliosheath the pressure of the PUIs is much higher and the pressure of the cold protons much lower in the simulation with channeling most of the non-adiabatic shock heating to the PUIs.

\begin{figure}[ht!]
\centering
\includegraphics[width=0.45\textwidth]{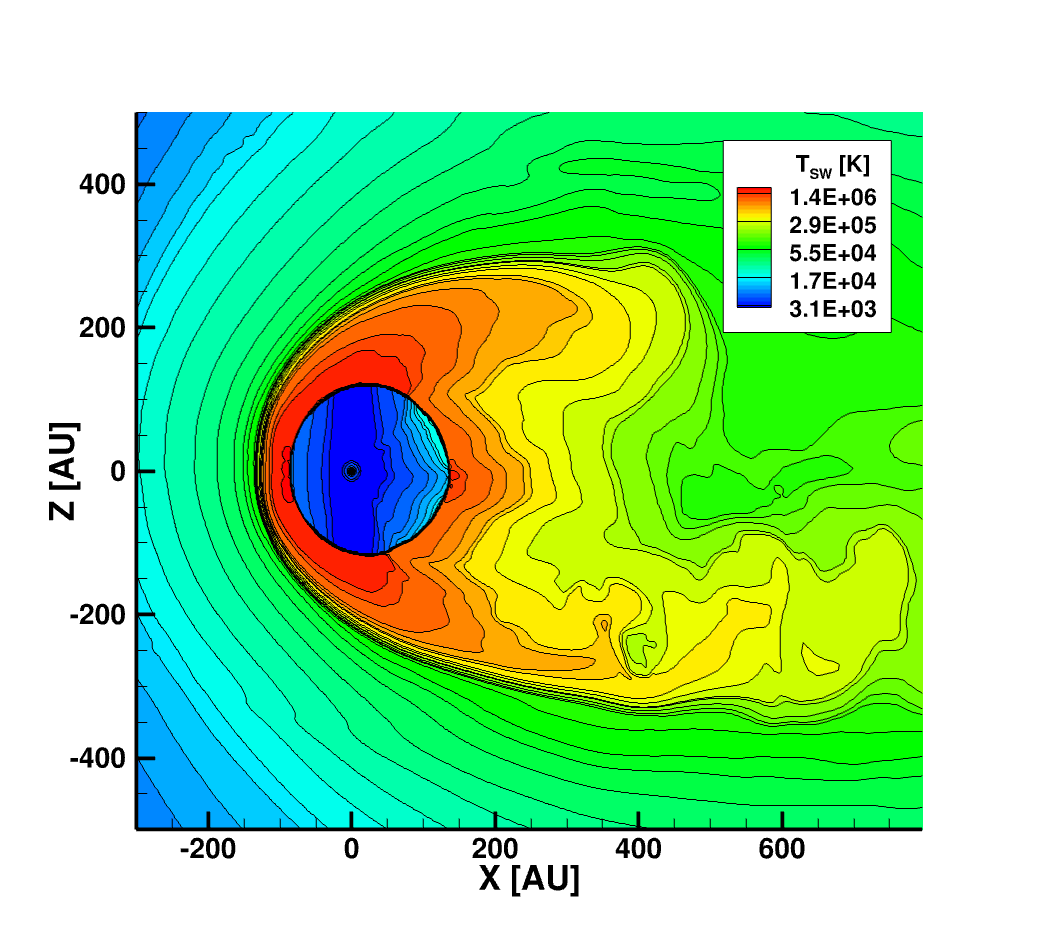}
\includegraphics[width=0.45\textwidth]{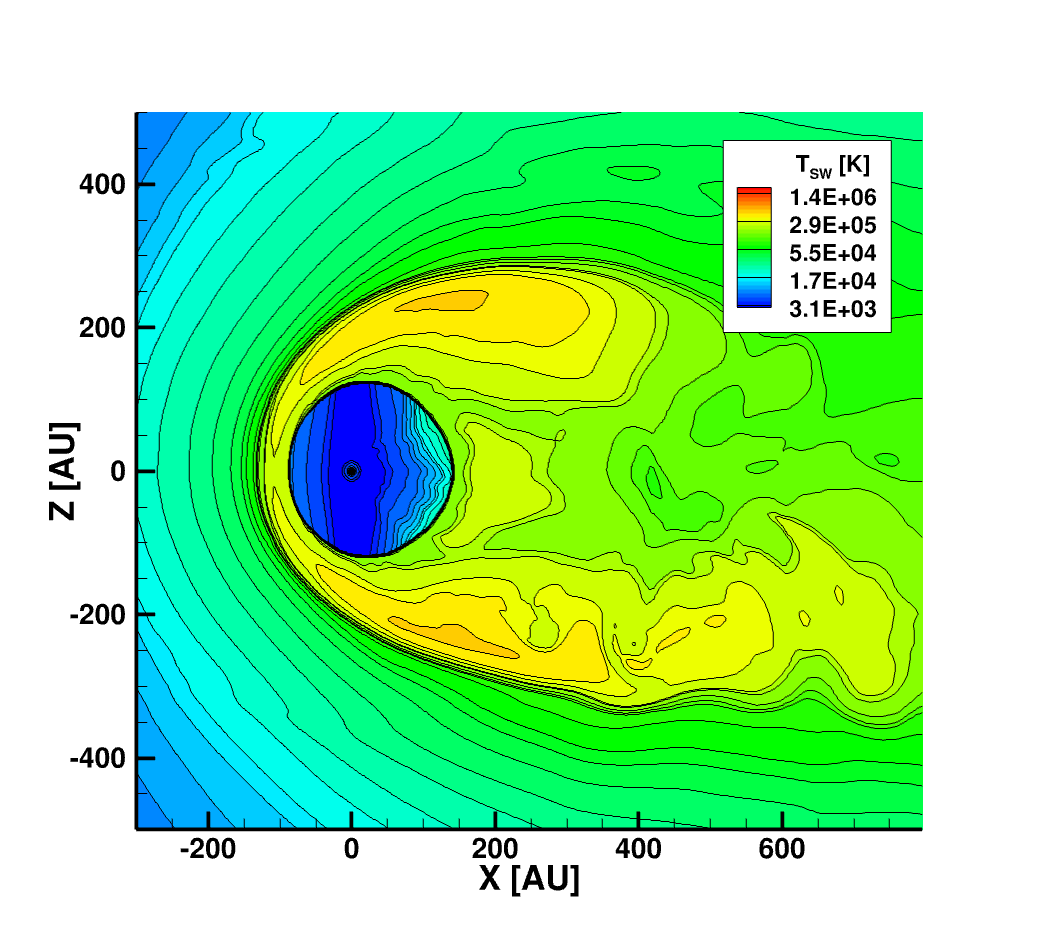}\\
\includegraphics[width=0.45\textwidth]{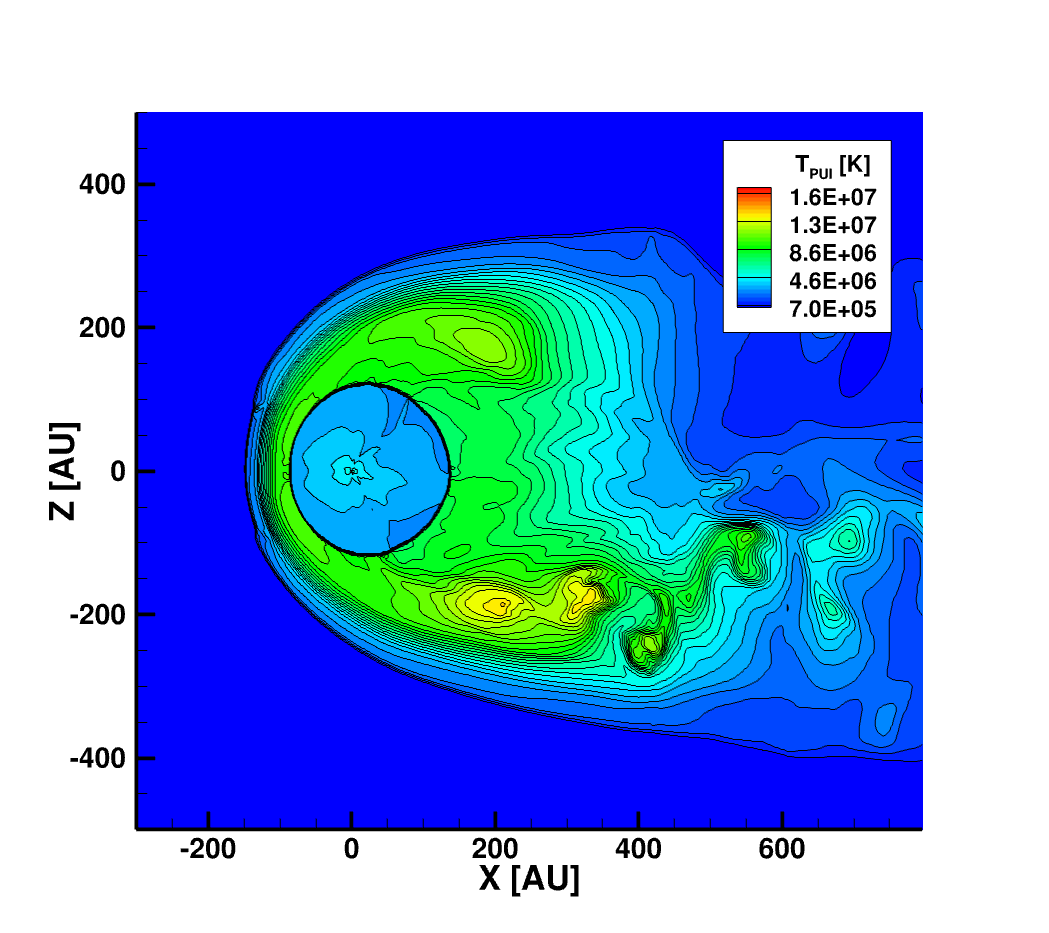}
\includegraphics[width=0.45\textwidth]{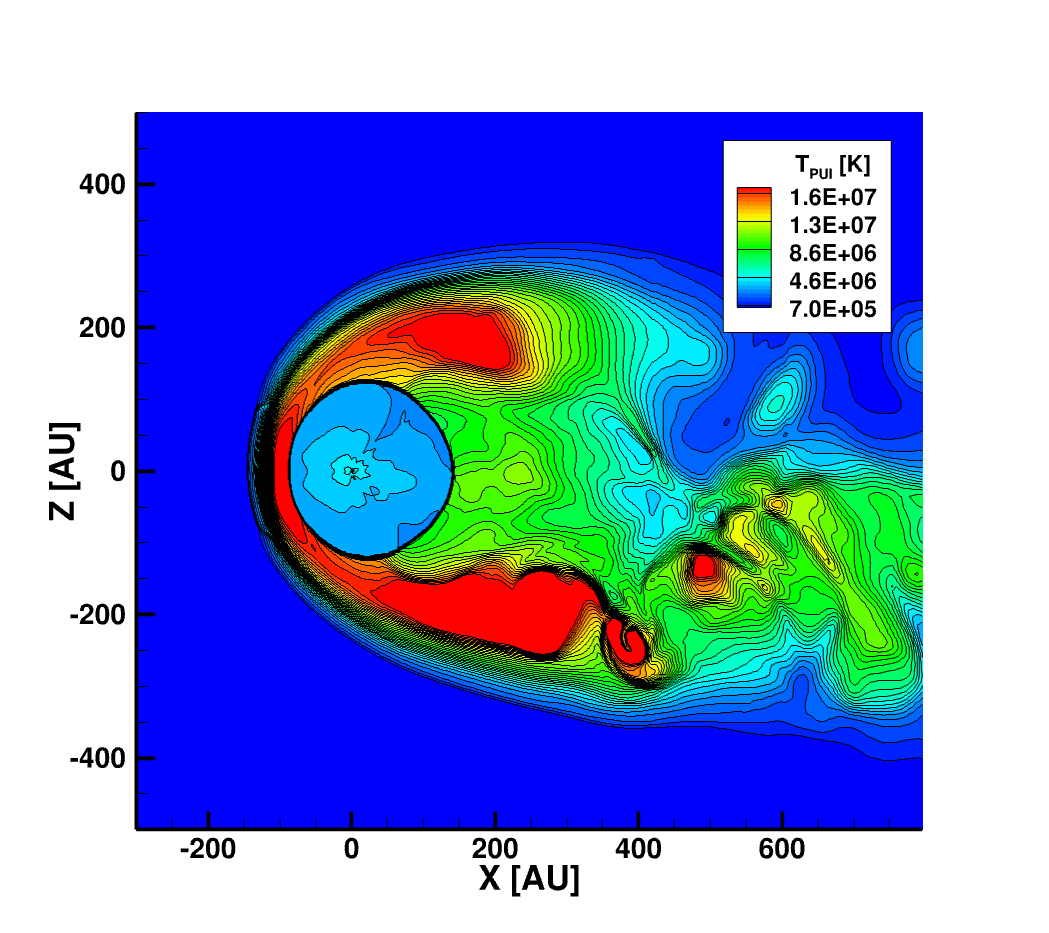}\\
\includegraphics[width=0.45\textwidth]{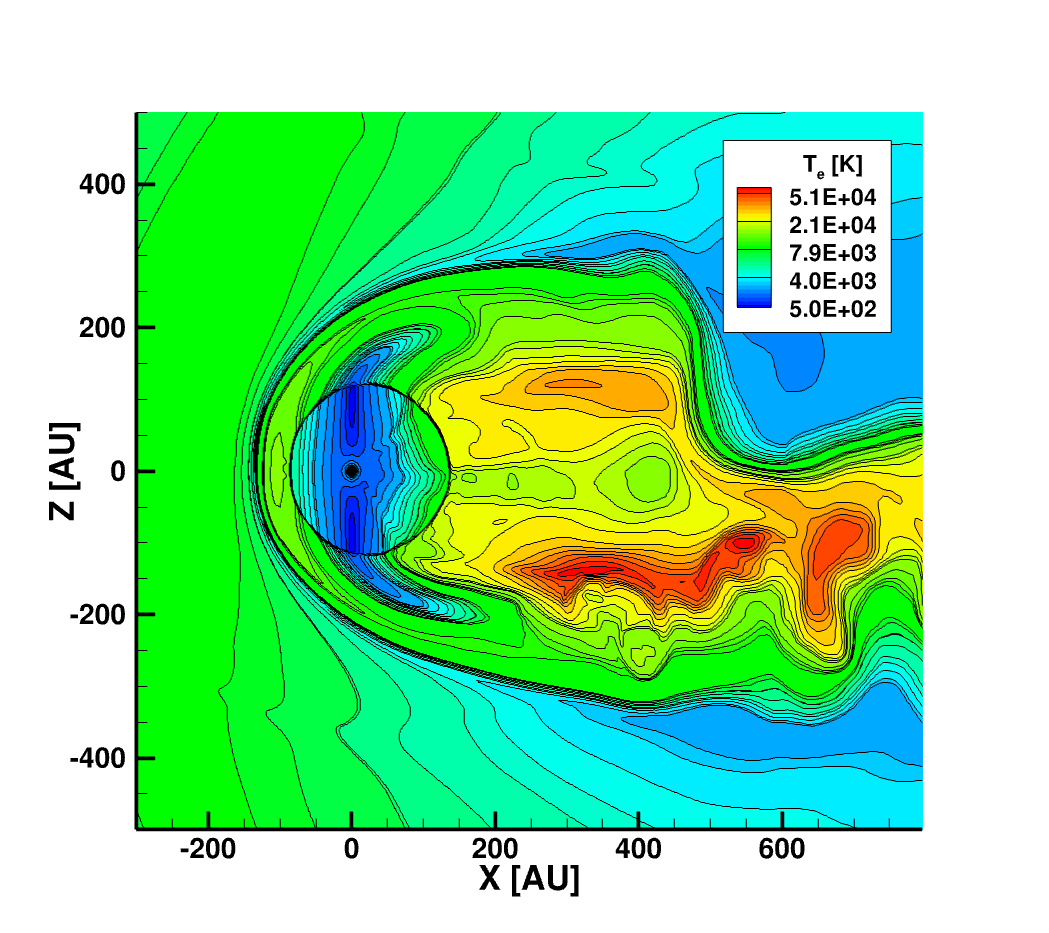}
\includegraphics[width=0.45\textwidth]{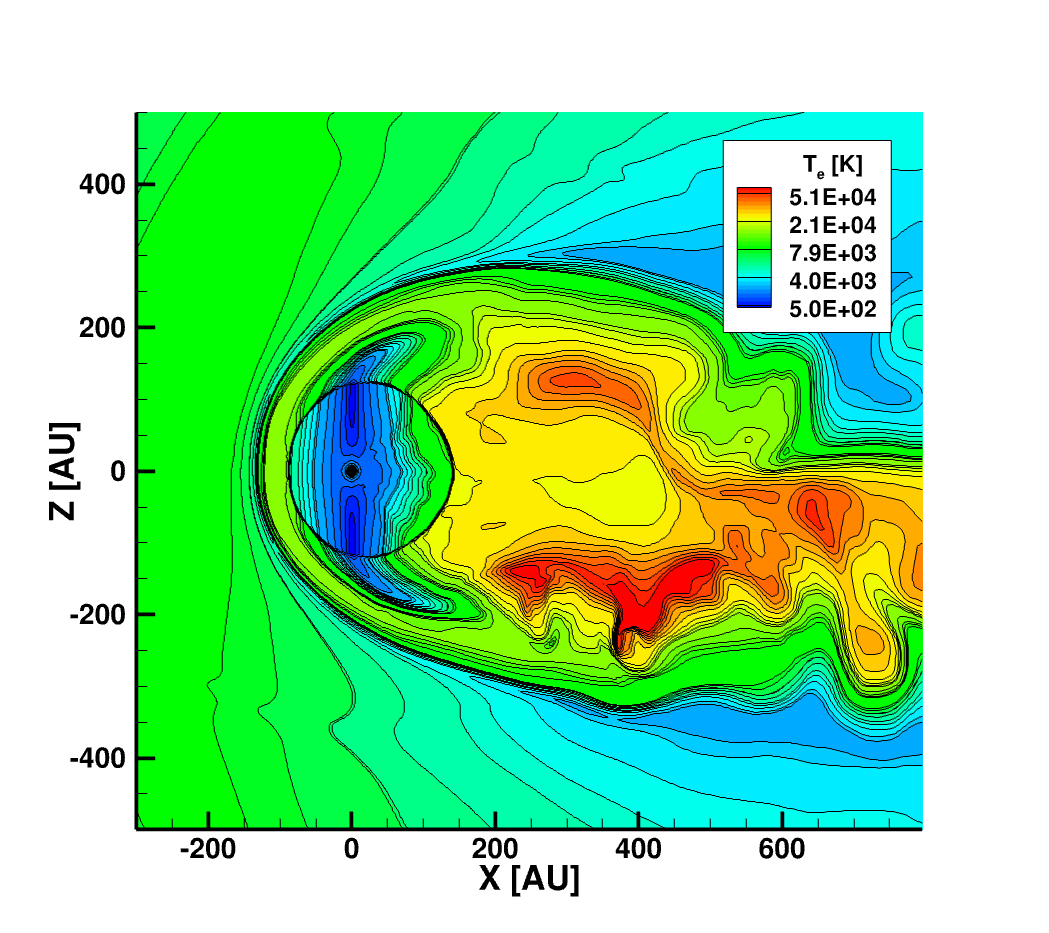}\\
\vspace{-2.5ex}
\caption{Temperature distribution in the meridional plane without (left) and with (right) non-adiabatic shock heating of the PUIs population. Downwards: cold solar wind, PUIs, and electron temperatures.}
\label{fig:t_2d}
\end{figure}

A similar trend can be found for the temperatures, shown in Figure \ref{fig:t_2d}. In the heliosheath, the PUI temperature is much higher, whereas the thermal proton is much lower in the simulation with channeling most of the non-adiabatic shock heating to the PUIs. In the old model it was not possible to add non-adiabatic shock heating to the electrons; in the new model we have set the weights by choice so that non-adiabatic shock heating is not channeled to the electrons, hence as a result the electron temperature did not change much for these two simulations. The supersonic solar wind cools nearly adiabatically over the poles more than at lower latitudes. This is because the PUIs-driven turbulence in the solar wind at lower latitudes provides additional heating. This is similar to Figure~1 from \citet{Usmanov:2016} and Figure~2 from \citet{vanderHolst:2026}. In case of the PUIs: less latitudinal dependence is observable in the supersonic solar wind, since we do not apply turbulence heating to the PUIs in Equation \ref{eq:swpressure}, similar to \citet{Adhikari2023,vanderHolst:2026}.

\begin{figure}
    \centering
    \includegraphics[width=0.49\linewidth]{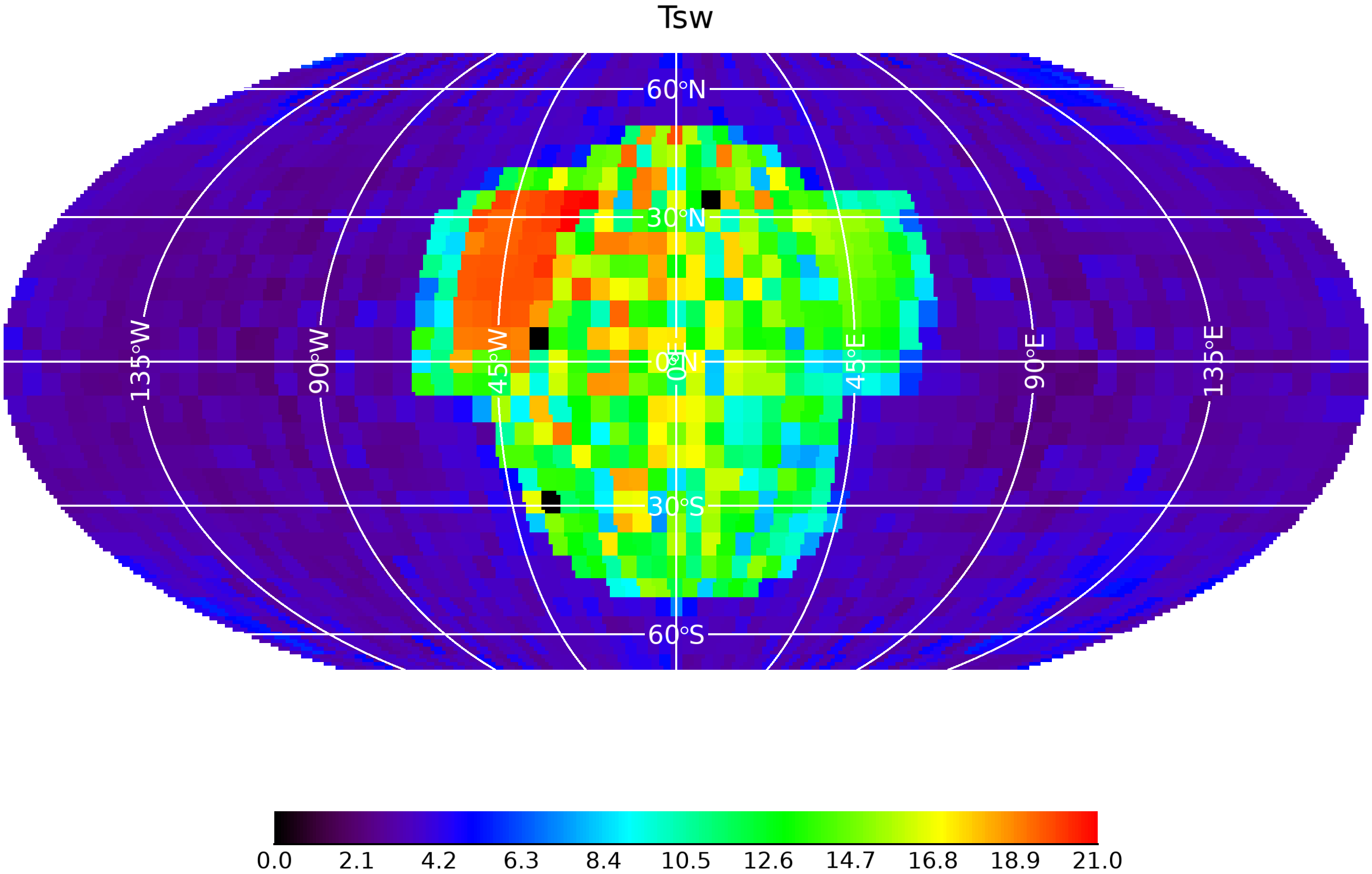}
    \includegraphics[width=0.49\linewidth]{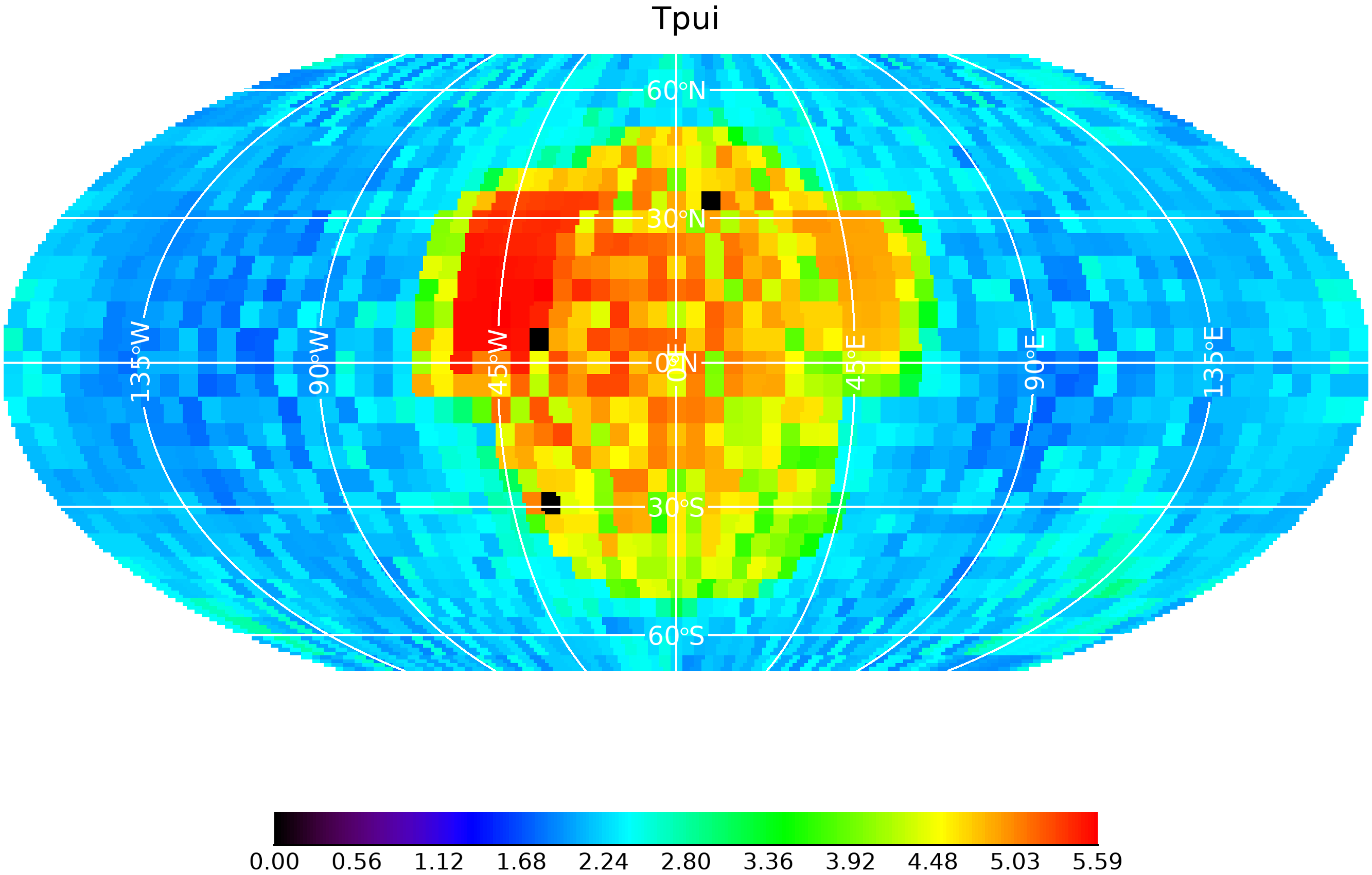}\\
        \includegraphics[width=0.49\linewidth]{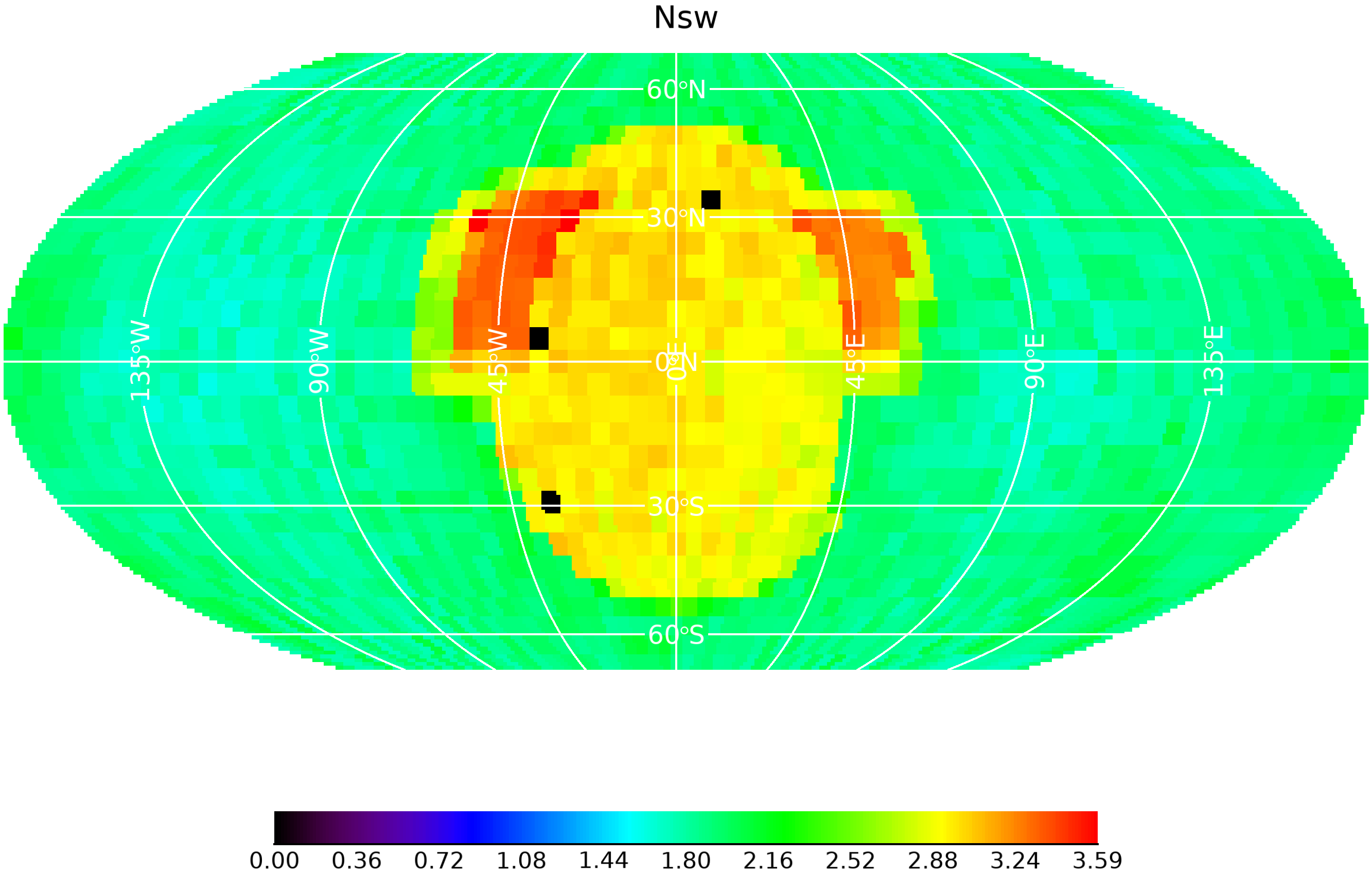}
        \includegraphics[width=0.49\linewidth]{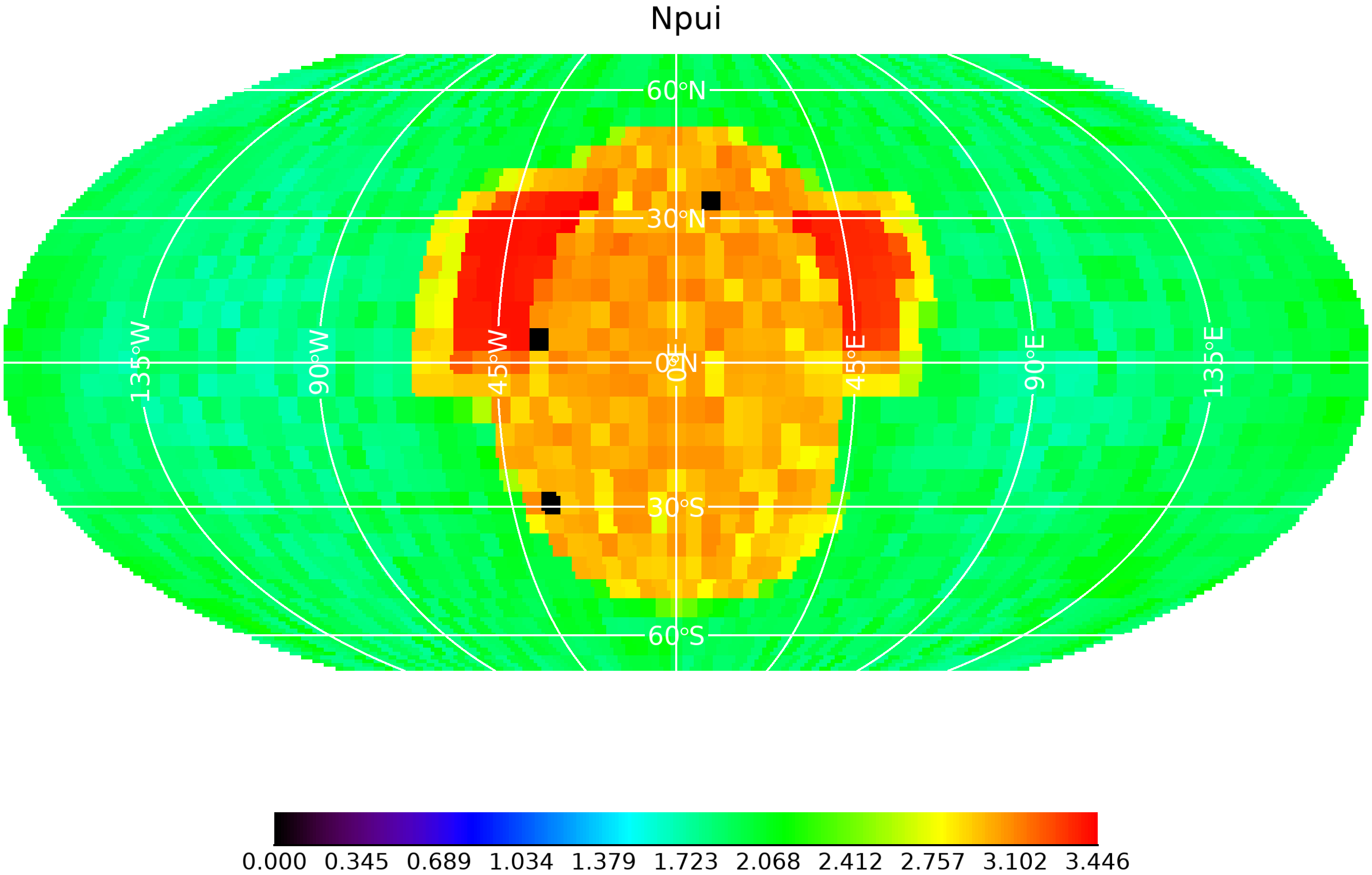}
    \caption{Mollweide-projection maps of shock jump conditions (DOWNstream/UPstream ratios)} of the solar wind proton temperature (left) and PUI temperature (right). Plasma conditions were taken 1~AU upstream and downstream along radial direction of the shock normal, where the shock central was defined by solar wind speed 250~km/s. The black markings are the crossing locations for V1 (34.1$^\circ$ north), NH (near equator), and V2 (27.5$^\circ$ south), respectively.
    \label{fig:Mollweide}
\end{figure}

The Mollweide-projection maps of the jump conditions (DOWNstream/UPstream ratios) at the TS are shown in Figure \ref{fig:Mollweide}. The termination shock location is identified by finding the solar wind speed 250 km/s beyond 70 AU in radial direction. For the sake of producing the maps we extracted radial data lines of the 3D solution starting 70 AU distance from the domain center (the Sun) in a 5x5 degree resolution. Then we identified the location 1 AU before and after the locations provided by the 250 km/s solar wind speed identification, and marked the plasma data for UPstream and DOWNstream accordingly. Then we took the ratios of these plasma variables' values at both locations and plotted the values in the Mollweide projection using IDL's internal functions. The black dots indicate the termination shock crossings of V1 (highest latitude), NH (near equator), and V2 (lowest latitude). We note that the highest temperature jumps are in the nose direction.

The compression (number density; DOWN/UP) ratio for protons and PUIs is shown in Figure~\ref{fig:Mollweide}. The solar wind number density ratios are below what was reported by \citet{Zirnstein:2025NatAs...9.1495Z} but at V2 crossing it is consistent with the observed 2.5 ratio. The pixel-to-pixel variations in the maps result from sampling a Cartesian grid at finite resolution, so the maps should be read as qualitative.

\begin{figure}[h!]
    \centering
    \includegraphics[width=0.89\linewidth]{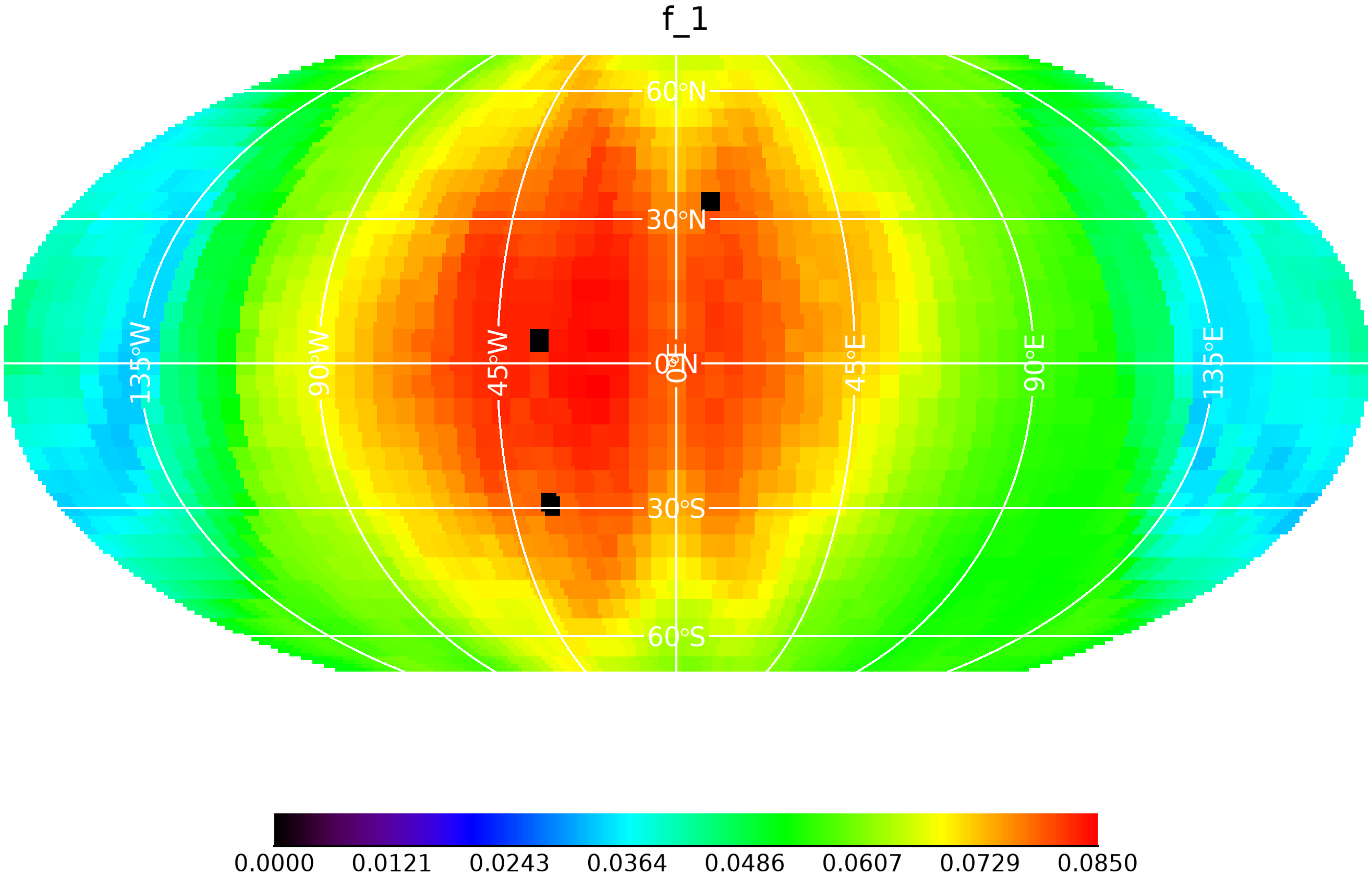}
    \caption{Mollweide-projection map of the $f_1$ from Equation~\ref{eq:f} from 1AU before the termination shock. The black markings are the crossing locations for V1 (34.1$^\circ$ north), NH (near equator), and V2 (27.5$^\circ$ south), respectively.}
    \label{fig:f1}
\end{figure}

In Figure \ref{fig:f1}, we show a Mollweide-projection map of the fraction $f_1$, defined in Equation \ref{eq:f}, of non-adiabatic heating channeled towards the thermal solar wind protons. The spatial variation of this fraction up to 8.5\% is due to the thermal proton and PUI number densities. This fraction is in the range with the value of 0.048 found by \citet{Zank:2010ApJ...708.1092Z}. Also note that \citet{Zank:2010ApJ...708.1092Z} defines the energy partitioning in terms of ratios of densities and temperatures, while our energy partitioning weights determine how the non-adiabatic shock heating is distributed among the various species. We also note that \citet{Richardson:2022} reported a higher fraction of 0.2.

\begin{figure}[ht!]
\centering
\includegraphics[width=0.45\textwidth]{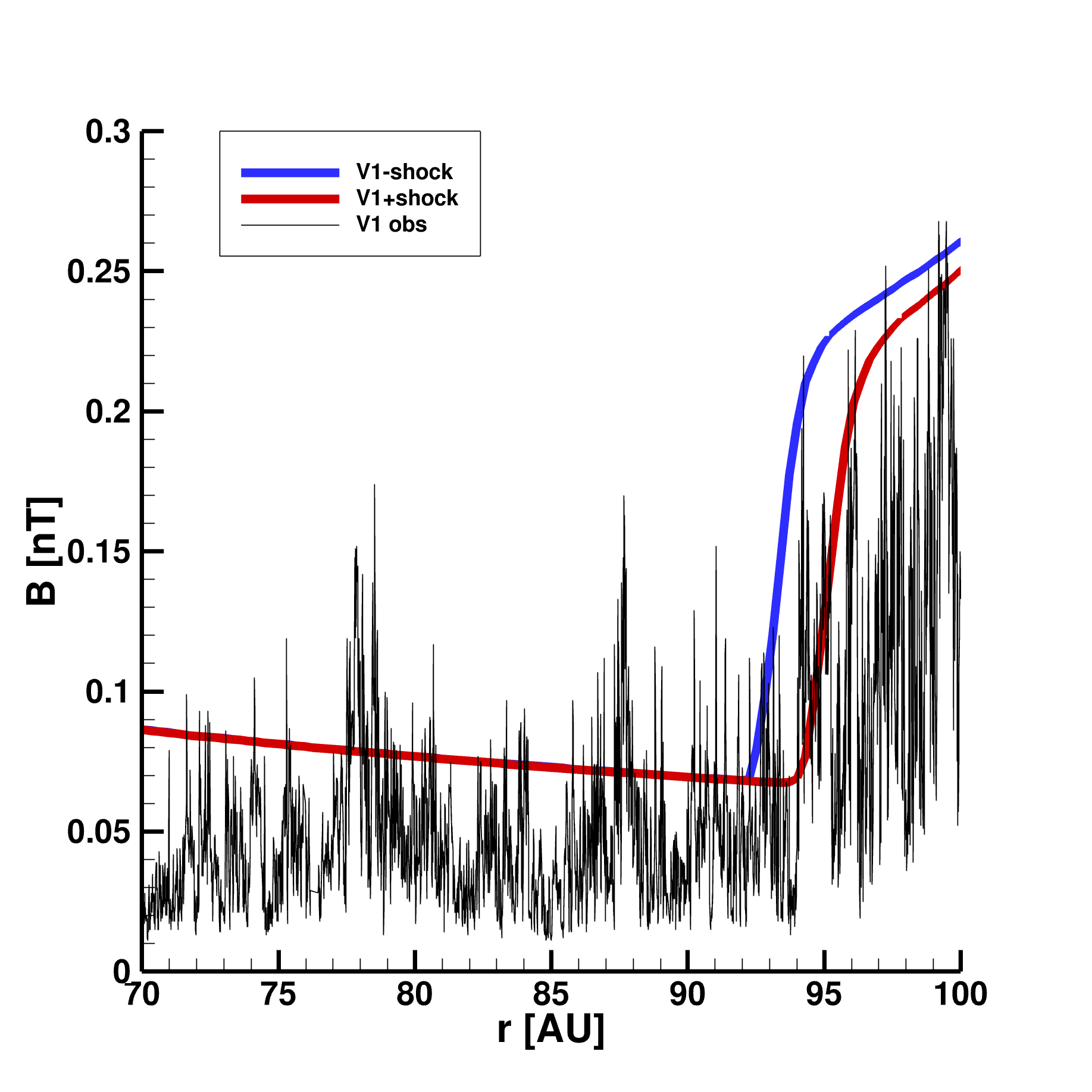}
\vspace{-2.5ex}
\caption{Termination shock observed by V1 spacecraft via the magnetic field measurements (black). Blue line shows the simulation results without non-adiabatic PUI shock heating, and red lines show the updated MHD solution with non-adiabatic PUI shock heating.}
\label{fig:linev1}
\end{figure}

In Figure~\ref{fig:linev1}, we examine the magnetic jump conditions at the upstream axis along the V1. Because our simulations have only MHD variables and since the plasma instrument measuring the cold solar wind is not working on V1, we can only compare the magnetic field. The jump in the magnetic field strength did not change much between the simulations with and without non-adiabatic PUI shock heating, while the TS location moved outward by 2\,AU with non-adiabatic PUI shock heating.

\begin{figure}[ht!]
\centering
\includegraphics[width=0.45\textwidth]{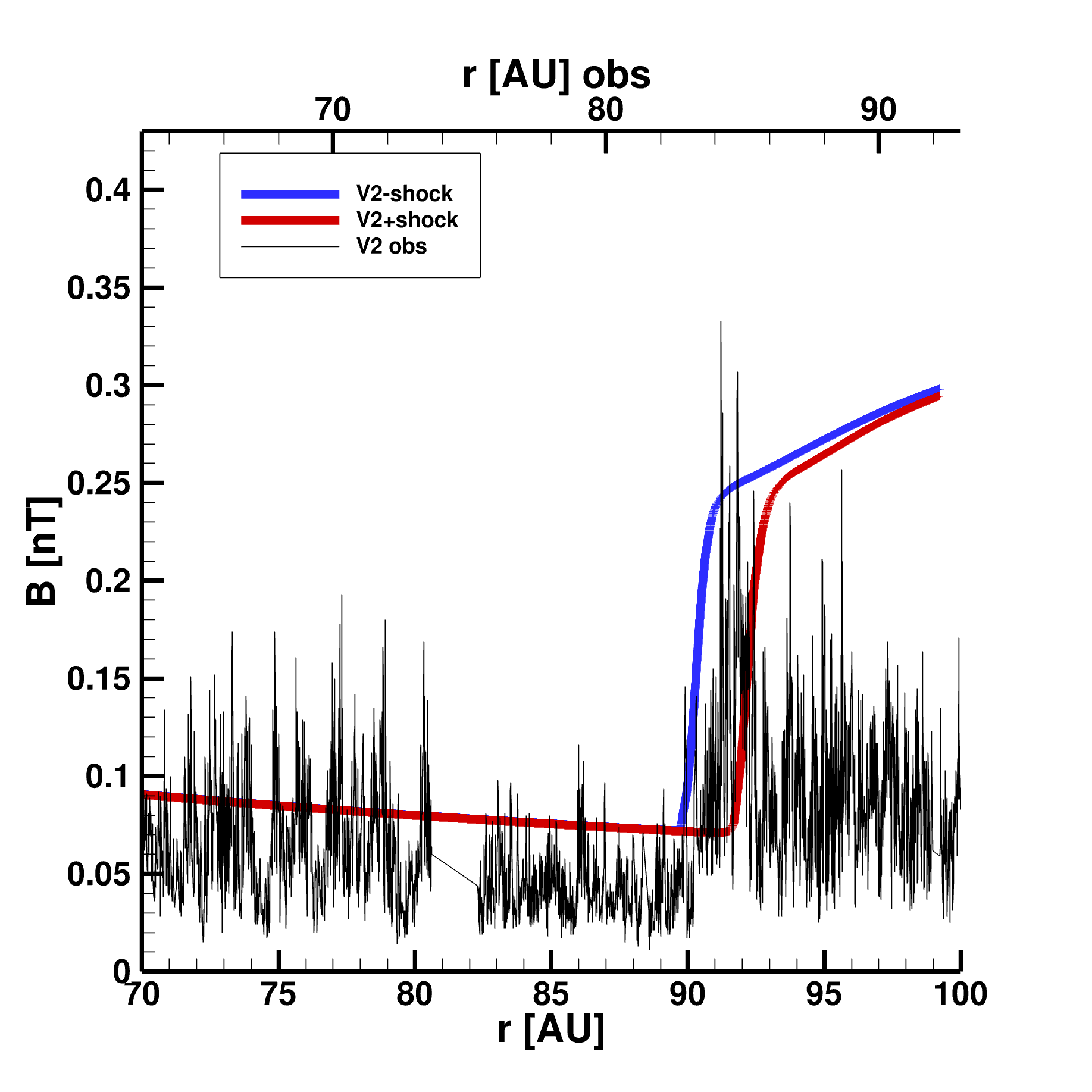}
\includegraphics[width=0.45\textwidth]{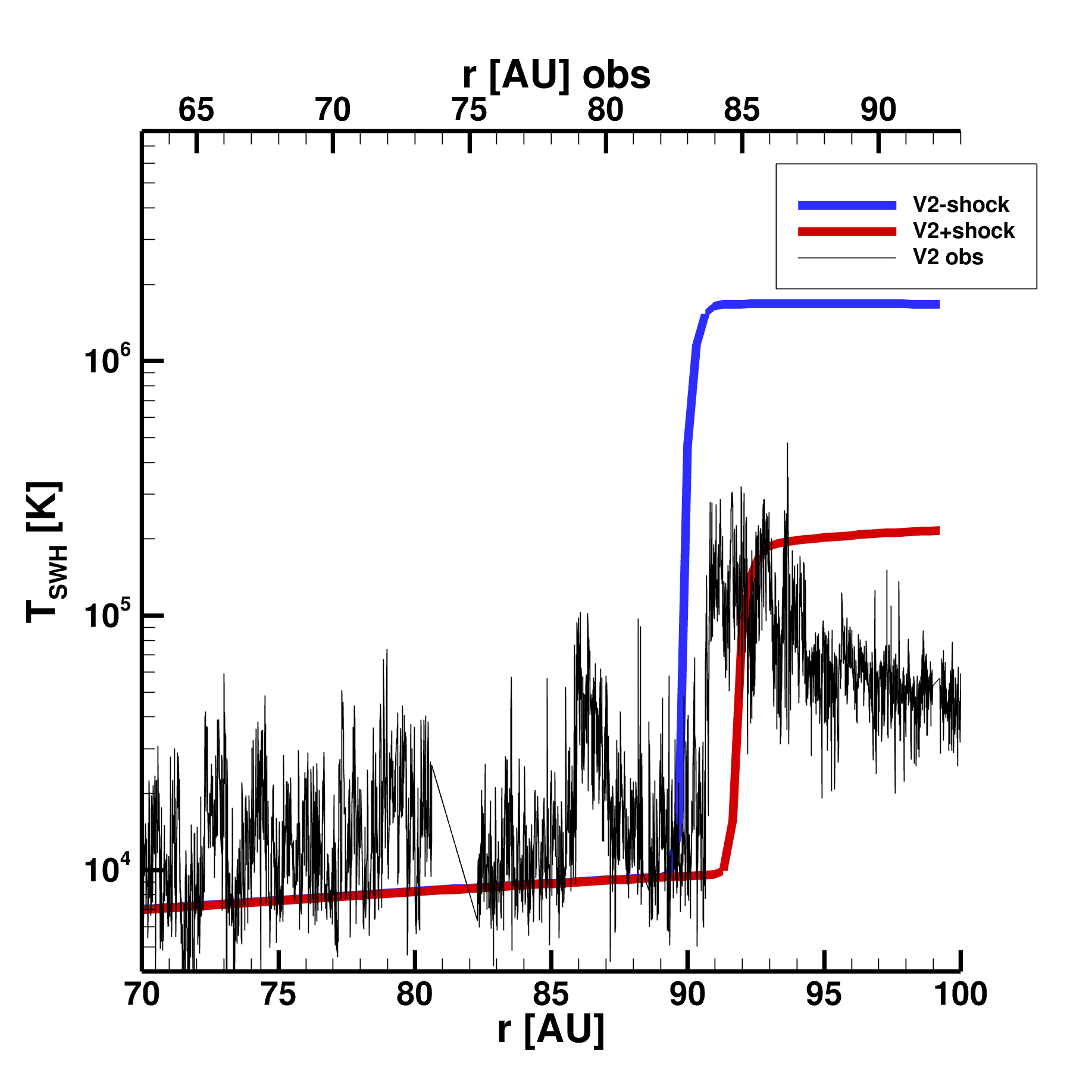}
\includegraphics[width=0.45\textwidth]{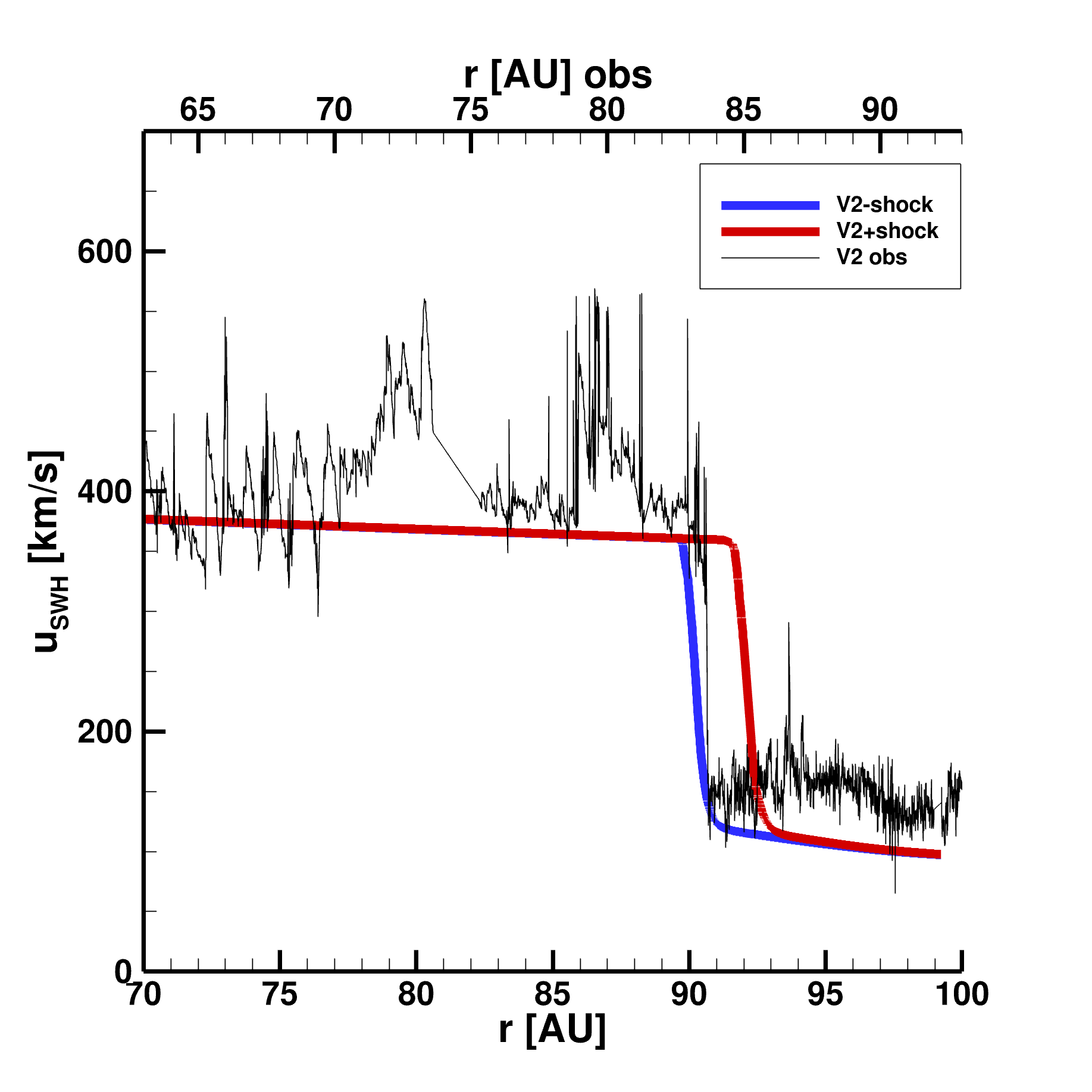}
\includegraphics[width=0.45\textwidth]{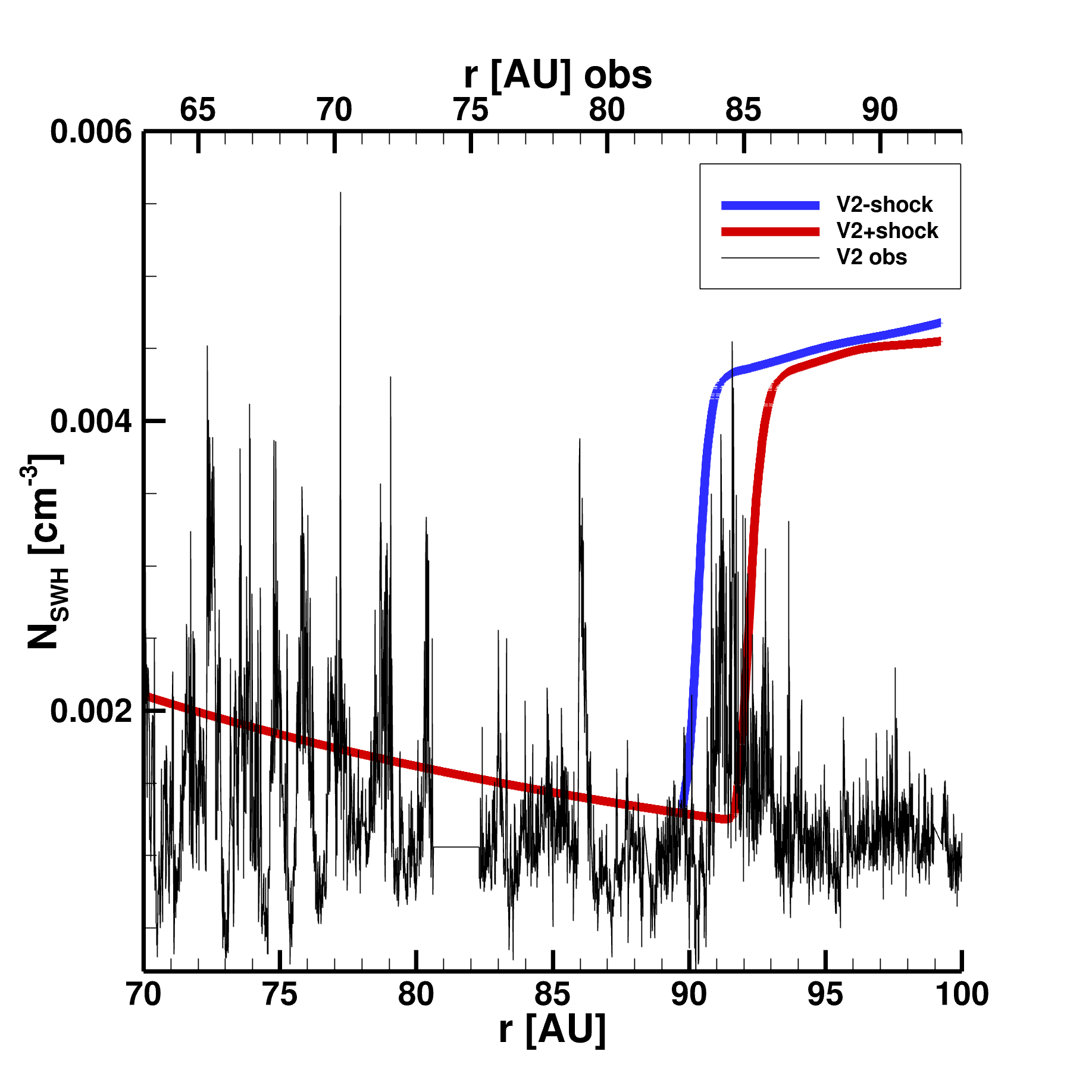}
\vspace{-2.5ex}
\caption{V2 crossing of the TS. The panels show the magnetic field (top left), the solar wind temperature (top right), the solar wind speed (bottom left) and the solar wind proton number density (bottom right). Black is observed values, blue is the solution without non-adiabatic PUI shock heating and red is with the newest version. The upper (observed) radial axis is shifted relative to the lower one so that the observed and simulated shock positions coincide. The temperature jump has clearly improved compared to the previous results.}
\label{fig:linev2}
\end{figure}

Figure~\ref{fig:linev2} shows the jump conditions along the V2 spacecraft trajectory. While V2 crossed the TS three times due to the variations in the solar wind or ripples in the shock front, in our simulation we have a near steady state solution and therefore only one crossing is simulated. All three observed crossings showed a quasi-perpendicular shock. The change in magnetic field arises from the displacement of the shock itself, while the actual behavior and structure of the magnetic field remain the same as before. Most importantly, the solar wind proton temperature experiences from upstream (UP) to downstream (DOWN), instead of a too large temperature jump (DOWN/UP ratio) of $\sim 200$, a temperature (DOWN/UP ratio) jump of 20 with the new approach, which is closer to V2 data. There is no significant change in the solar wind speed and mass density jump conditions as compared with previous simulation results; only the termination shock moved a few AU outward compared to previous simulation results.

\begin{figure}[ht!]
\centering
\includegraphics[width=0.31\textwidth]{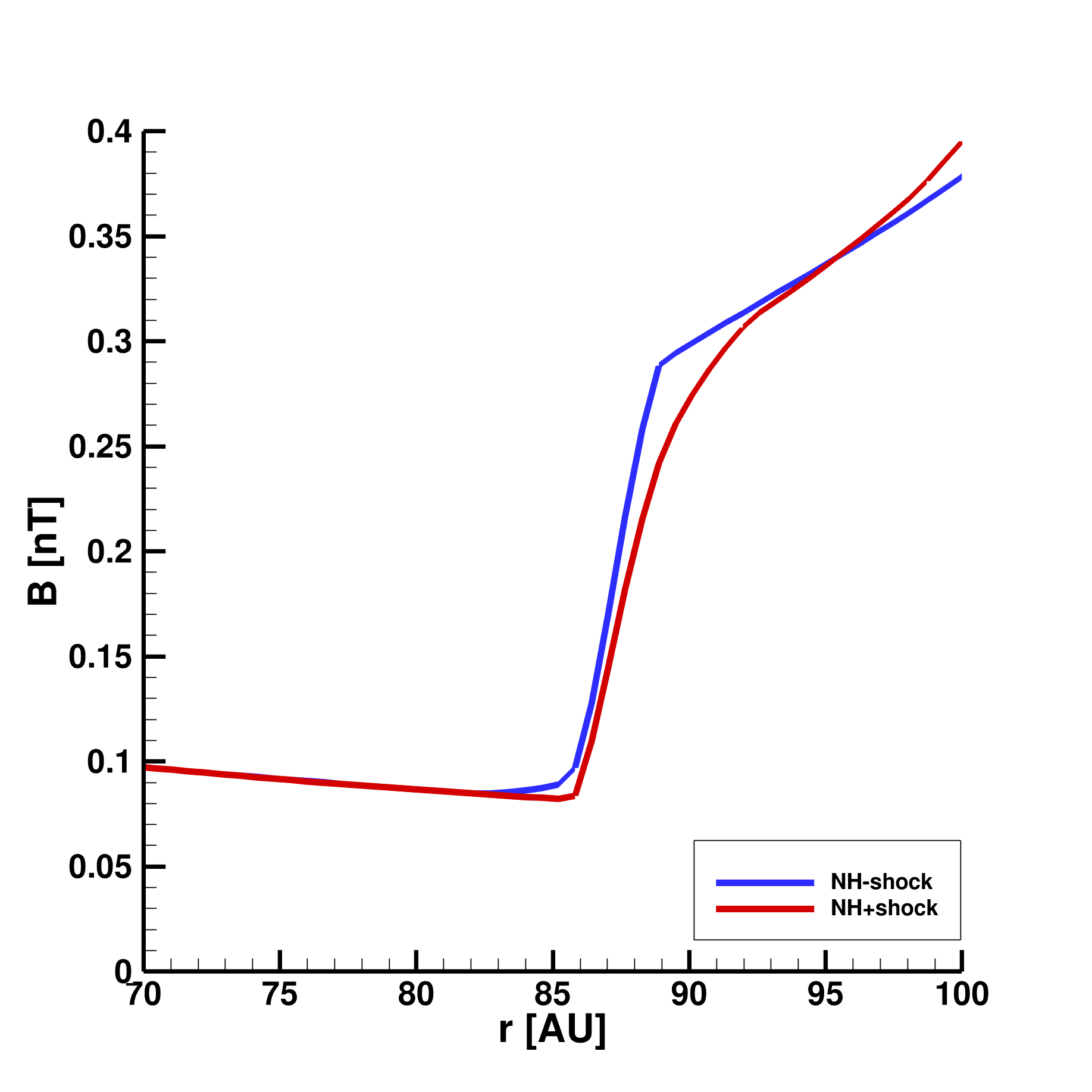}
\includegraphics[width=0.31\textwidth]{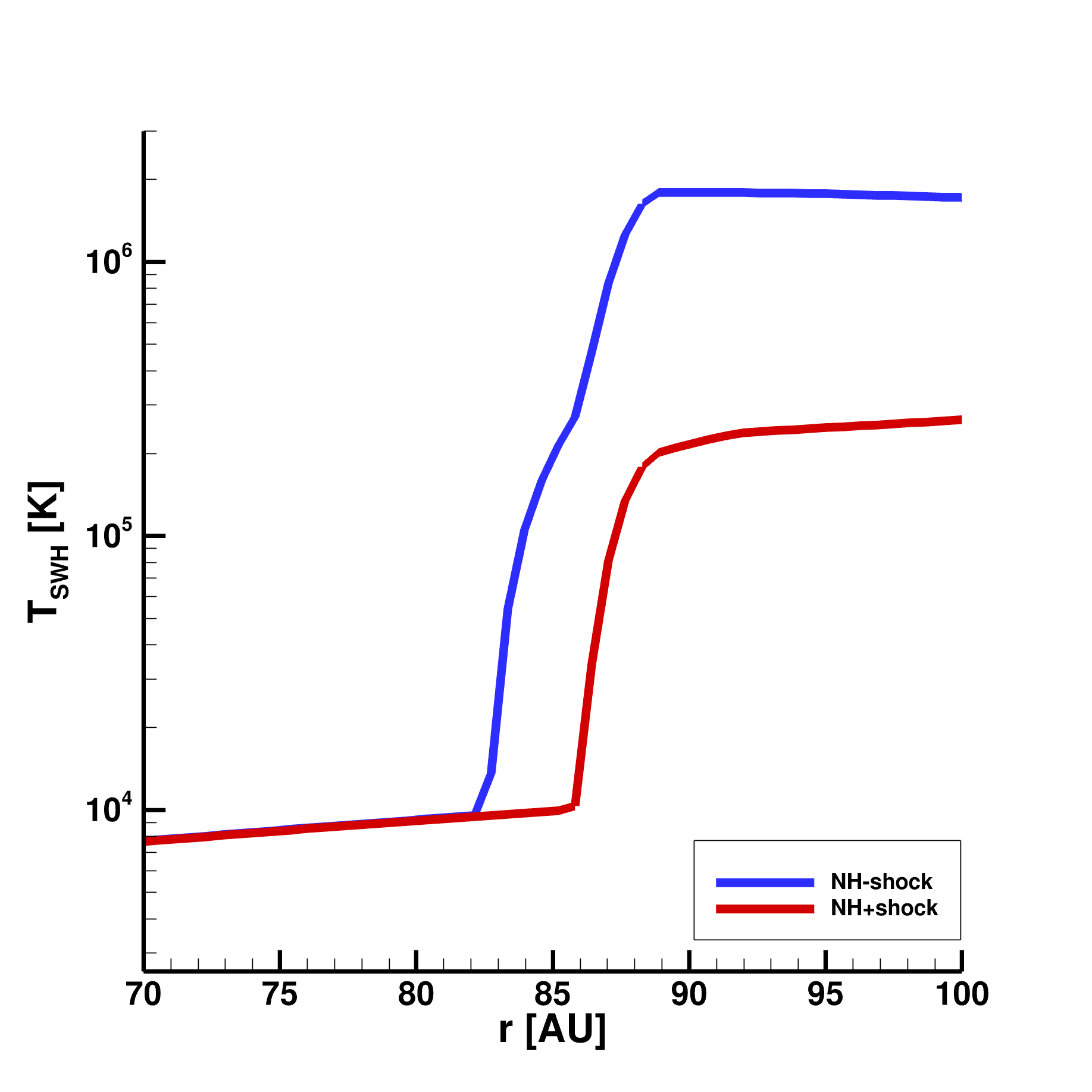}
\includegraphics[width=0.31\textwidth]{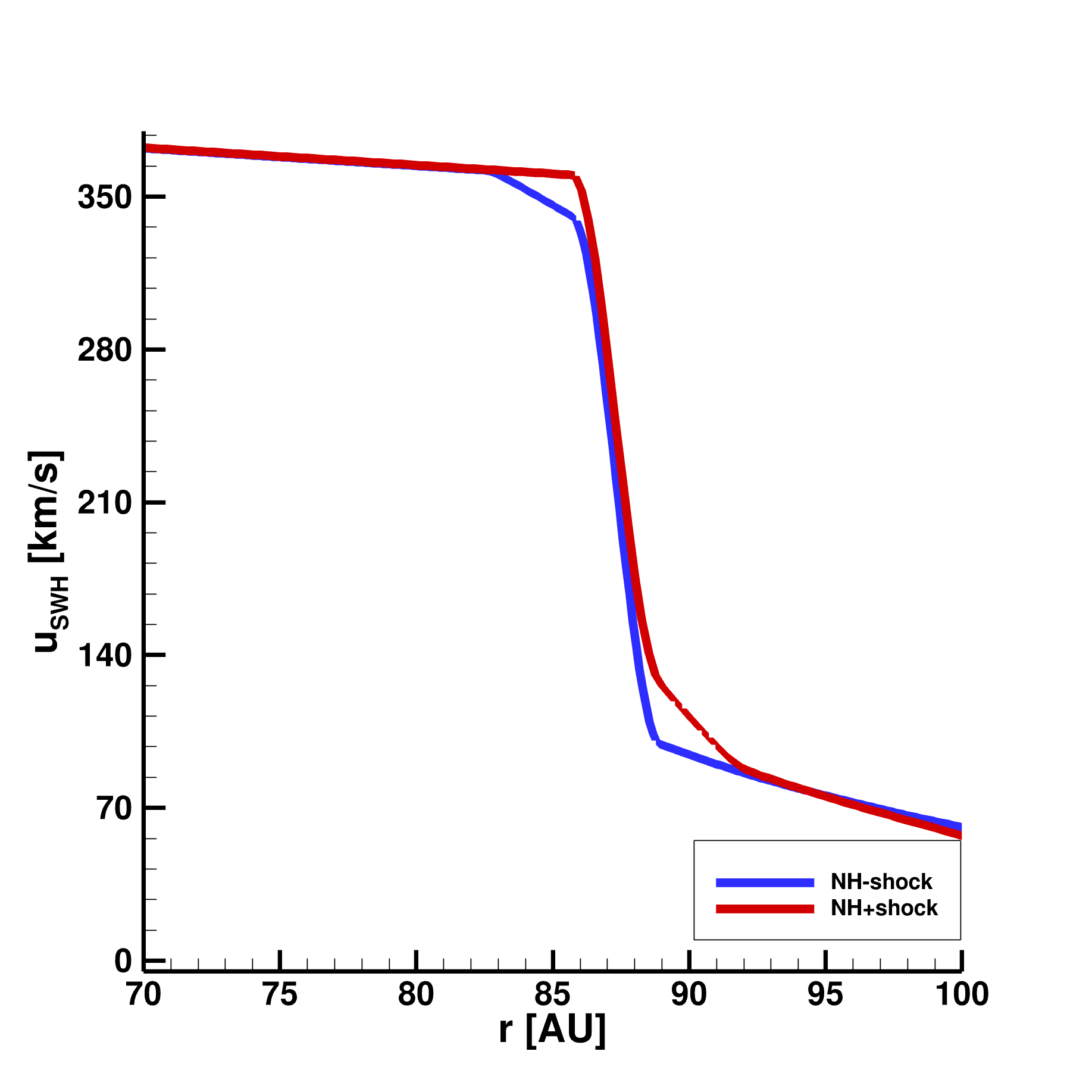}
\includegraphics[width=0.31\textwidth]{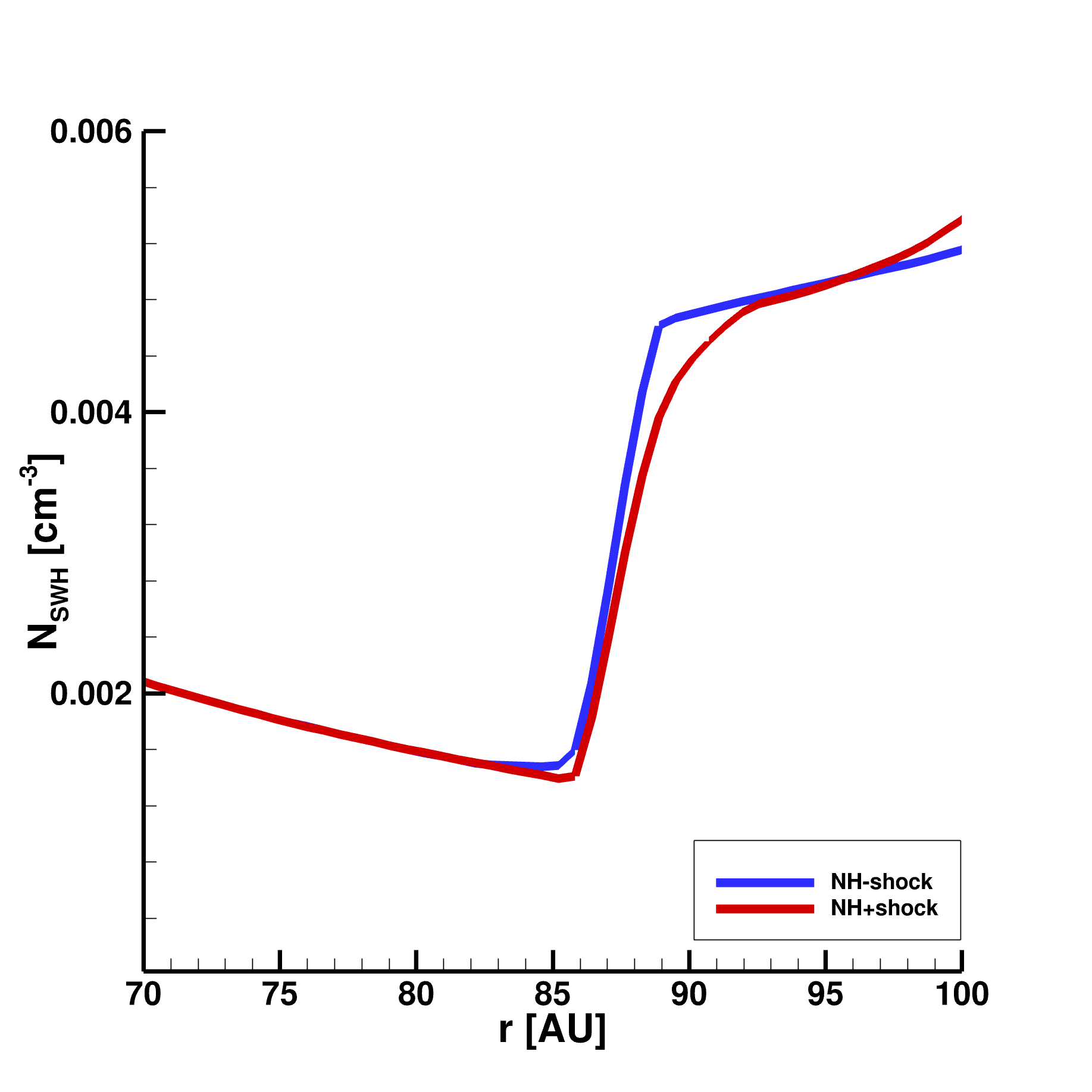}
\includegraphics[width=0.31\textwidth]{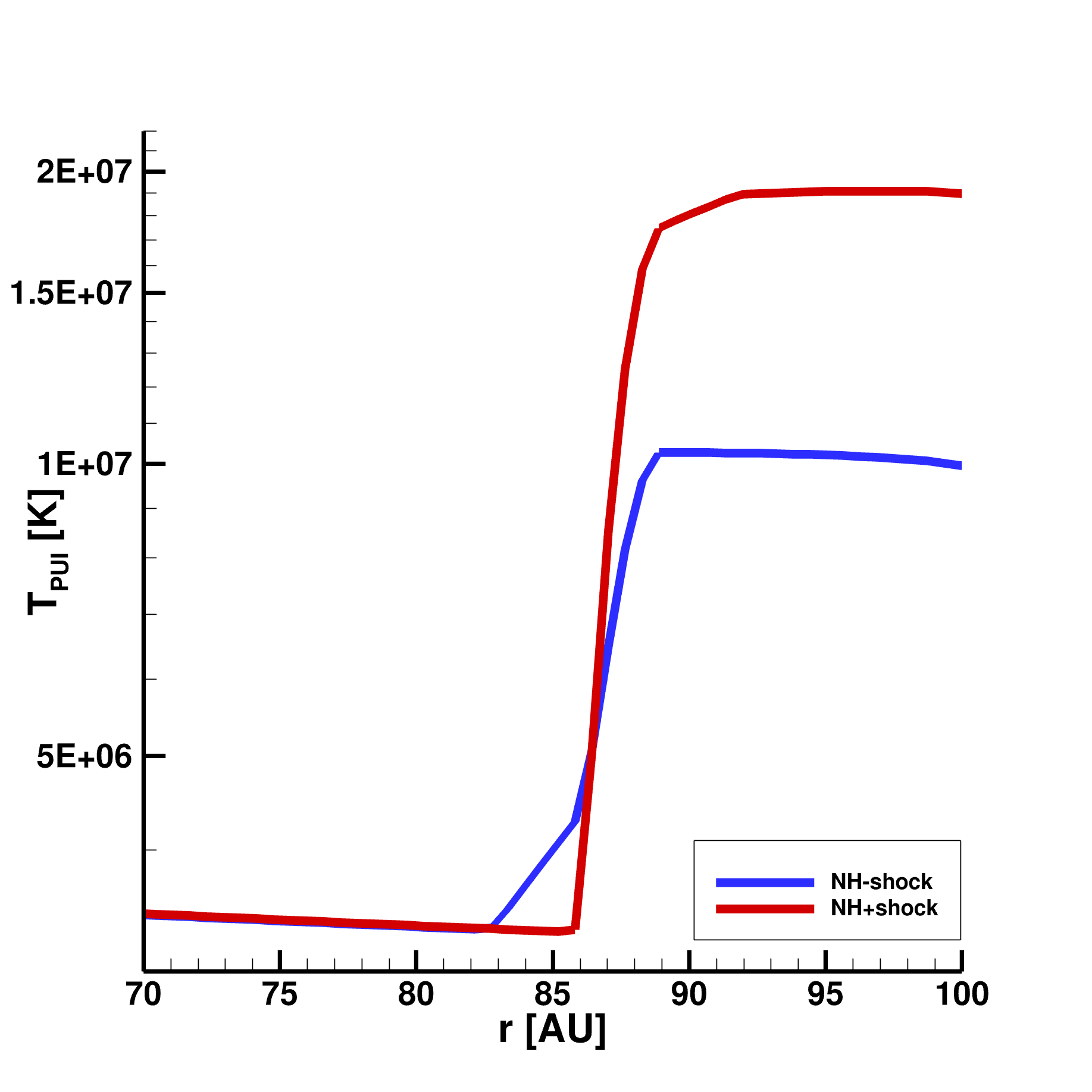}
\includegraphics[width=0.31\textwidth]{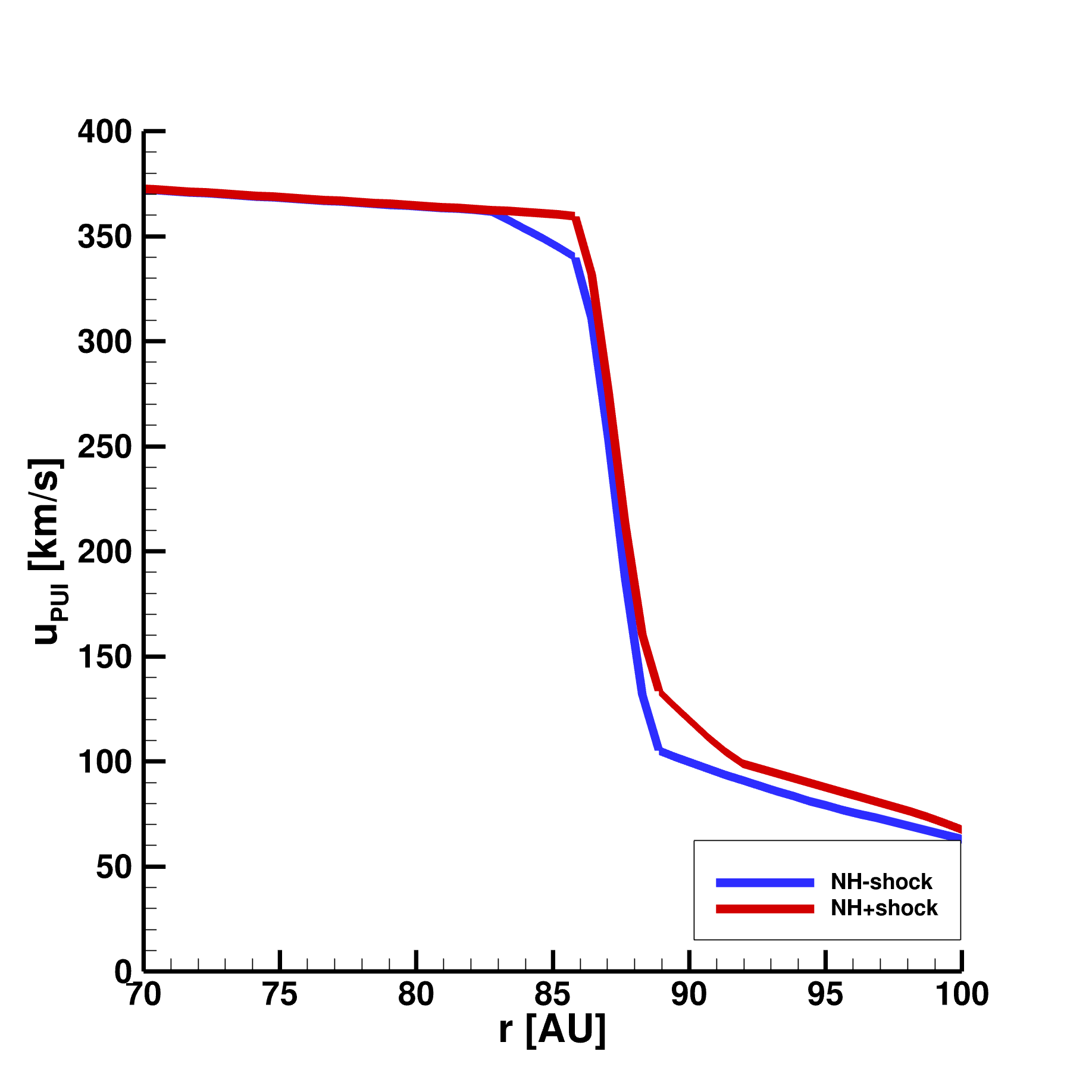}
\includegraphics[width=0.31\textwidth]{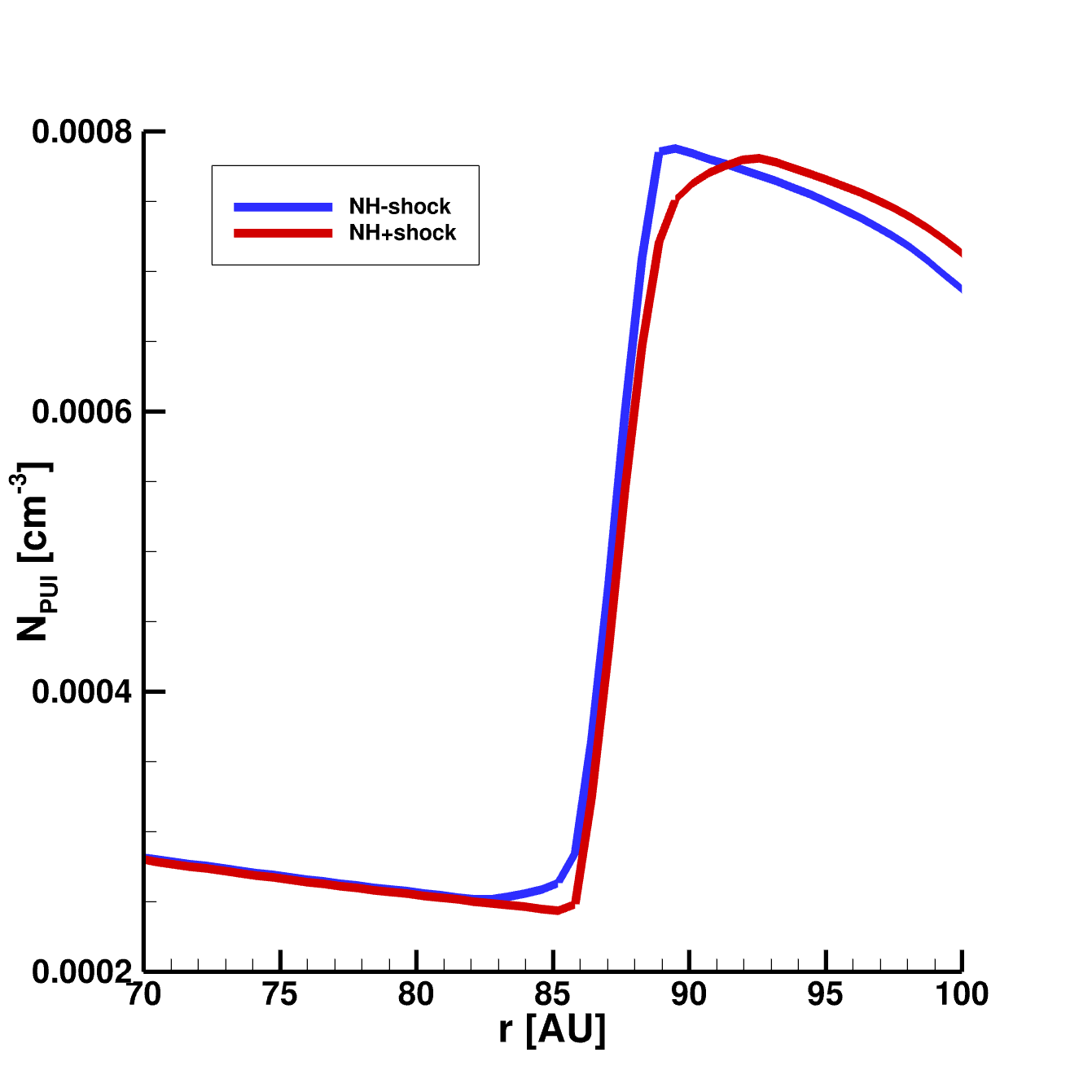}
\includegraphics[width=0.31\textwidth]{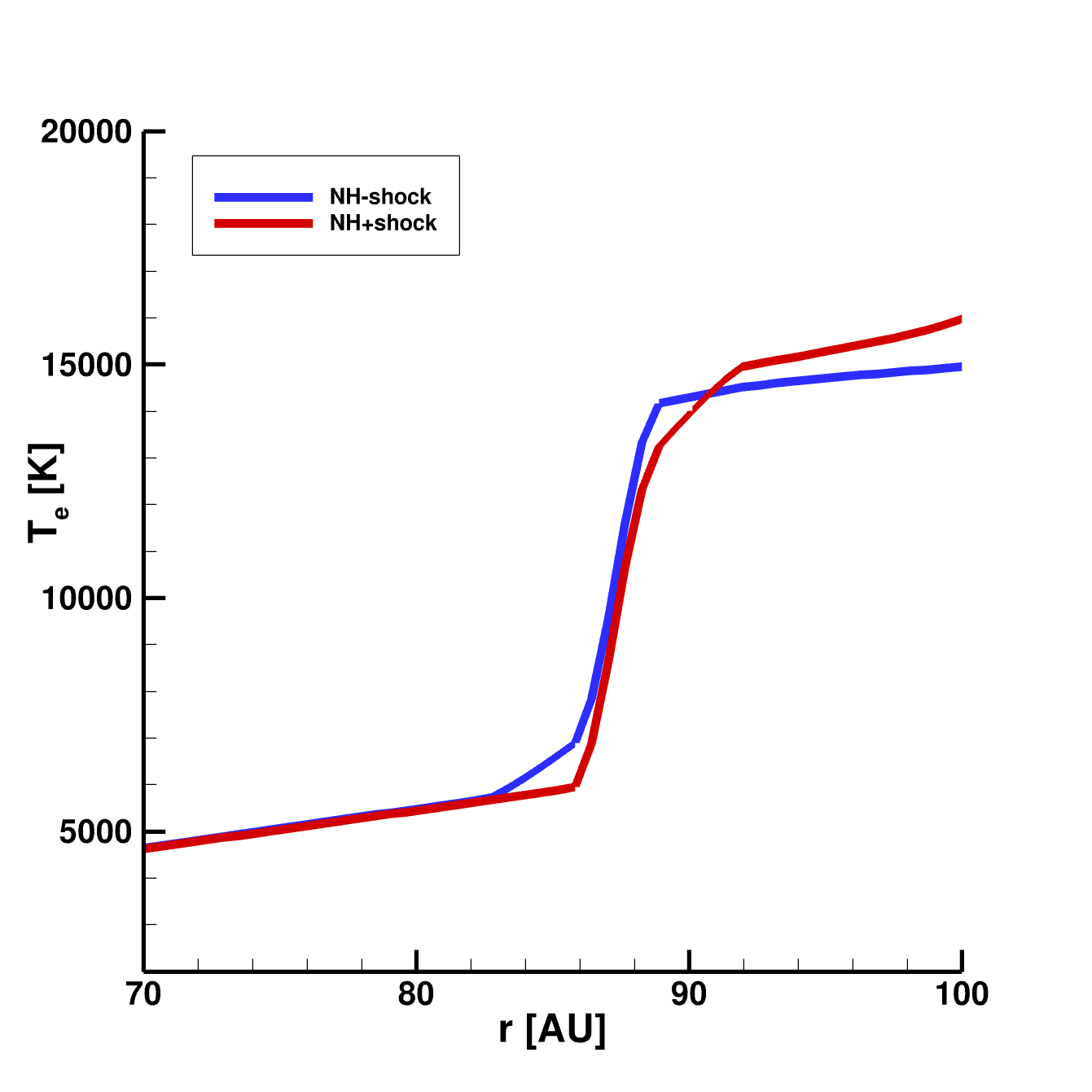}
\vspace{-2.5ex}
\caption{Prediction for the NH crossing of the TS. The panels show the magnetic field (top left), the solar wind proton temperature (top center), the solar wind speed (top right), the solar wind proton number density (second row, left), the PUI temperature (second row, center) and speed (second row, right), and PUI density (third row, left) and electron temperature (third row, right) respectively. Blue line is the solution without non-adiabatic PUI shock heating and red is with.}
\label{fig:linenh}
\end{figure}

Figure \ref{fig:linenh} shows the predictions along the NH trajectory for its upcoming TS crossing. We show how this model would predict jump conditions as opposed to the one without updated shock heating. The present work can be used to incorporate kinetic models that treat the micro-physics of the shock; into MHD models. In Appendix \ref{app:heliosheath}, we have extended the domain of the one-dimensional plots along the upstream direction (negative $x$-direction) to show the effect of the energy partitioning at the termination shock on the heliosheath.

\section{Summary}\label{sec:summary}
This paper presented a new method for partitioning the non-adiabatic shock heating among the cold protons, PUIs, and electrons, while rigorously maintaining total energy conservations. We provided an estimation of the heat-distribution between solar wind protons and PUIs across the TS. The results show that having a fraction of up to 8.5\% of the non-adiabatic shock heating going towards thermal solar wind protons at the TS reproduces the jump conditions observed along the V2 trajectory well, whereas in the original numerical scheme the thermal proton temperature jump was too large. The rest of the plasma conditions do not change significantly, meaning that the observed and simulated magnetic field jump, mass density, and speed properties of the plasma remain very similar to what was simulated in \citet{vanderHolst:2026}. We also provided predictions for solar wind proton and PUIs temperature jumps for NH's upcoming TS crossing event. Our results on the heating being mainly non-adiabatic and preferentially heating PUIs at the TS are consistent with the shock conditions described by \citet{Gedalin:2021ApJ...916...57G}. \citet{Richardson:2008GeoRL..3523104R} showed that the average increase in temperature across the TS as observed by V2 was a factor of 13. It is consistent with our results, and based on that we expect a jump about a factor 20 in case of NH.

To further improve the termination shock jumps, we could in future simulations use hybrid simulations, such as described in \citet{Giacalone:2021,Swisdak:2023,Giacalone:2025ApJ...980...29G}, to obtain the weights that represent the partitioning of non-adiabatic shock heating. It is possible that the weights are not uniform considering the non-homogeneous nature of the termination shock. In future work, we could explore how plasma parameters, such as magnetic field obliquity, the upstream Alfvén Mach number, and other factors determine the heat partitioning using lookup tables generated with a hybrid model, similar to \citet{Bera:2023}, but instead of varying the jump conditions, we will focus on varying the weights in the energy partitioning. Also, a study of grid convergence around the termination shock could be carried out. If shocks would appear in the heliotail, then this energy partitioning would also be applied for those shocks. This could also be a study for the future. For an extension to time-dependent solar wind modeling, interplanetary shocks might also appear if the variations are strong enough. The exact percentage of transfer of non-adiabatic shock heating to PUIs will depend on the number densities, since the weights in the energy partitioning contain the ratio of PUI to thermal proton number density. Near 1 AU, most of the non-adiabatic shock heating will, for instance, be transferred to the thermal protons instead of PUIs, due to the small PUI number density. Additionally, at interplanetary shocks, enhanced turbulence reflection can cause more counter propagating waves, resulting in enhanced turbulence cascade, which will increase the plasma heating of thermal protons and electrons. The amount of turbulent cascade and plasma heating will also depend on the changes of turbulence energies due to gradients in the bulk velocity at the shocks. This new method of energy partitioning has a wider application than the outer heliosphere, such as modeling of multi-fluid solar wind \citep{2026ApJ..1008...16V}, coronal mass ejections, and magnetospheres \citep{2009JGRA..11412203G}.

\appendix
\section{Heliosheath} \label{app:heliosheath}
Here, we present the extended domain of the one-dimensional plots to show the effect of the change in the energy partitioning at the termination shock on the heliosheath region. We extracted values along the negative X-axis towards the nose region. The heliosheath is between the termination shock (about 90 AU) and the heliopause (about 130 AU). The current model presented in this paper does not discuss the physical processes in the heliosheath such as magnetic field diffusion, compressible turbulence, etc. Therefore, the similarities and differences presented in these figures in the sheath region are not topics of our work. 
\begin{figure}
    \centering
    \includegraphics[trim={0.1cm 0.1cm 0.1cm 0.1cm}, clip, width=0.32\linewidth]{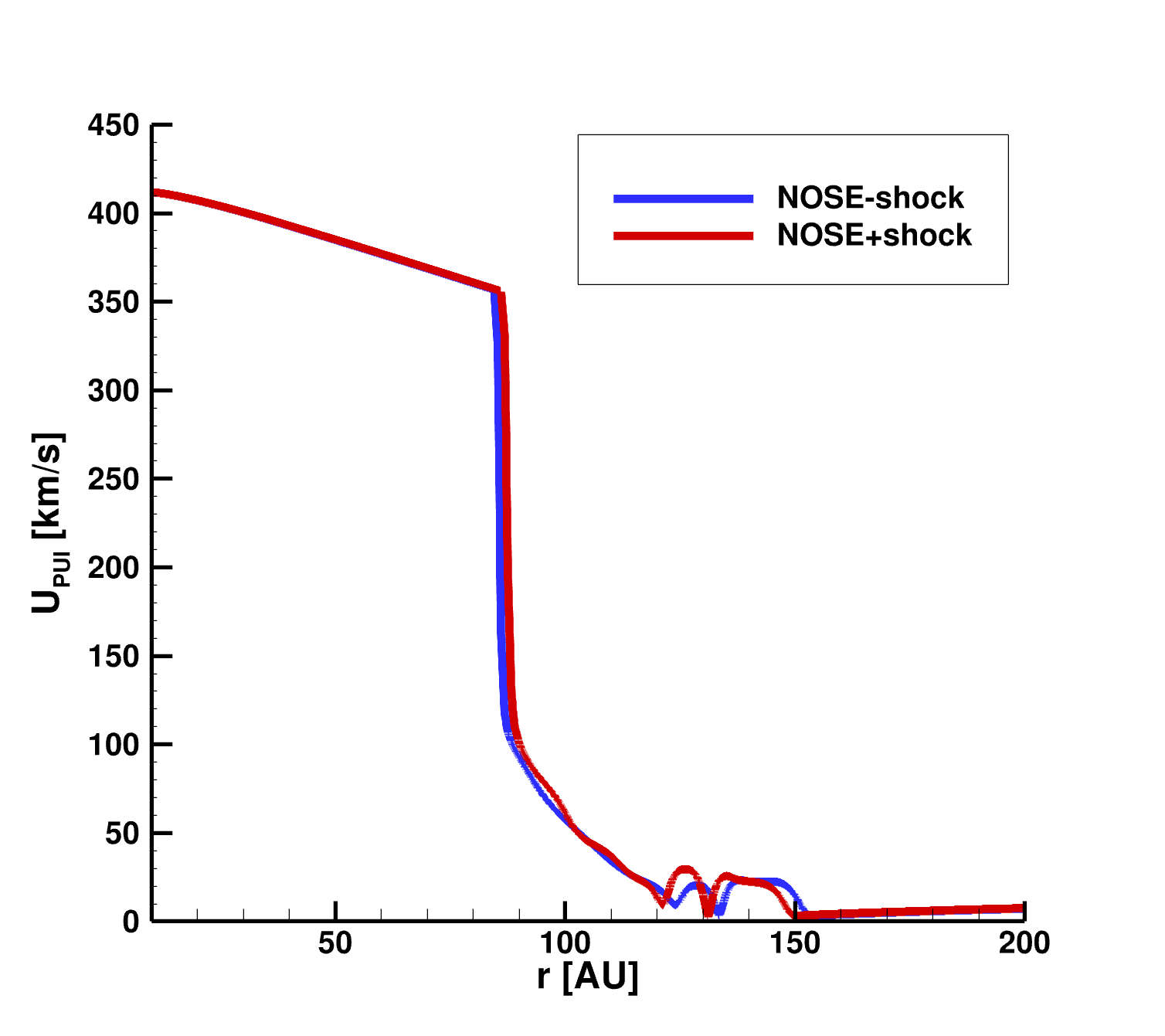}
    \includegraphics[trim={0.1cm 0.1cm 0.1cm 0.1cm}, clip, width=0.32\linewidth]{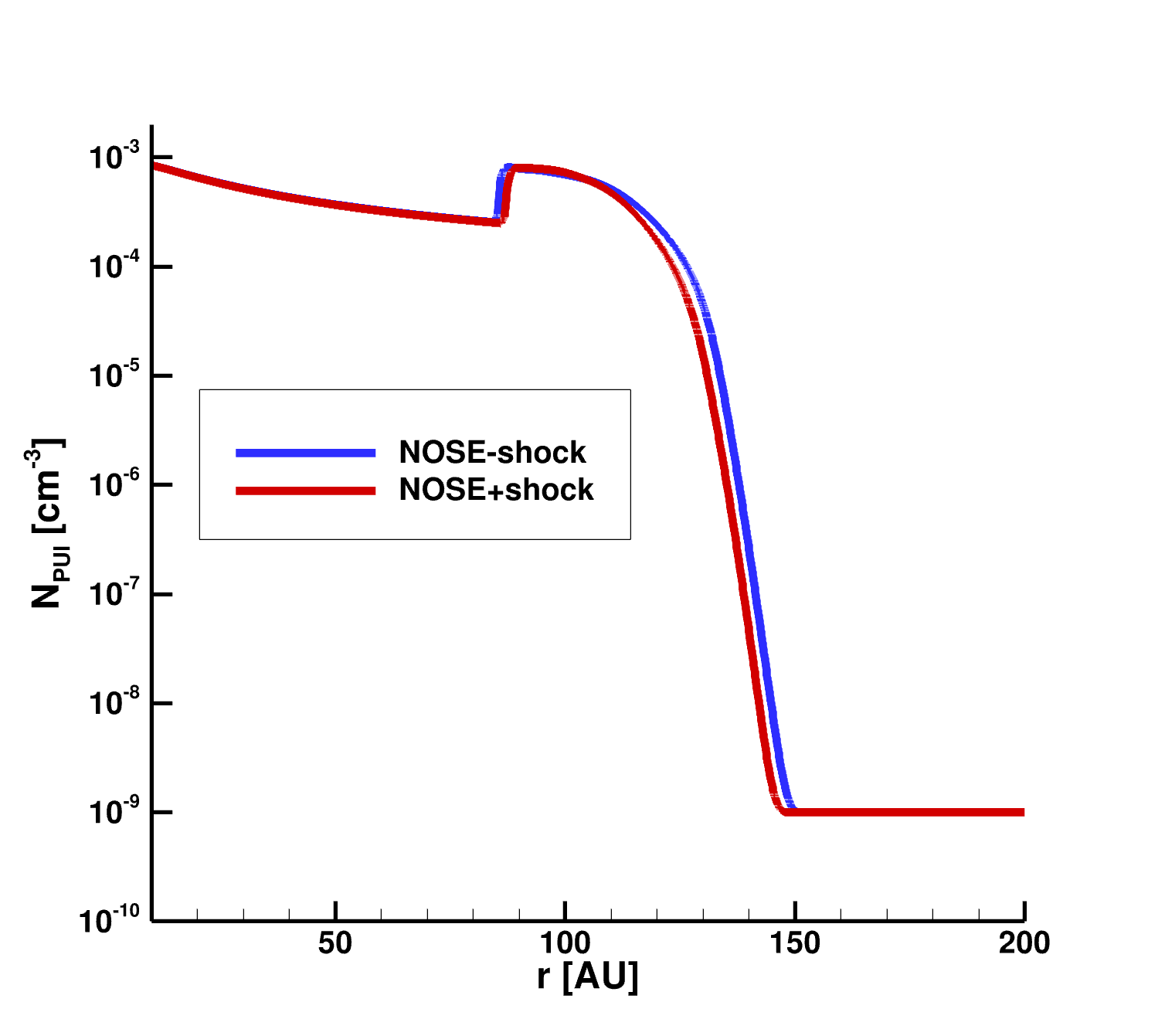}
    \includegraphics[trim={0.1cm 0.1cm 0.1cm 0.1cm}, clip, width=0.32\linewidth]{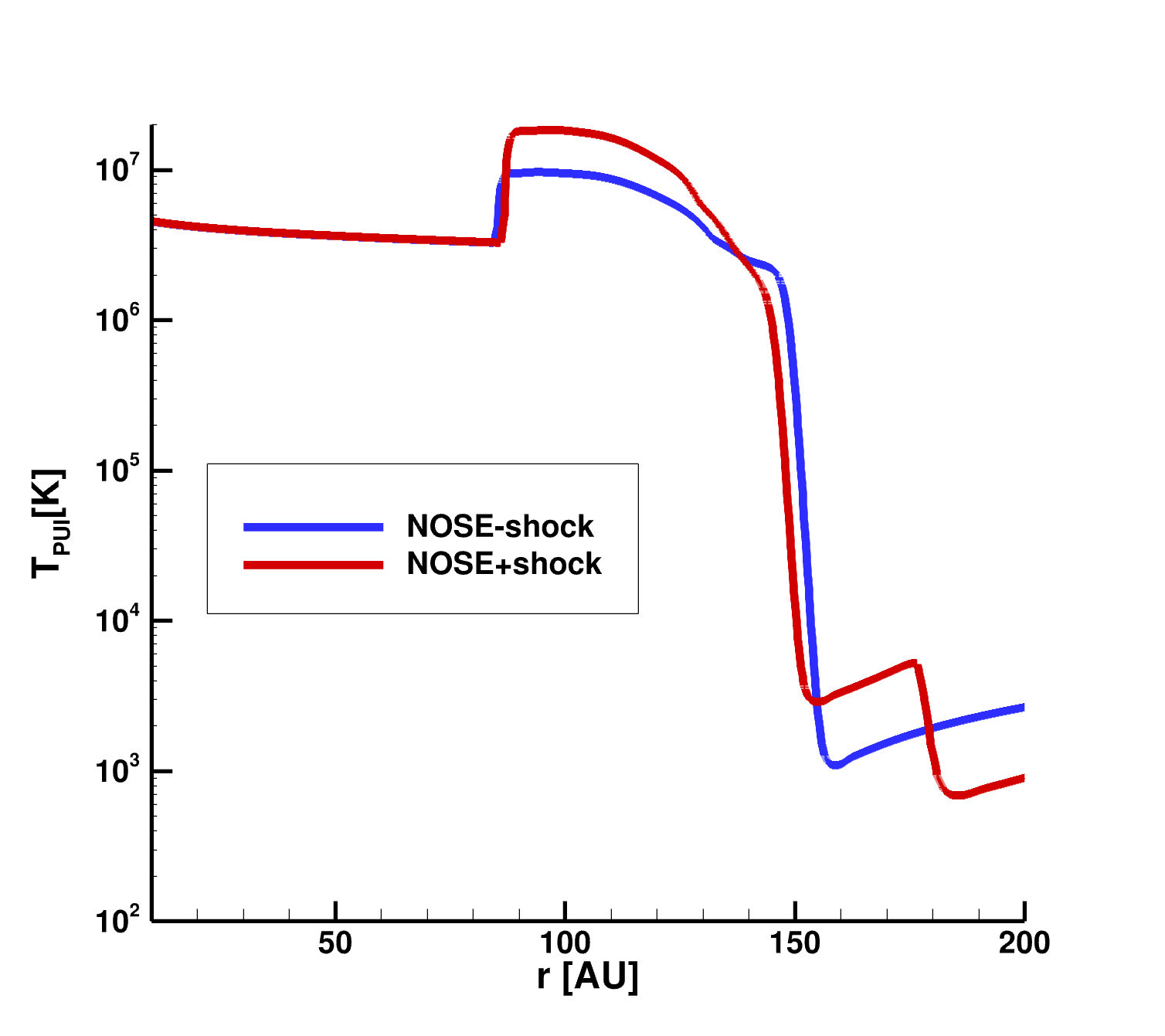}\\
    \includegraphics[trim={0.1cm 0.1cm 0.1cm 0.1cm}, clip, width=0.32\linewidth]{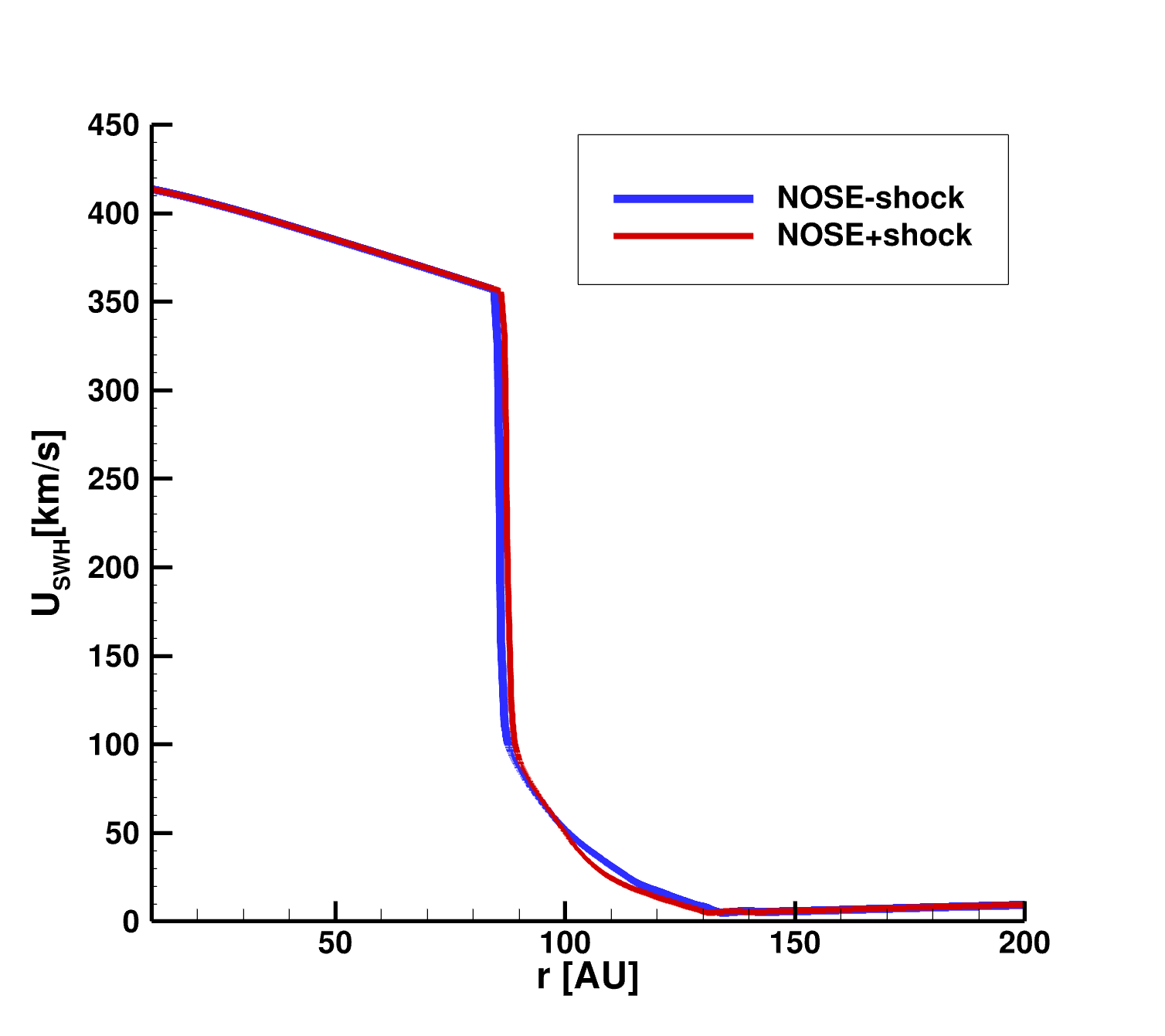}
    \includegraphics[trim={0.1cm 0.1cm 0.1cm 0.1cm}, clip, width=0.32\linewidth]{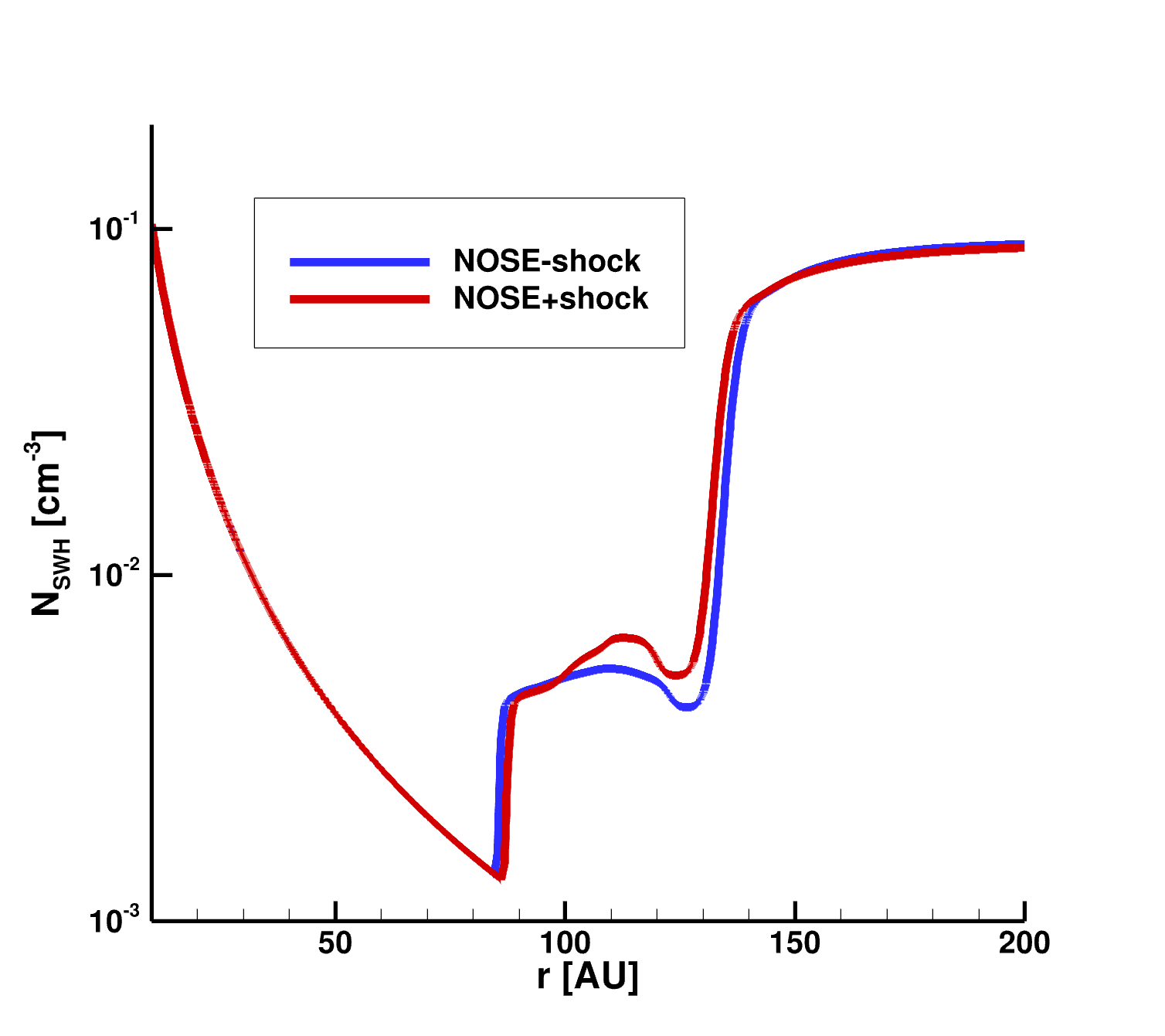}
    \includegraphics[trim={0.1cm 0.1cm 0.1cm 0.1cm}, clip, width=0.32\linewidth]{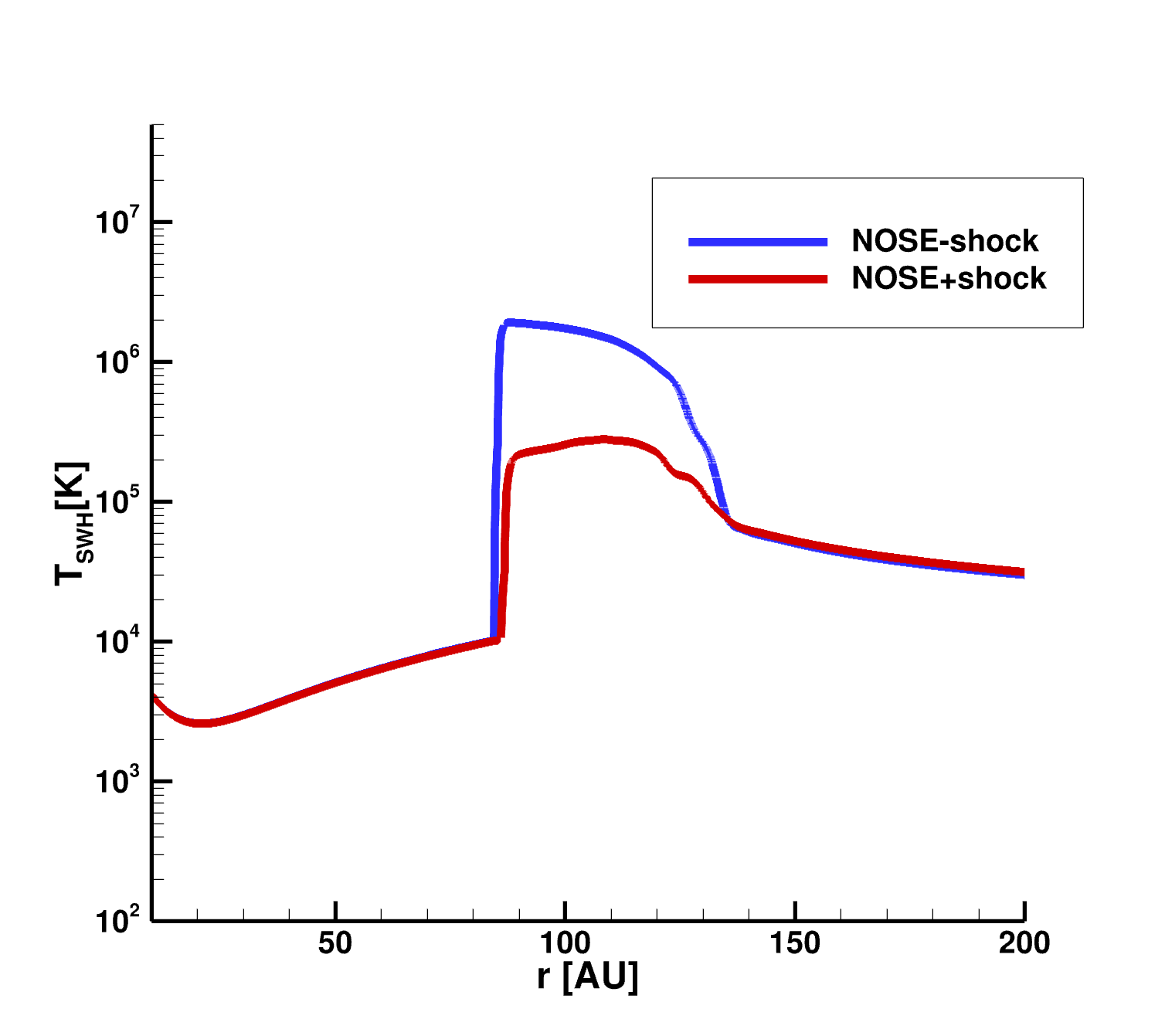}\\
    \includegraphics[trim={0.1cm 0.1cm 0.1cm 0.1cm}, clip, width=0.32\linewidth]{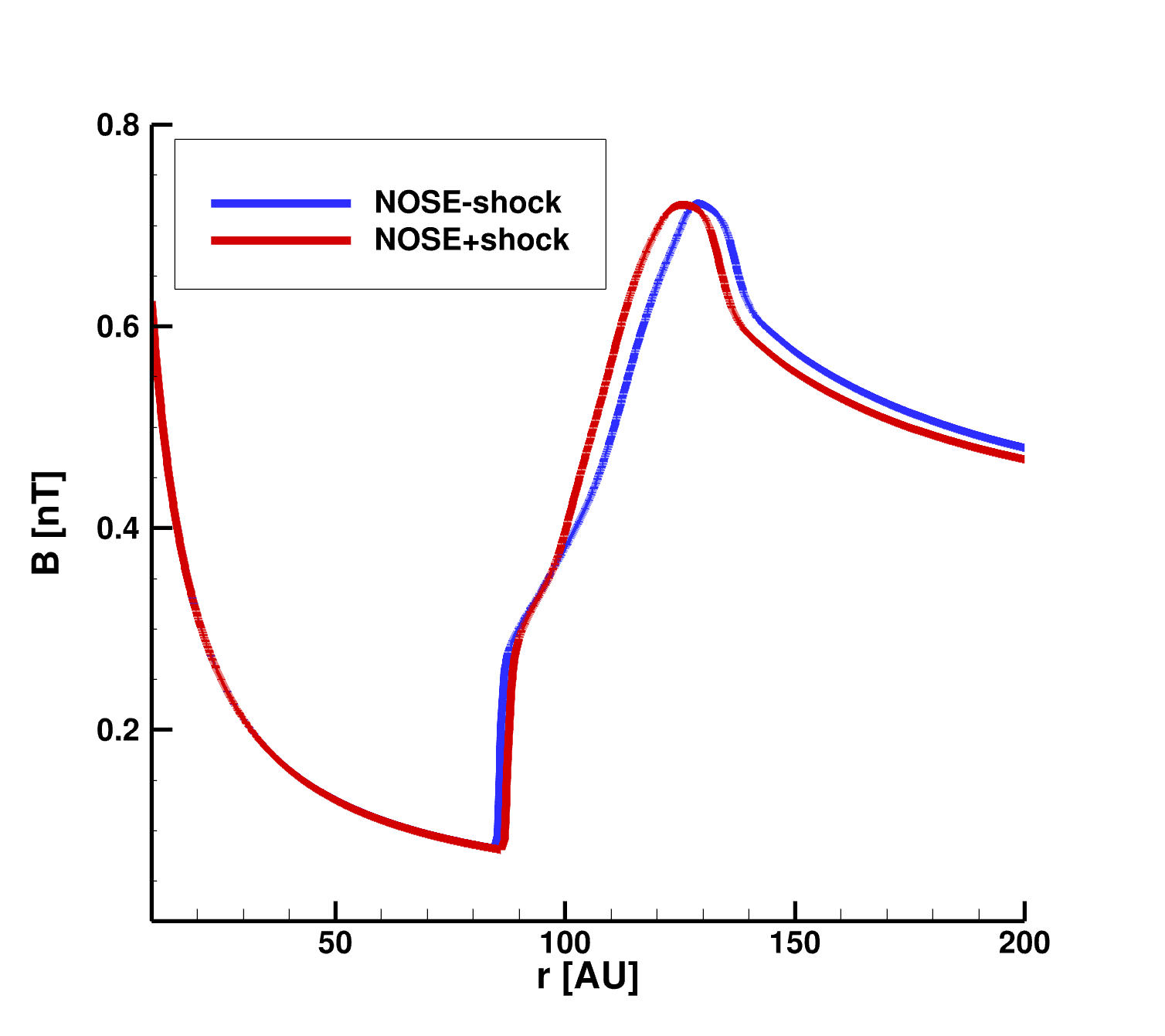}
    \includegraphics[trim={0.1cm 0.1cm 0.1cm 0.1cm}, clip, width=0.32\linewidth]{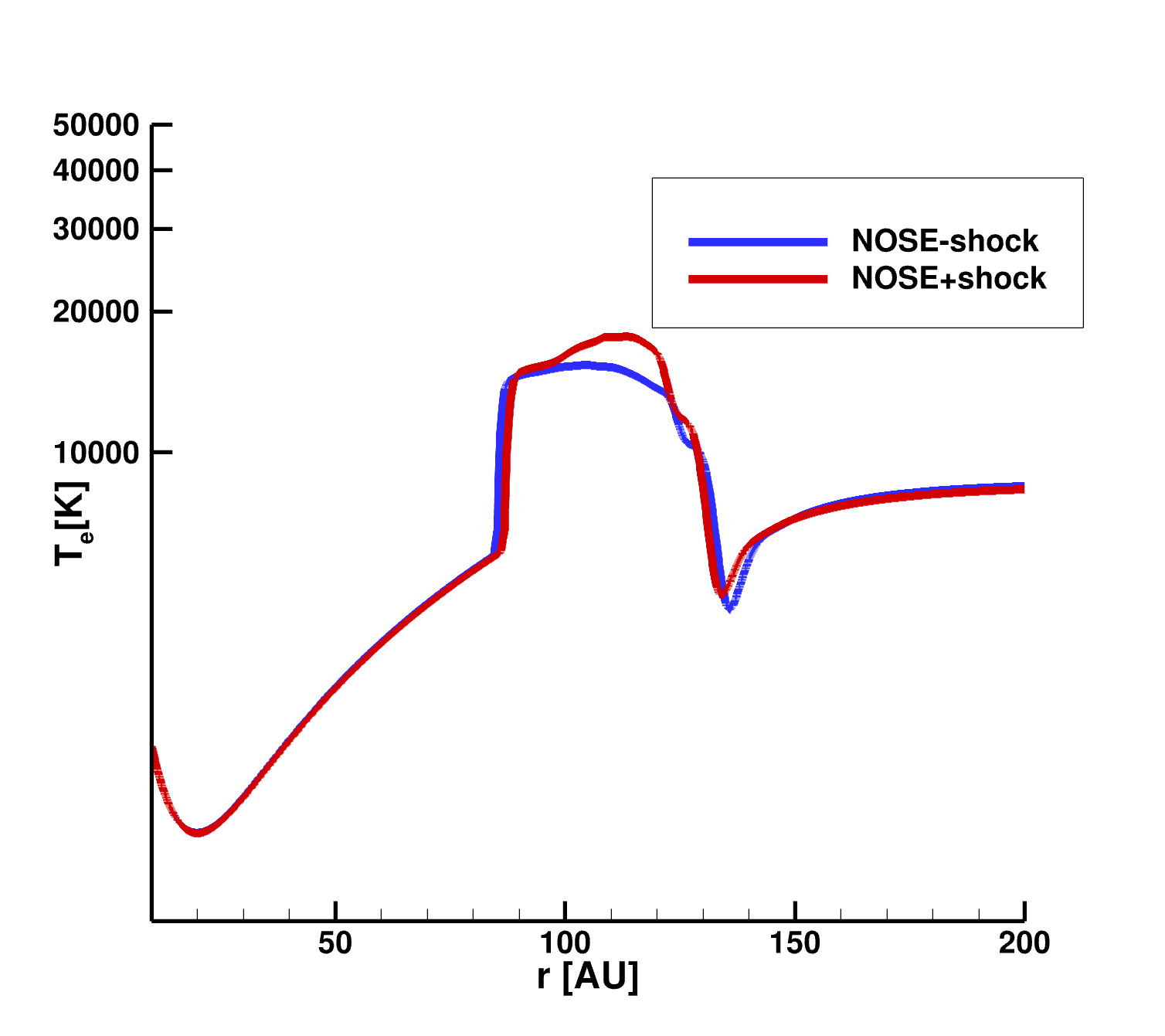}
    \caption{1D solutions regarding the heliosheath extracted along the negative X-axis, towards the nose region: PUI speed, density and temperatures \emph{in the first row left to right}, solar wind thermal proton speed, density and temperature \emph{in the middle row left to right}, and magnetic field strength and electron temperature \emph{in the third row left to right}.}
    \label{fig:appx}
\end{figure}

\begin{acknowledgments}
This work is supported by NASA grant 18-DRIVE18\_2-0029, “Our Heliospheric Shield.” For more information about this center, visit https://shielddrivecenter.com.
\end{acknowledgments}

\bibliography{references}{}

@ARTICLE{Richardson:2008Natur.454...63R,
       author = {{Richardson}, John D. and {Kasper}, Justin C. and {Wang}, Chi and {Belcher}, John W. and {Lazarus}, Alan J.},
        title = "{Cool heliosheath plasma and deceleration of the upstream solar wind at the termination shock}",
      journal = {Nature},
         year = 2008,
        month = jul,
       volume = {454},
       number = {7200},
        pages = {63-66},
          doi = {10.1038/nature07024},
       adsurl = {https://ui.adsabs.harvard.edu/abs/2008Natur.454...63R}
}

@ARTICLE{Zank:2010ApJ...708.1092Z,
       author = {{Zank}, G.~P. and {Heerikhuisen}, J. and {Pogorelov}, N.~V. and {Burrows}, R. and {McComas}, D.},
        title = "{Microstructure of the Heliospheric Termination Shock: Implications for Energetic Neutral Atom Observations}",
      journal = {\apj},
         year = 2010,
        month = jan,
       volume = {708},
       number = {2},
        pages = {1092-1106},
          doi = {10.1088/0004-637X/708/2/1092},
       adsurl = {https://ui.adsabs.harvard.edu/abs/2010ApJ...708.1092Z}
}

@article{Parker1958,
  author  = {Parker, E. N.},
  title   = {Dynamics of the Interplanetary Gas and Magnetic Fields},
  journal = {The Astrophysical Journal},
  year    = {1958},
  volume  = {128},
  pages   = {664},
  doi     = {10.1086/146579}
}

@ARTICLE{Burlaga:2008Natur.454...75B,
       author = {{Burlaga}, L.~F. and {Ness}, N.~F. and {Acu{\~n}a}, M.~H. and {Lepping}, R.~P. and {Connerney}, J.~E.~P. and {Richardson}, J.~D.},
        title = "{Magnetic fields at the solar wind termination shock}",
      journal = {\nat},
         year = 2008,
        month = jul,
       volume = {454},
       number = {7200},
        pages = {75-77},
          doi = {10.1038/nature07029},
       adsurl = {https://ui.adsabs.harvard.edu/abs/2008Natur.454...75B}
}

@ARTICLE{Opher:2020,
       author = {{Opher}, Merav and {Loeb}, Abraham and {Drake}, James and {Toth}, Gabor},
        title = "{A small and round heliosphere suggested by magnetohydrodynamic modelling of pick-up ions}",
      journal = {Nature Astronomy},
         year = 2020,
        month = mar,
       volume = {4},
        pages = {675-683},
          doi = {10.1038/s41550-020-1036-0},
archivePrefix = {arXiv},
       eprint = {1808.06611},
 primaryClass = {physics.space-ph},
       adsurl = {https://ui.adsabs.harvard.edu/abs/2020NatAs...4..675O}
}

@ARTICLE{Toth:2024,
       author = {{T{\'o}th}, G{\'a}bor and {van der Holst}, Bart},
        title = "{Weak solutions for extended magnetohydrodynamics using linear combination of entropies}",
      journal = {Journal of Computational Physics},
         year = 2024,
        month = jul,
       volume = {508},
          eid = {113036},
        pages = {113036},
          doi = {10.1016/j.jcp.2024.113036},
       adsurl = {https://ui.adsabs.harvard.edu/abs/2024JCoPh.50813036T}
}

@ARTICLE{Zieger:2015,
       author = {{Zieger}, Bertalan and {Opher}, Merav and {T{\'o}th}, G{\'a}bor and {Decker}, Robert B. and {Richardson}, John D.},
        title = "{Constraining the pickup ion abundance and temperature through the multifluid reconstruction of the Voyager 2 termination shock crossing}",
      journal = {Journal of Geophysical Research (Space Physics)},
         year = 2015,
        month = sep,
       volume = {120},
       number = {9},
        pages = {7130-7153},
          doi = {10.1002/2015JA021437},
       adsurl = {https://ui.adsabs.harvard.edu/abs/2015JGRA..120.7130Z}
}

@ARTICLE{Bera:2023,
       author = {{Bera}, R.~K. and {Fraternale}, F. and {Pogorelov}, N.~V. and {Roytershteyn}, V. and {Gedalin}, M. and {McComas}, D.~J. and {Zank}, G.~P.},
        title = "{The Role of Pickup Ions in the Interaction of the Solar Wind with the Local Interstellar Medium. I. Importance of Kinetic Processes at the Heliospheric Termination Shock}",
      journal = {The Astrophysical Journal},
         year = 2023,
        month = sep,
       volume = {954},
       number = {2},
          eid = {147},
        pages = {147},
          doi = {10.3847/1538-4357/acea7d},
       adsurl = {https://ui.adsabs.harvard.edu/abs/2023ApJ...954..147B}
}

@ARTICLE{Toth:2012,
       author = {{T{\'o}th}, G{\'a}bor and {van der Holst}, Bart and {Sokolov}, Igor V. and {De Zeeuw}, Darren L. and {Gombosi}, Tamas I. and {Fang}, Fang and {Manchester}, Ward B. and {Meng}, Xing and {Najib}, Dalal and {Powell}, Kenneth G. and {Stout}, Quentin F. and {Glocer}, Alex and {Ma}, Ying-Juan and {Opher}, Merav},
        title = "{Adaptive numerical algorithms in space weather modeling}",
      journal = {Journal of Computational Physics},
         year = 2012,
        month = feb,
       volume = {231},
       number = {3},
        pages = {870-903},
          doi = {10.1016/j.jcp.2011.02.006},
       adsurl = {https://ui.adsabs.harvard.edu/abs/2012JCoPh.231..870T}
}

@ARTICLE{Swisdak:2023,
       author = {{Swisdak}, M. and {Giacalone}, J. and {Drake}, J.~F. and {Opher}, M. and {Zank}, G.~P. and {Zieger}, B.},
        title = "{A Comparison of Particle-in-cell and Hybrid Simulations of the Heliospheric Termination Shock}",
      journal = {The Astrophysical Journal},
         year = 2023,
        month = dec,
       volume = {959},
       number = {1},
          eid = {4},
        pages = {4},
          doi = {10.3847/1538-4357/ad03e2},
       adsurl = {https://ui.adsabs.harvard.edu/abs/2023ApJ...959....4S}
}

@ARTICLE{Giacalone:2021,
       author = {{Giacalone}, J. and {Nakanotani}, M. and {Zank}, G.~P. and {K{\`o}ta}, J. and {Opher}, M. and {Richardson}, J.~D.},
        title = "{Hybrid Simulations of Interstellar Pickup Protons Accelerated at the Solar-wind Termination Shock at Multiple Locations}",
      journal = {The Astrophysical Journal},
         year = 2021,
        month = apr,
       volume = {911},
       number = {1},
          eid = {27},
        pages = {27},
          doi = {10.3847/1538-4357/abe93a},
       adsurl = {https://ui.adsabs.harvard.edu/abs/2021ApJ...911...27G}
}

@ARTICLE{vanderHolst:2026,
       author = {{van der Holst}, Bart and {Opher}, Merav and {Adhikari}, Laxman and {Zank}, Gary P.},
        title = "{Pickup Ion Production and Turbulence in the Global Heliosphere}",
      journal = {\apj},
         year = 2026,
        month = jun,
       volume = {1004},
       number = {2},
          eid = {141},
        pages = {141},
          doi = {10.3847/1538-4357/ae6cd9},
       adsurl = {https://ui.adsabs.harvard.edu/abs/2026ApJ..1004..141V}
}

@ARTICLE{Decker:2008,
       author = {{Decker}, R.~B. and {Krimigis}, S.~M. and {Roelof}, E.~C. and {Hill}, M.~E. and {Armstrong}, T.~P. and {Gloeckler}, G. and {Hamilton}, D.~C. and {Lanzerotti}, L.~J.},
        title = "{Mediation of the solar wind termination shock by non-thermal ions}",
      journal = {\nat},
         year = 2008,
        month = jul,
       volume = {454},
       number = {7200},
        pages = {67-70},
          doi = {10.1038/nature07030},
       adsurl = {https://ui.adsabs.harvard.edu/abs/2008Natur.454...67D}
}

@ARTICLE{Usmanov:2016,
       author = {{Usmanov}, Arcadi V. and {Goldstein}, Melvyn L. and {Matthaeus}, William H.},
        title = "{A Four-fluid MHD Model of the Solar Wind/Interstellar Medium Interaction with Turbulence Transport and Pickup Protons as Separate Fluid}",
      journal = {\apj},
         year = 2016,
        month = mar,
       volume = {820},
       number = {1},
          eid = {17},
        pages = {17},
          doi = {10.3847/0004-637X/820/1/17},
       adsurl = {https://ui.adsabs.harvard.edu/abs/2016ApJ...820...17U}
}

@ARTICLE{Burlaga:2005Sci...309.2027B,
       author = {{Burlaga}, L.~F. and {Ness}, N.~F. and {Acu{\~n}a}, M.~H. and {Lepping}, R.~P. and {Connerney}, J.~E.~P. and {Stone}, E.~C. and {McDonald}, F.~B.},
        title = "{Crossing the Termination Shock into the Heliosheath: Magnetic Fields}",
      journal = {Science},
         year = 2005,
        month = sep,
       volume = {309},
       number = {5743},
        pages = {2027-2029},
          doi = {10.1126/science.1117542},
       adsurl = {https://ui.adsabs.harvard.edu/abs/2005Sci...309.2027B}
}

@ARTICLE{Bair:2025,
       author = {{Bair}, Ethan Schuyler and {Opher}, Merav and {Kornbleuth}, Marc Zachary and {Zieger}, Bertalan and {Toth}, Gabor and {van der Holst}, Bart},
        title = "{Consequences of Hot Electrons for the Structure of the Outer Heliosphere}",
      journal = {\apj},
         year = 2025,
        month = aug,
       volume = {988},
       number = {2},
          eid = {223},
        pages = {223},
          doi = {10.3847/1538-4357/ade437},
       adsurl = {https://ui.adsabs.harvard.edu/abs/2025ApJ...988..223B}
}

@ARTICLE{Zirnstein:2025NatAs...9.1495Z,
       author = {{Zirnstein}, E.~J. and {Kumar}, R. and {Shrestha}, B.~L. and {Swaczyna}, P. and {Dayeh}, M.~A. and {Heerikhuisen}, J. and {Szalay}, J.~R.},
        title = "{Global heliospheric termination shock strength in the solar-interstellar interaction}",
      journal = {Nature Astronomy},
         year = 2025,
        month = aug,
       volume = {9},
        pages = {1495-1510},
          doi = {10.1038/s41550-025-02634-3},
archivePrefix = {arXiv},
       eprint = {2501.15004},
 primaryClass = {astro-ph.SR},
       adsurl = {https://ui.adsabs.harvard.edu/abs/2025NatAs...9.1495Z}
}

@ARTICLE{Gedalin:2021ApJ...916...57G,
       author = {{Gedalin}, Michael and {Pogorelov}, Nikolai V. and {Roytershteyn}, Vadim},
        title = "{Boundary Conditions at the Heliospheric Termination Shock with Pickup Ions}",
      journal = {\apj},
         year = 2021,
        month = jul,
       volume = {916},
       number = {1},
          eid = {57},
        pages = {57},
          doi = {10.3847/1538-4357/ac05b7},
       adsurl = {https://ui.adsabs.harvard.edu/abs/2021ApJ...916...57G}
}

@ARTICLE{Richardson:2008GeoRL..3523104R,
       author = {{Richardson}, J.~D.},
        title = "{Plasma temperature distributions in the heliosheath}",
      journal = {\grl},
         year = 2008,
        month = dec,
       volume = {35},
       number = {23},
          eid = {L23104},
        pages = {L23104},
          doi = {10.1029/2008GL036168},
       adsurl = {https://ui.adsabs.harvard.edu/abs/2008GeoRL..3523104R}
}

@ARTICLE{Giacalone:2025ApJ...980...29G,
       author = {{Giacalone}, Joe and {Kornbleuth}, M. and {Opher}, M. and {Gkioulidou}, M. and {K{\"o}ta}, J. and {Puzzoni}, E. and {Richardson}, J.~D. and {Zank}, G.~P.},
        title = "{Hybrid Simulations of Interstellar Pickup Ions at the Solar Wind Termination Shock Revisited}",
      journal = {\apj},
         year = 2025,
        month = feb,
       volume = {980},
       number = {1},
          eid = {29},
        pages = {29},
          doi = {10.3847/1538-4357/ada89c},
       adsurl = {https://ui.adsabs.harvard.edu/abs/2025ApJ...980...29G}
}

@ARTICLE{Zhao:2019ApJ...879...32Z,
       author = {{Zhao}, L.-L. and {Zank}, G.~P. and {Adhikari}, L.},
        title = "{Generation Mechanisms for Low-energy Interstellar Pickup Ions}",
      journal = {\apj},
         year = 2019,
        month = jul,
       volume = {879},
       number = {1},
          eid = {32},
        pages = {32},
          doi = {10.3847/1538-4357/ab2381},
       adsurl = {https://ui.adsabs.harvard.edu/abs/2019ApJ...879...32Z}
}

@INPROCEEDINGS{Burlaga:2015JPhCS.642a2003B,
       author = {{Burlaga}, L.},
        title = "{Voyager observations of the magnetic field in the heliosheath and the local interstellar medium}",
    booktitle = {Journal of Physics Conference Series},
         year = 2015,
       series = {Journal of Physics Conference Series},
       volume = {642},
        month = sep,
    publisher = {IOP},
          eid = {012003},
        pages = {012003},
          doi = {10.1088/1742-6596/642/1/012003},
       adsurl = {https://ui.adsabs.harvard.edu/abs/2015JPhCS.642a2003B}
}

@ARTICLE{Zirnstein:2022SSRv..218...28Z,
       author = {{Zirnstein}, E.~J. and {M{\"o}bius}, E. and {Zhang}, M. and {Bower}, J. and {Elliott}, H.~A. and {McComas}, D.~J. and {Pogorelov}, N.~V. and {Swaczyna}, P.},
        title = "{In Situ Observations of Interstellar Pickup Ions from 1 au to the Outer Heliosphere}",
      journal = {\ssr},
         year = 2022,
        month = jun,
       volume = {218},
       number = {4},
          eid = {28},
        pages = {28},
          doi = {10.1007/s11214-022-00895-2},
       adsurl = {https://ui.adsabs.harvard.edu/abs/2022SSRv..218...28Z}
}

@ARTICLE{Livadiotis:2024ApJ...968...66L,
       author = {{Livadiotis}, G. and {McComas}, D.~J. and {Shrestha}, Bishwas. L.},
        title = "{Thermodynamics of Pickup Ions in the Heliosphere}",
      journal = {\apj},
         year = 2024,
        month = jun,
       volume = {968},
       number = {2},
          eid = {66},
        pages = {66},
          doi = {10.3847/1538-4357/ad3e79},
       adsurl = {https://ui.adsabs.harvard.edu/abs/2024ApJ...968...66L}
}

@ARTICLE{Giacalone:2010GeoRL..3719104G,
       author = {{Giacalone}, Joe and {Burgess}, David},
        title = "{Interaction between inclined current sheets and the heliospheric termination shock}",
      journal = {\grl},
         year = 2010,
        month = oct,
       volume = {37},
       number = {19},
          eid = {L19104},
        pages = {L19104},
          doi = {10.1029/2010GL044656},
       adsurl = {https://ui.adsabs.harvard.edu/abs/2010GeoRL..3719104G}
}

@ARTICLE{Opher:2003ApJ...591L..61O,
       author = {{Opher}, Merav and {Liewer}, Paulett C. and {Gombosi}, Tamas I. and {Manchester}, Ward and {DeZeeuw}, Darren L. and {Sokolov}, Igor and {Toth}, Gabor},
        title = "{Probing the Edge of the Solar System: Formation of an Unstable Jet-Sheet}",
      journal = {\apjl},
         year = 2003,
        month = jul,
       volume = {591},
       number = {1},
        pages = {L61-L65},
          doi = {10.1086/376960},
archivePrefix = {arXiv},
       eprint = {astro-ph/0305420},
 primaryClass = {astro-ph},
       adsurl = {https://ui.adsabs.harvard.edu/abs/2003ApJ...591L..61O}
}

@ARTICLE{Opher:2009Natur.462.1036O,
       author = {{Opher}, M. and {Bibi}, F. Alouani and {Toth}, G. and {Richardson}, J.~D. and {Izmodenov}, V.~V. and {Gombosi}, T.~I.},
        title = "{A strong, highly-tilted interstellar magnetic field near the Solar System}",
      journal = {\nat},
         year = 2009,
        month = dec,
       volume = {462},
       number = {7276},
        pages = {1036-1038},
          doi = {10.1038/nature08567},
       adsurl = {https://ui.adsabs.harvard.edu/abs/2009Natur.462.1036O}
}

@ARTICLE{Opher:2015ApJ...800L..28O,
       author = {{Opher}, M. and {Drake}, J.~F. and {Zieger}, B. and {Gombosi}, T.~I.},
        title = "{Magnetized Jets Driven By the Sun: the Structure of the Heliosphere Revisited}",
      journal = {\apjl},
         year = 2015,
        month = feb,
       volume = {800},
       number = {2},
          eid = {L28},
        pages = {L28},
          doi = {10.1088/2041-8205/800/2/L28},
archivePrefix = {arXiv},
       eprint = {1412.7687},
 primaryClass = {astro-ph.SR},
       adsurl = {https://ui.adsabs.harvard.edu/abs/2015ApJ...800L..28O}
}

@ARTICLE{Usmanov:2014,
       author = {{Usmanov}, Arcadi V. and {Goldstein}, Melvyn L. and {Matthaeus}, William H.},
        title = "{Three-fluid, Three-dimensional Magnetohydrodynamic Solar Wind Model with Eddy Viscosity and Turbulent Resistivity}",
      journal = {\apj},
         year = 2014,
        month = jun,
       volume = {788},
       number = {1},
          eid = {43},
        pages = {43},
          doi = {10.1088/0004-637X/788/1/43},
       adsurl = {https://ui.adsabs.harvard.edu/abs/2014ApJ...788...43U}
}

@ARTICLE{Usmanov:2011,
       author = {{Usmanov}, Arcadi V. and {Matthaeus}, William H. and {Breech}, Benjamin A. and {Goldstein}, Melvyn L.},
        title = "{Solar Wind Modeling with Turbulence Transport and Heating}",
      journal = {\apj},
         year = 2011,
        month = feb,
       volume = {727},
       number = {2},
          eid = {84},
        pages = {84},
          doi = {10.1088/0004-637X/727/2/84},
       adsurl = {https://ui.adsabs.harvard.edu/abs/2011ApJ...727...84U}
}

@ARTICLE{Usmanov:2018,
       author = {{Usmanov}, Arcadi V. and {Matthaeus}, William H. and {Goldstein}, Melvyn L. and {Chhiber}, Rohit},
        title = "{The Steady Global Corona and Solar Wind: A Three-dimensional MHD Simulation with Turbulence Transport and Heating}",
      journal = {\apj},
         year = 2018,
        month = sep,
       volume = {865},
       number = {1},
          eid = {25},
        pages = {25},
          doi = {10.3847/1538-4357/aad687},
       adsurl = {https://ui.adsabs.harvard.edu/abs/2018ApJ...865...25U}
}

@ARTICLE{Williams1995,
       author = {{Williams}, L.~L. and {Zank}, G.~P. and {Matthaeus}, W.~H.},
        title = "{Dissipation of pickup-induced waves: A solar wind temperature increase in the outer heliosphere?}",
      journal = {\jgr},
         year = 1995,
        month = sep,
       volume = {100},
       number = {A9},
        pages = {17059-17068},
          doi = {10.1029/95JA01261},
       adsurl = {https://ui.adsabs.harvard.edu/abs/1995JGR...10017059W}
}

@ARTICLE{Zank:1996,
       author = {{Zank}, G.~P. and {Matthaeus}, W.~H. and {Smith}, C.~W.},
        title = "{Evolution of turbulent magnetic fluctuation power with heliospheric distance}",
      journal = {\jgr},
         year = 1996,
        month = aug,
       volume = {101},
       number = {A8},
        pages = {17093-17108},
          doi = {10.1029/96JA01275},
       adsurl = {https://ui.adsabs.harvard.edu/abs/1996JGR...10117093Z}
}

@ARTICLE{Isenberg2003,
       author = {{Isenberg}, Philip A. and {Smith}, Charles W. and {Matthaeus}, William H.},
        title = "{Turbulent Heating of the Distant Solar Wind by Interstellar Pickup Protons}",
      journal = {\apj},
         year = 2003,
        month = jul,
       volume = {592},
       number = {1},
        pages = {564-573},
          doi = {10.1086/375584},
       adsurl = {https://ui.adsabs.harvard.edu/abs/2003ApJ...592..564I}
}

@INCOLLECTION{Koren:1993,
       author = {{Koren}, B. and {van der Maarel}, H.~T.~M.},
        title = "{Monotone, higher-order accurate, multi-dimensional upwinding}",
    booktitle = {Numerical Methods in Fluid Dynamics},
    publisher = "Springer Berlin Heidelberg",
         year = 1993,
       editor = {{Napolitano}, M. and {Sabetta}, F.},
       volume = {414},
        pages = {110-114},
          doi = {10.1007/3-540-56394-6_198},
       adsurl = {https://ui.adsabs.harvard.edu/abs/1993LNP...414..110K}
}

@ARTICLE{Toth:2008,
       author = {{T{\'o}th}, G{\'a}bor and {Ma}, Yingjuan and {Gombosi}, Tamas I.},
        title = "{Hall magnetohydrodynamics on block-adaptive grids}",
      journal = {Journal of Computational Physics},
         year = 2008,
        month = jul,
       volume = {227},
       number = {14},
        pages = {6967-6984},
          doi = {10.1016/j.jcp.2008.04.010},
       adsurl = {https://ui.adsabs.harvard.edu/abs/2008JCoPh.227.6967T}
}

@ARTICLE{Richardson:2022,
       author = {{Richardson}, J.~D. and {Burlaga}, L.~F. and {Elliott}, H. and {Kurth}, W.~S. and {Liu}, Y.~D. and {von Steiger}, R.},
        title = "{Observations of the Outer Heliosphere, Heliosheath, and Interstellar Medium}",
      journal = {\ssr},
         year = 2022,
        month = jun,
       volume = {218},
       number = {4},
          eid = {35},
        pages = {35},
          doi = {10.1007/s11214-022-00899-y},
       adsurl = {https://ui.adsabs.harvard.edu/abs/2022SSRv..218...35R}
}

@ARTICLE{Bale:2005SSRv..118..161B,
       author = {{Bale}, S.~D. and {Balikhin}, M.~A. and {Horbury}, T.~S. and {Krasnoselskikh}, V.~V. and {Kucharek}, H. and {M{\"o}bius}, E. and {Walker}, S.~N. and {Balogh}, A. and {Burgess}, D. and {Lemb{\`e}ge}, B. and {Lucek}, E.~A. and {Scholer}, M. and {Schwartz7 10}, S.~J. and {Thomsen}, M.~F.},
        title = "{Quasi-perpendicular Shock Structure and Processes}",
      journal = {\ssr},
         year = 2005,
        month = jun,
       volume = {118},
       number = {1-4},
        pages = {161-203},
          doi = {10.1007/s11214-005-3827-0},
       adsurl = {https://ui.adsabs.harvard.edu/abs/2005SSRv..118..161B}
}

@ARTICLE{Malama:2006A&A...445..693M,
       author = {{Malama}, Y.~G. and {Izmodenov}, V.~V. and {Chalov}, S.~V.},
        title = "{Modeling of the heliospheric interface: multi-component nature of the heliospheric plasma}",
      journal = {\aap},
         year = 2006,
        month = jan,
       volume = {445},
       number = {2},
        pages = {693-701},
          doi = {10.1051/0004-6361:20053646},
       adsurl = {https://ui.adsabs.harvard.edu/abs/2006A&A...445..693M}
}

@ARTICLE{Yang:2015,
       author = {{Yang}, Zhongwei and {Liu}, Ying D. and {Richardson}, John D. and {Lu}, Quanming and {Huang}, Can and {Wang}, Rui},
        title = "{Impact of Pickup Ions on the Shock Front Nonstationarity and Energy Dissipation of the Heliospheric Termination Shock: Two-dimensional Full Particle Simulations and Comparison with Voyager 2 Observations}",
      journal = {\apj},
         year = 2015,
        month = aug,
       volume = {809},
       number = {1},
          eid = {28},
        pages = {28},
          doi = {10.1088/0004-637X/809/1/28},
archivePrefix = {arXiv},
       eprint = {1506.08492},
 primaryClass = {physics.space-ph},
       adsurl = {https://ui.adsabs.harvard.edu/abs/2015ApJ...809...28Y}
}

@ARTICLE{Gasser:2026,
       author = {{Gasser}, Jonathan and {Sok{\'o}{\l}}, Justyna M. and {Elliott}, Heather A. and {Dayeh}, Maher A. and {Singer}, Kelsi N. and {Stern}, S. Alan and {Baliukin}, Igor and {Korolkov}, Sergey and {Brandt}, Pontus C. and {Parker}, Joel W. and {Poppe}, Andrew R. and {Verbiscer}, Anne J. and {Opher}, Merav},
        title = "{Predictions of New Horizons' termination shock crossing}",
      journal = {Advances in Space Research},
         year = 2026,
        month = jun,
       volume = {77},
       number = {12},
        pages = {12741-12753},
          doi = {10.1016/j.asr.2026.04.074},
       adsurl = {https://ui.adsabs.harvard.edu/abs/2026AdSpR..7712741G}
}

@ARTICLE{2005A&A...439.1171A,
       author = {{Alexashov}, D. and {Izmodenov}, V.},
        title = "{Kinetic vs. multi-fluid models of H atoms in the heliospheric interface: a comparison}",
      journal = {\aap},
         year = 2005,
        month = sep,
       volume = {439},
       number = {3},
        pages = {1171-1181},
          doi = {10.1051/0004-6361:20052821},
       adsurl = {https://ui.adsabs.harvard.edu/abs/2005A&A...439.1171A}
}

@ARTICLE{2011ApJ...734...45A,
       author = {{Alouani-Bibi}, Fathallah and {Opher}, Merav and {Alexashov}, Dimitry and {Izmodenov}, Vladislav and {Toth}, Gabor},
        title = "{Kinetic versus Multi-fluid Approach for Interstellar Neutrals in the Heliosphere: Exploration of the Interstellar Magnetic Field Effects}",
      journal = {\apj},
         year = 2011,
        month = jun,
       volume = {734},
       number = {1},
          eid = {45},
        pages = {45},
          doi = {10.1088/0004-637X/734/1/45},
archivePrefix = {arXiv},
       eprint = {1103.3202},
 primaryClass = {astro-ph.SR},
       adsurl = {https://ui.adsabs.harvard.edu/abs/2011ApJ...734...45A}
}

@ARTICLE{2006JGRA..111.6110H,
       author = {{Heerikhuisen}, J. and {Florinski}, V. and {Zank}, G.~P.},
        title = "{Interaction between the solar wind and interstellar gas: A comparison between Monte Carlo and fluid approaches}",
      journal = {Journal of Geophysical Research (Space Physics)},
         year = 2006,
        month = jun,
       volume = {111},
       number = {A6},
          eid = {A06110},
        pages = {A06110},
          doi = {10.1029/2006JA011604},
       adsurl = {https://ui.adsabs.harvard.edu/abs/2006JGRA..111.6110H}
}

@ARTICLE{2026MNRAS.548ag664K,
       author = {{Korolkov}, Sergey D.},
        title = "{Volumetric and surface current densities in the global heliosphere: implications for magnetic energy storage and release}",
      journal = {\mnras},
         year = 2026,
        month = may,
       volume = {548},
       number = {2},
          eid = {stag664},
        pages = {stag664},
          doi = {10.1093/mnras/stag664},
       adsurl = {https://ui.adsabs.harvard.edu/abs/2026MNRAS.548ag664K}
}

@ARTICLE{2024ApJ...973...46C,
       author = {{Chen}, Yuxi and {Toth}, Gabor and {Powell}, Erick and {Arshad}, Talha and {Bair}, Ethan and {Kornbleuth}, Marc and {Opher}, Merav},
        title = "{A Kinetic-magnetohydrodynamic Model with Adaptive Mesh Refinement for Modeling Heliosphere Neutral-plasma Interaction}",
      journal = {\apj},
         year = 2024,
        month = sep,
       volume = {973},
       number = {1},
          eid = {46},
        pages = {46},
          doi = {10.3847/1538-4357/ad6323},
archivePrefix = {arXiv},
       eprint = {2403.12395},
 primaryClass = {physics.space-ph},
       adsurl = {https://ui.adsabs.harvard.edu/abs/2024ApJ...973...46C}
}

@ARTICLE{Adhikari2023,
       author = {{Adhikari}, Laxman and {Zank}, Gary P. and {Wang}, Bingbing and {Zhao}, Lingling and {Telloni}, Daniele and {Pitna}, Alex and {Opher}, Merav and {Shrestha}, Bishwas and {McComas}, David J. and {Nykyri}, Katariina},
        title = "{Theory and Transport of Nearly Incompressible Magnetohydrodynamic Turbulence: High Plasma Beta Regime}",
      journal = {\apj},
         year = 2023,
        month = aug,
       volume = {953},
       number = {1},
          eid = {44},
        pages = {44},
          doi = {10.3847/1538-4357/acde57},
       adsurl = {https://ui.adsabs.harvard.edu/abs/2023ApJ...953...44A}
}

@ARTICLE{2026ApJ..1008...16V,
       author = {{van der Holst}, Bart and {Szente}, Judit and {Landi}, Enrico},
        title = "{Global Multi-ion Solar Wind Model. I. Ion Temperatures}",
      journal = {\apj},
         year = 2026,
        month = sep,
       volume = {1008},
       number = {1},
          eid = {16},
        pages = {16},
          doi = {10.3847/1538-4357/ae939d},
archivePrefix = {arXiv},
       eprint = {2606.19232},
 primaryClass = {astro-ph.SR},
       adsurl = {https://ui.adsabs.harvard.edu/abs/2026ApJ..1008...16V}
}

@ARTICLE{2009JGRA..11412203G,
       author = {{Glocer}, A. and {T{\'o}th}, G. and {Ma}, Y. and {Gombosi}, T. and {Zhang}, J.-C. and {Kistler}, L.~M.},
        title = "{Multifluid Block-Adaptive-Tree Solar wind Roe-type Upwind Scheme: Magnetospheric composition and dynamics during geomagnetic storms{\textemdash}Initial results}",
      journal = {Journal of Geophysical Research (Space Physics)},
         year = 2009,
        month = dec,
       volume = {114},
       number = {A12},
          eid = {A12203},
        pages = {A12203},
          doi = {10.1029/2009JA014418},
       adsurl = {https://ui.adsabs.harvard.edu/abs/2009JGRA..11412203G}
}

@article{DalMaso:1995,
author    = {Dal Maso, Gianni and LeFloch, Philippe G. and Murat, Fran{\c{c}}ois},
  title     = {Definition and weak stability of nonconservative products},
  journal   = {Journal de Math{\'e}matiques Pures et Appliqu{\'e}es},
  volume    = {74},
  number    = {6},
  pages     = {483--548},
  year      = {1995},
  issn      = {0021-7824}
}
\bibliographystyle{aasjournalv7}

\end{document}